\documentclass[10pt,a4paper]{article}
\usepackage{amssymb}
\usepackage{amsmath}
\usepackage{graphicx}
\usepackage[dvipsnames]{xcolor}
\usepackage{tikz}
\usepackage{xspace}
\usepackage{subcaption}
\usepackage[margin=1in]{geometry}
\usetikzlibrary{shapes.geometric, arrows.meta}
\usetikzlibrary{decorations.pathreplacing}
\usetikzlibrary{calc}

\usepackage{amsthm}

\theoremstyle{definition}
\newtheorem{definition}{Definition}
\theoremstyle{plain}
\newtheorem{theorem}[definition]{Theorem}
\newtheorem{corollary}[definition]{Corollary}
\newtheorem{lemma}[definition]{Lemma}
\newtheorem{problem}[definition]{Problem}

 \definecolor{pincolor}{RGB}{0, 70, 200}
\definecolor{framecolor}{RGB}{0, 140, 70}
\newcommand{\legendpin}[2]{\filldraw[draw = pincolor, fill = pincolor] (#1, #2) circle (.22);}
\providecolor{Magenta}{cmyk}{0, 1, 0, 0}
\newcommand{\uncoveredpin}[2]{
  \filldraw [draw = Magenta, fill = Magenta] (#1, #2) circle (.3);
  \draw [Magenta, uwidth = .3] (#1, #2) circle (.72);}

\definecolor{Ofcolor}{RGB}{253, 225, 0}

\makeatletter
\tikzset{
  uwidth/.code={
\pgfpointxy{1}{0}\pgfmathsetmacro{\@unitlength}{veclen(\the\pgf@x,\the\pgf@y)}\pgfkeysalso{line width = #1 * \@unitlength pt}}
}
\makeatother

\makeatletter
\newcommand{\cell}[4]{\filldraw [line width = .125\pgf@xx, draw = #3, fill = #4] (#1 - .4375, #2 - .4375) rectangle ++ (.875, .875);}
\makeatother

\newcommand{\whitebg}[4]{
    \pgfmathsetmacro{\xmin}{#1 - 0.5}
    \pgfmathsetmacro{\xmax}{#2 + 0.5}
    \pgfmathsetmacro{\ymin}{#3 - 0.5}
    \pgfmathsetmacro{\ymax}{#4 + 0.5}

    \fill[white] (\xmin, \ymin) rectangle (\xmax, \ymax);

    \pgfmathsetmacro{\xinc}{\xmin + 1}
    \pgfmathsetmacro{\yinc}{\ymin + 1}
    \foreach \x in {\xmin, \xinc, ..., \xmax}{
        \draw [darkgray, opacity = .3] (\x, \ymin) -- (\x, \ymax);
    }
    \foreach \y in {\ymin, \yinc, ..., \ymax}{
        \draw [darkgray, opacity = .3] (\xmin, \y) -- (\xmax, \y);
    }
}

\newcommand{\filler}[2]{
    \cell{#1}{#2}{darkgray}{gray}
}

\newcommand{\omino}[3]{
    \ifnum #3>2
        \def\dcolor{black}
        \def\fcolor{Ofcolor}
    \else \ifnum #3>1
        \def\dcolor{gray}
        \def\fcolor{olive!40}
    \else \ifnum #3>0
        \def\dcolor{violet}
        \def\fcolor{yellow}
    \else
        \def\dcolor{teal}
        \def\fcolor{lime}
    \fi\fi\fi

    \ifnum #3=3
\begin{scope}
            \clip (#1 - 0.5, #2 - 0.5) rectangle (#1 + 1.5, #2 + 1.5);
            \cell{#1}{#2}{gray}{\fcolor}
            \cell{#1 + 1}{#2}{gray}{\fcolor}
            \cell{#1}{#2 + 1}{gray}{\fcolor}
            \cell{#1 + 1}{#2 + 1}{gray}{\fcolor}
            \draw [draw = \dcolor, line width = 8pt] (#1 - 0.5, #2 - 0.5) -- (#1 + 1.5, #2 - 0.5) -- (#1 + 1.5, #2 + 1.5) -- (#1 - 0.5, #2 + 1.5) -- cycle;
        \end{scope}
    \else
\filldraw [draw = \dcolor, fill = \fcolor, uwidth = .2] (#1 - .4, #2 - .4) rectangle ++ (1.8, 1.8);
        \draw [draw = gray, opacity = .5, uwidth = .05] (#1 + .5, #2 - .5) -- ++ (0, 2);
        \draw [draw = gray, opacity = .5, uwidth = .05] (#1 - .5, #2 + .5) -- ++ (2, 0);
    \fi
}

\newcommand{\varblockframe}[4]{{
    \pgfmathsetmacro{\x}{#1}
    \pgfmathsetmacro{\y}{#2}
    \pgfmathsetmacro{\left}{#3}
    \pgfmathsetmacro{\right}{#4}

    \filler{\x}{\y}
    \filler{\x - 3}{\y}
    \filler{\x + 3}{\y}

    \filler{\x}{\y + 5}
    \filler{\x - 3}{\y + 5}
    \filler{\x + 3}{\y + 5}

    \filler{\x - 4}{\y + 1}
    \filler{\x + 4}{\y + 1}

    \ifnum \left>0
        \foreach \i in {1, ..., \left}{
            \filler{\x - 2 * \i - 3}{\y + 2}
            \filler{\x - 2 * \i - 3}{\y + 5}
            \filler{\x - 2 * \i - 2}{\y + 2}
            \filler{\x - 2 * \i - 2}{\y + 5}
        }
    \else
        \foreach \i in {1, ..., 4}{
            \filler{\x - 4}{\y + \i}
        }
    \fi
    \ifnum \right>0
        \foreach \i in {1, ..., \right}{
            \filler{\x + 2 * \i + 3}{\y + 2}
            \filler{\x + 2 * \i + 3}{\y + 5}
            \filler{\x + 2 * \i + 2}{\y + 2}
            \filler{\x + 2 * \i + 2}{\y + 5}
        }
    \else
        \foreach \i in {1, ..., 4}{
            \filler{\x + 4}{\y + \i}
        }
    \fi

}}

\newcommand{\varblock}[5]{{
    \pgfmathsetmacro{\x}{#1}
    \pgfmathsetmacro{\y}{#2}
    \pgfmathsetmacro{\left}{int(#3)}
    \pgfmathsetmacro{\right}{int(#4)}
    \pgfmathsetmacro{\bool}{#5}

    \varblockframe{\x}{\y}{\left}{\right}
    \ifnum \bool=0
        \omino{\x - 2}{\y}{1}
        \omino{\x }{\y + 1}{1}
        \omino{\x + 2}{\y + 1}{1}
        \omino{\x - 2}{\y + 2}{1}
        \omino{\x }{\y + 3}{1}
        \omino{\x + 2}{\y + 3}{1}
        \omino{\x - 2}{\y + 4}{1}
        \omino{\x + 1}{\y + 5}{1}

        \ifnum \left>0
            \foreach \i in {1, ..., \left}{
                \omino{\x - 2 * \i - 2}{\y + 3}{1}
            }
        \fi
        \ifnum \right>0
            \foreach \i in {1, ..., \right}{
                \omino{\x + 2 * \i + 2}{\y + 3}{1}
            }
        \fi

    \else\ifnum \bool=1
        \omino{\x + 1}{\y}{1}
        \omino{\x - 3}{\y + 1}{1}
        \omino{\x - 1}{\y + 1}{1}
        \omino{\x + 1}{\y + 2}{1}
        \omino{\x - 3}{\y + 3}{1}
        \omino{\x - 1}{\y + 3}{1}
        \omino{\x + 1}{\y + 4}{1}
        \omino{\x - 2}{\y + 5}{1}

        \ifnum \left > 0
            \foreach \i in {1, ..., \left}{
                \omino{\x - 2 * \i - 3}{\y + 3}{1}
            }
        \fi
        \ifnum \right > 0
            \foreach \i in {1, ..., \right}{
                \omino{\x + 2 * \i + 1}{\y + 3}{1}
            }
        \fi
    \else
    \fi\fi
}}

\newcommand{\varbot}[3]{{
    \pgfmathsetmacro{\x}{#1}
    \pgfmathsetmacro{\y}{#2}
    \pgfmathsetmacro{\bool}{#3}

    \filler{\x - 3}{\y - 1}
    \filler{\x + 3}{\y - 1}
    \filler{\x - 3}{\y - 2}
    \filler{\x + 3}{\y - 2}
    \filler{\x - 2}{\y - 3}
    \filler{\x + 2}{\y - 3}

    \ifnum \bool = 1
        \omino{\x + 1}{\y - 2}{1}
        \omino{\x - 1}{\y - 3}{1}
        \omino{\x - 2}{\y - 1}{1}
    \else \ifnum \bool = 0
        \omino{\x - 2}{\y - 2}{1}
        \omino{\x }{\y - 3}{1}
        \omino{\x + 1}{\y - 1}{1}
    \fi\fi
}}

\newcommand{\vartop}[3]{{
    \pgfmathsetmacro{\x}{#1}
    \pgfmathsetmacro{\y}{#2}
    \pgfmathsetmacro{\bool}{#3}

    \filler{\x - 3}{\y}
    \filler{\x + 3}{\y}
    \filler{\x - 3}{\y + 1}
    \filler{\x + 3}{\y + 1}
    \filler{\x - 2}{\y + 2}
    \filler{\x + 2}{\y + 2}

    \ifnum \bool=1
        \omino{\x - 1}{\y + 1}{1}
        \omino{\x + 1}{\y}{1}
    \else\ifnum \bool=0
        \omino{\x - 2}{\y}{1}
        \omino{\x }{\y + 1}{1}
    \fi\fi
}}

\newcommand{\var}[4]{{
    \pgfmathsetmacro{\x}{int(#1)}
    \pgfmathsetmacro{\y}{int(#2)}
    \pgfmathsetmacro{\bool}{int(#3)}
    \def\widthlist{#4}

    \varbot{\x}{\y}{\bool}
    \xdef\h{\y}
    \foreach \l/\r in #4{

            \varblock{\x}{\h}{\l}{\r}{\bool}

            \pgfmathsetmacro{\tmph}{\h + 6}
            \xdef\h{\tmph}
    }
    \vartop{\x}{\h}{\bool}

}}

\newcommand{\vars}[2]{{
    \pgfmathsetmacro{\n}{int(#1)}

    \xdef\x{0}
    \xdef\y{0}

    \foreach \bool/\w in{#2}{
        \var{\x}{\y}{\bool}{\w}

        \foreach \l/\r in \w{
            \pgfmathsetmacro{\tmpy}{\y + 6}
            \xdef\y{\tmpy}
        }

        \pgfmathtruncatemacro{\tmpy}{\y + 6}
        \xdef\y{\tmpy}
    }
}}

\newcommand{\clauseblockframe}[6]{{
    \pgfmathsetmacro{\i}{int(#1)}
    \pgfmathsetmacro{\h}{int(#2)}
    \pgfmathsetmacro{\lr}{int(#3)}
    \pgfmathsetmacro{\pos}{int(#4)}
    \pgfmathsetmacro{\previ}{int(#5)}
    \pgfmathsetmacro{\prevh}{int(#6)}

    \ifnum \lr<0
        \pgfmathsetmacro{\x}{2 * \lr - 4}
        \pgfmathsetmacro{\y}{\i * 6 * 4 + \h * 6 + 2}

        \filler{\x}{\y}
        \filler{\x - 4}{\y + 1}

        \filler{\x }{\y + 3}
        \filler{\x - 1}{\y + 3}
        \filler{\x - 2}{\y + 3}

        \ifnum \pos=0
            \filler{\x - 1}{\y}
            \filler{\x - 2}{\y}
            \filler{\x - 3}{\y}
        \else
            \pgfmathsetmacro{\prevy}{int(\previ * 6 * 4 + \prevh * 6 + 12)}

            \filler{\x - 4}{\y}

            \foreach \i in {\y, ..., \prevy}{
                \filler{\x}{\i - 1}
                \filler{\x - 3}{\i - 1}
            }

            \filler{\x}{\prevy - 2}
            \filler{\x}{\prevy - 3}
            \filler{\x - 4}{\prevy - 2}
            \filler{\x - 4}{\prevy - 3}
            \filler{\x - 5}{\prevy - 3}
            \filler{\x - 1}{\prevy - 4}
            \filler{\x - 1}{\prevy - 5}
            \filler{\x - 1}{\prevy - 6}
            \filler{\x - 6}{\prevy - 4}
            \filler{\x - 6}{\prevy - 5}
            \filler{\x - 6}{\prevy - 6}
            \filler{\x - 5}{\prevy - 7}
            \filler{\x - 5}{\prevy - 8}
        \ifnum \pos=2
            \filler{\x - 3}{\y + 2}
        \fi\fi
    \else
        \pgfmathsetmacro{\x}{2 * \lr + 4}
        \pgfmathsetmacro{\y}{\i * 6 * 4 + \h * 6 + 2}

        \filler{\x}{\y}
        \filler{\x + 4}{\y + 1}

        \filler{\x }{\y + 3}
        \filler{\x + 1}{\y + 3}
        \filler{\x + 2}{\y + 3}

        \ifnum \pos=0
            \filler{\x + 1}{\y}
            \filler{\x + 2}{\y}
            \filler{\x + 3}{\y}
        \else
            \pgfmathsetmacro{\prevy}{int(\previ * 6 * 4 + \prevh * 6 + 12)}

            \filler{\x + 4}{\y}

            \foreach \i in {\y, ..., \prevy}{
                \filler{\x}{\i - 1}
                \filler{\x + 3}{\i - 1}
            }

            \filler{\x}{\prevy - 2}
            \filler{\x}{\prevy - 3}
            \filler{\x + 4}{\prevy - 2}
            \filler{\x + 4}{\prevy - 3}
            \filler{\x + 5}{\prevy - 3}
            \filler{\x + 1}{\prevy - 4}
            \filler{\x + 1}{\prevy - 5}
            \filler{\x + 1}{\prevy - 6}
            \filler{\x + 6}{\prevy - 4}
            \filler{\x + 6}{\prevy - 5}
            \filler{\x + 6}{\prevy - 6}
            \filler{\x + 5}{\prevy - 7}
            \filler{\x + 5}{\prevy - 8}
        \ifnum \pos=2
            \filler{\x + 3}{\y + 2}
        \fi\fi
    \fi
}}

\newcommand{\clauseblock}[7]{{
    \pgfmathsetmacro{\i}{int(#1)}
    \pgfmathsetmacro{\h}{int(#2)}
    \pgfmathsetmacro{\lr}{int(#3)}
    \pgfmathsetmacro{\pos}{int(#4)}
    \pgfmathsetmacro{\previ}{int(#5)}
    \pgfmathsetmacro{\prevh}{int(#6)}
    \pgfmathsetmacro{\state}{int(#7)}

    \clauseblockframe{\i}{\h}{\lr}{\pos}{\previ}{\prevh}

    \ifnum \lr < 0
        \pgfmathsetmacro{\x}{2 * \lr - 5}
        \pgfmathsetmacro{\y}{\i * 6 * 4 + \h * 6 + 2}

        \ifnum \state=0
            \omino{\x + 1}{\y + 1}{1}
            \omino{\x - 1}{\y}{1}
        \fi \ifnum \state=1
            \omino{\x}{\y + 1}{1}
            \omino{\x - 2}{\y}{1}
        \fi \ifnum \state=2
            \omino{\x + 1}{\y + 1}{1}
            \omino{\x - 1}{\y + 1}{1}
        \fi \ifnum \state=3
            \omino{\x}{\y + 1}{1}
            \omino{\x - 2}{\y + 1}{1}
        \fi

        \ifnum \pos>0
        \pgfmathsetmacro{\prevy}{int(\previ * 6 * 4 + \prevh * 6 + 11)}
            \ifnum \state<4
                \ifnum \state>1
                    \foreach \i in {\y, ..., \prevy}{
                        \ifodd \i
                            \omino{\x - 1}{\i}{1}
                        \fi
                    }
                    \omino{\x - 2}{\prevy - 4}{1}
                    \omino{\x - 4}{\prevy - 4}{1}
                    \omino{\x - 3}{\prevy - 6}{1}
                    \omino{\x - 2}{\prevy - 2}{1}

                \else
                    \foreach \i in {\y, ..., \prevy}{
                        \ifodd \i
                            \omino{\x - 1}{\i - 1}{1}
                        \fi
                    }

                    \omino{\x - 2}{\prevy - 5}{1}
                    \omino{\x - 4}{\prevy - 5}{1}
                    \omino{\x - 3}{\prevy - 7}{1}
                    \omino{\x - 2}{\prevy - 3}{1}
                \fi
            \fi
        \fi
    \else
        \pgfmathsetmacro{\x}{2 * \lr + 4}
        \pgfmathsetmacro{\y}{\i * 6 * 4 + \h * 6 + 2}

        \ifnum \state=0
            \omino{\x - 1}{\y + 1}{1}
            \omino{\x + 1}{\y}{1}
        \fi \ifnum \state=1
            \omino{\x}{\y + 1}{1}
            \omino{\x + 2}{\y}{1}
        \fi \ifnum \state=2
            \omino{\x - 1}{\y + 1}{1}
            \omino{\x + 1}{\y + 1}{1}
        \fi \ifnum \state=3
            \omino{\x}{\y + 1}{1}
            \ifnum \pos<2
                \omino{\x + 2}{\y + 1}{1}
            \fi
        \fi

        \ifnum \pos>0
        \pgfmathsetmacro{\prevy}{int(\previ * 6 * 4 + \prevh * 6 + 11)}
            \ifnum \state<4
                \ifnum \state>1
                    \foreach \i in {\y, ..., \prevy}{
                        \ifodd \i
                            \omino{\x + 1}{\i}{1}
                        \fi
                    }
                    \omino{\x + 2}{\prevy - 4}{1}
                    \omino{\x + 4}{\prevy - 4}{1}
                    \omino{\x + 3}{\prevy - 6}{1}
                    \omino{\x + 2}{\prevy - 2}{1}

                \else
                    \foreach \i in {\y, ..., \prevy}{
                        \ifodd \i
                            \omino{\x + 1}{\i - 1}{1}
                        \fi
                    }

                    \omino{\x + 2}{\prevy - 5}{1}
                    \omino{\x + 4}{\prevy - 5}{1}
                    \omino{\x + 3}{\prevy - 7}{1}
                    \omino{\x + 2}{\prevy - 3}{1}
                \fi
            \fi
        \fi
    \fi
}}

\newcommand{\ilo}{\resizebox{10pt}{!}{\begin{tikzpicture}[baseline = {(0, -0.2)}]
        \omino{0}{0}{3}
\end{tikzpicture}}\xspace}
\newcommand{\True}{\textup{\textsc{True}}\xspace}
\newcommand{\False}{\textup{\textsc{False}}\xspace}

\usepackage{titlesec}

\makeatletter

\def\ps@headings{
	\def\@oddhead{		}
	\def\@evenhead{}
	\def\@evenfoot{\hfil \,\,\thepage\hfill}
	\def\@oddfoot{\hfil \,\,\thepage\hfil}
}

\makeatother

\title{
	 \textbf{Tetris Survival is Hard with Only O Pieces}
	}
\author{
	Takehiro Ishibashi\thanks{Tohoku University} \and
	Ryo Yoshinaka\footnotemark[1] \and
	Ayumi Shinohara\footnotemark[1]
}
\date{\empty}

\begin{document}
\maketitle

\begin{abstract}
    We prove that Tetris survival is NP-complete when the piece sequence consists only of \ilo~pieces.
    This answers an open problem left by the MIT Hardness Group~\cite{MBDHL26}, who proved NP-hardness of Tetris clearing and survival for each of the other six piece types.
    Their proofs rely on rotating pieces, but \ilo~pieces do not rotate, so we need a different construction.
    Like Abrahamsen and Stade~\cite{AS26} for packing $2 \times 2$ squares, we reduce from a planar monotone variant of SAT.
    Unlike in packing, we must also show an order of placing the pieces in which every piece can reach its position from the top of the board.
\end{abstract}
 \section{Introduction}
\label{sec:intro}

Tetris was created by Alexey Pajitnov in 1984.
Breukelaar et al.~\cite{BDH04} studied the complexity of Tetris as a perfect-information one-player game: the player is given an initial board with some filled cells and the entire sequence of pieces in advance.
Such initial boards are not an artificial assumption: essentially every configuration can be built from an empty board by a suitable sequence of pieces~\cite{HK04}.
The two objectives that have received the most attention are \emph{Tetris clearing}, emptying the board entirely, and \emph{Tetris survival}, placing every piece of the sequence without stacking above the top of the board.
Both are NP-complete, and various associated optimization problems are hard to approximate~\cite{BDH04}; hardness persists with only $8$ columns or $4$ rows~\cite{ACD20}, with unrotatable $k$-ominoes for small $k$~\cite{DDE17}, and when the pieces are restricted to any two of the seven types~\cite{MDHL24}.

Among these parameters, we focus on the set of piece types given to the player.
The original proof uses five of the seven tetrominoes~\cite[Section~6.2]{BDH04}.
Twenty years later this was reduced to any two of the seven~\cite{MDHL24}, and then to a single type~\cite{MBDHL26} -- for every type except \ilo.
The proofs in~\cite{MBDHL26} rely on rotating pieces, but \ilo~pieces do not rotate.

The MIT Hardness Group~\cite{MBDHL26} names the complexity of clearing and of survival with \ilo~pieces as the main remaining open problems for tetromino piece types.
For clearing, the authors expect a polynomial-time algorithm, since \ilo~pieces do not rotate, but note that the row-clearing mechanic makes this awkward to prove.
For survival, they conjecture NP-hardness, because survival is close to a packing problem that is NP-hard.

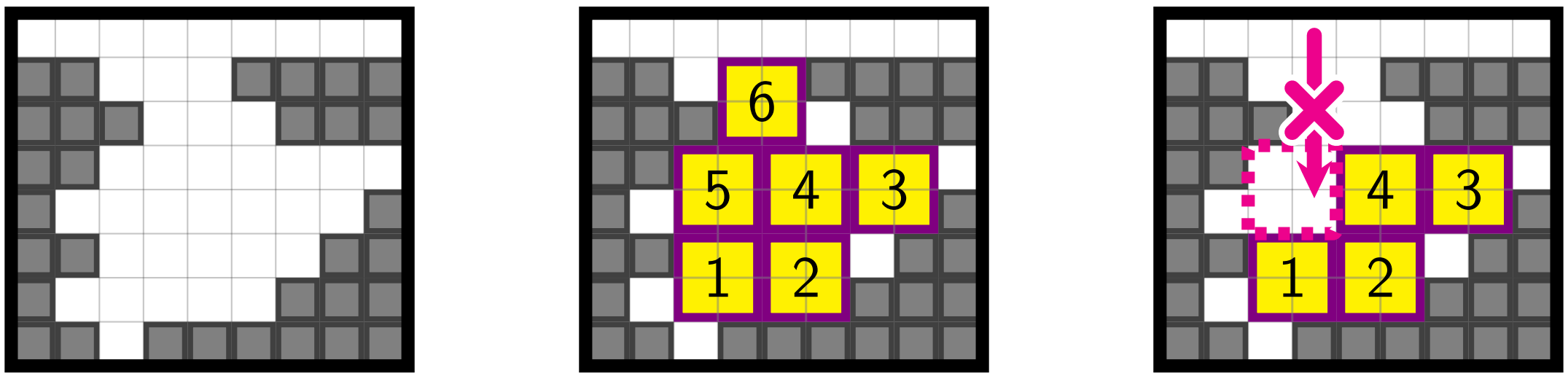
\begin{figure}[t]
\centering
\newcommand{\notpackingboard}{
  \whitebg{-2}{6}{-1}{6}
  \filler{-2}{-1}\filler{-1}{-1}\filler{1}{-1}\filler{2}{-1}\filler{3}{-1}
  \filler{4}{-1}\filler{5}{-1}\filler{6}{-1}\filler{-2}{0}\filler{4}{0}
  \filler{5}{0}\filler{6}{0}\filler{-2}{1}\filler{-1}{1}\filler{5}{1}
  \filler{6}{1}\filler{-2}{2}\filler{6}{2}\filler{-2}{3}\filler{-1}{3}
  \filler{-2}{4}\filler{-1}{4}\filler{0}{4}\filler{4}{4}\filler{5}{4}
  \filler{6}{4}\filler{-2}{5}\filler{-1}{5}\filler{3}{5}\filler{4}{5}
  \filler{5}{5}\filler{6}{5}
  \draw [draw = black, uwidth = .3] (-2.5, -1.5) rectangle ++ (9, 8);
}
\begin{subfigure}[b]{0.32\linewidth}
  \centering
\begin{tikzpicture}[x = .4cm, y = .4cm]
    \notpackingboard
  \end{tikzpicture}
  \caption{The initial board.}
  \label{fig:notpacking-initial}
\end{subfigure}
\begin{subfigure}[b]{0.32\linewidth}
  \centering
\begin{tikzpicture}[x = .4cm, y = .4cm]
    \notpackingboard

    \omino{0}{0}{1}
    \omino{2}{0}{1}
    \omino{0}{2}{1}
    \omino{2}{2}{1}
    \omino{4}{2}{1}
    \omino{1}{4}{1}

    \node [scale = 1.5, font = \sffamily] at (.5, .5) {1};
    \node [scale = 1.5, font = \sffamily] at (2.5, .5) {2};
    \node [scale = 1.5, font = \sffamily] at (0.5, 2.5) {5};
    \node [scale = 1.5, font = \sffamily] at (2.5, 2.5) {4};
    \node [scale = 1.5, font = \sffamily] at (4.5, 2.5) {3};
    \node [scale = 1.5, font = \sffamily] at (1.5, 4.5) {6};
  \end{tikzpicture}
  \caption{A packing of six $2 \times 2$ squares.}
  \label{fig:notpacking-packing}
\end{subfigure}
\begin{subfigure}[b]{0.32\linewidth}
  \centering
\begin{tikzpicture}[x = .4cm, y = .4cm]
    \notpackingboard

    \omino{0}{0}{1}
    \node [scale = 1.5, font = \sffamily] at (.5, .5) {1};
    \omino{2}{0}{1}
    \node [scale = 1.5, font = \sffamily] at (2.5, .5) {2};
    \omino{4}{2}{1}
    \node [scale = 1.5, font = \sffamily] at (4.5, 2.5) {3};
    \omino{2}{2}{1}
    \node [scale = 1.5, font = \sffamily] at (2.5, 2.5) {4};

\draw [Magenta, uwidth = .3, dashed, rounded corners = 1pt]
      (-.5, 1.5) rectangle ++ (2, 2);

\draw [Magenta, uwidth = .34, line cap = round,
           -{Stealth[length = 3.4mm, width = 3.2mm]}] (1, 6.0) -- (1, 2.3);

\draw [white, uwidth = .62, line cap = round] (.5, 4.8) -- (1.5, 3.8);
    \draw [white, uwidth = .62, line cap = round] (.5, 3.8) -- (1.5, 4.8);
    \draw [Magenta, uwidth = .35, line cap = round] (.5, 4.8) -- (1.5, 3.8);
    \draw [Magenta, uwidth = .35, line cap = round] (.5, 3.8) -- (1.5, 4.8);
  \end{tikzpicture}
  \caption{Piece 4 blocks piece 5.}
  \label{fig:notpacking-board}
\end{subfigure}
\caption{Packing and playing on the board~\subref{fig:notpacking-initial}: grey cells are filled and white cells are empty.
Six $2 \times 2$ squares can be packed~\subref{fig:notpacking-packing}, but no play places every piece, because pieces~$4$ and~$5$ block each other's path~\subref{fig:notpacking-board}.}
\label{fig:notpacking}
\end{figure}

The packing problem asks whether a given number of $2 \times 2$ squares can be placed on the empty cells of a board without overlap.
It is NP-hard~\cite{BJLSS90}, even when the empty cells form an orthogonally convex polygon~\cite{AS26}.
If no row is cleared during a play, the pieces placed in the play form such a packing.
However, not every packing comes from a play, because each piece must reach its position from the top of the board through cells that are still empty.
The MIT Hardness Group regards this difference as the main obstacle.
Figure~\ref{fig:notpacking} shows a board whose empty cells can be packed, but on which no play places all the pieces.

This paper proves the conjecture: Tetris survival is NP-complete even if every piece in the sequence is an \ilo~piece.
From a formula of a variant of SAT, we build a board on which no row is ever cleared and a number~$N$ such that $N$ squares of size $2 \times 2$ can be packed into its empty cells if and only if the formula is satisfiable; this also gives another proof that the packing problem is NP-hard.
For a satisfying assignment, we further show that the corresponding packing of $N$ pieces can be realized by a play.

Section~\ref{sec:prelim} fixes the Tetris model and defines the source problem.
Section~\ref{sec:otris} contains the reduction.
Section~\ref{sec:open} lists open problems.

 \section{Preliminaries}
\label{sec:prelim}

\subsection{The Tetris model}
\label{sec:model}

A \emph{board} of width $W$ and height $H$ is a grid of $W \times H$ cells, each either \emph{filled} or \emph{empty}.
The player is given a sequence of pieces;
this paper uses only the \ilo~piece, which occupies a $2 \times 2$ block of cells.
A piece spawns at a designated position at the top of the board.
The player may repeatedly move it one cell left, one cell right, or one cell down, provided the destination cells are empty.
When the player attempts to move the piece down and one of the cells below is filled, the piece \emph{locks}: its four cells become filled.
Every completely filled row is then cleared, and each row above a cleared row moves down by the number of cleared rows below it.
Then the next piece of the sequence spawns.
The board has a designated row, called the \emph{losing row}.
The player \emph{loses} if a piece locks with any of its cells in or above the losing row.


\begin{problem}[\ilo-\textsc{Survival}]
\label{prob:survival}
Given a board $B$, with its losing row, and an integer $N$ written in unary, determine whether the player can place $N$ consecutive \ilo~pieces on $B$ without losing.
\end{problem}


\ilo-\textsc{Survival} is in NP, like Tetris in general~\cite{BDH04,ACD20}.

 \subsection{Monotone Rectilinear SAT}
\label{sec:mrsat}

We reduce from the following problem, which de Berg and Khosravi~\cite{dBK12} showed to be NP-complete.

\begin{problem}[Monotone Rectilinear SAT]
\label{prob:mrsat}
Given a CNF formula $\phi$ with $n$ variables and $m$ clauses, each of which is \emph{positive} (all its literals are positive) or \emph{negative} (all its literals are negative), and an arrangement of $\phi$ in the plane of the following form:
\begin{enumerate}
\item every variable is a vertical segment on the $y$-axis;
\item every clause is a vertical segment, to the right of the $y$-axis if it is positive and to the left if it is negative;
\item every clause is joined to each of its variables by a horizontal edge;
\item segments and edges intersect only at the ends of edges;
\end{enumerate}
determine whether there is a truth assignment under which every clause of $\phi$ has at least one true literal.
\end{problem}
Our definition rotates the drawing of~\cite{dBK12} by a quarter turn.
Without loss of generality, we assume that every clause contains at least two literals.

We illustrate the construction throughout on
\begin{equation}
\label{eq:running}
  \phi = (v_1 \vee v_2 \vee v_4 \vee v_5)
  \wedge (v_2 \vee v_3 \vee v_4)
  \wedge (\neg v_1 \vee \neg v_2)
  \wedge (\neg v_3 \vee \neg v_4 \vee \neg v_5),
\end{equation}
whose arrangement is shown in Figure~\ref{fig:mrsat}.
We write $c_1^+, c_2^+$ for its positive clauses and $c_1^-, c_2^-$ for its negative ones, in the order in which they appear in~\eqref{eq:running}.

 \section{Survival with O pieces}
\label{sec:otris}

This section proves our main theorem.

\begin{theorem}
\label{thm:main}
Tetris survival is NP-complete even if only \ilo~pieces are given to the player.
\end{theorem}

Since \ilo-\textsc{Survival} is in NP, it remains to reduce Monotone Rectilinear SAT, Problem~\ref{prob:mrsat}, to \ilo-\textsc{Survival}, Problem~\ref{prob:survival}.
Section~\ref{sec:structure} lays out the board and fixes the number $N$ of pieces.
The gadgets named there are described in Sections~\ref{sec:vargadget}--\ref{sec:clause}, and Section~\ref{sec:correctness} proves that the reduction is correct.

\subsection{General structure}
\label{sec:structure}

\begin{figure}[tp]
\centering
\begin{subfigure}[t]{0.27\linewidth}
  \centering
  \begin{tikzpicture}[x = .16cm, y = .16cm]
\path (-10, -5.5) rectangle (15, 69.5);
    \foreach \y/\xa/\xb in {3.5/2/11, 15.5/2/11, 51.5/2/11, 63.5/2/11,
                            21.5/2/6, 33.5/2/6, 45.5/2/6,
                            3.5/-2/-6, 21.5/-2/-6, 33.5/-2/-6, 45.5/-2/-6, 63.5/-2/-6}{
        \draw [line width = 2.2pt, draw = orange] (\xa, \y) -- (\xb, \y);
    }
    \foreach \i/\b/\t in {1/-3/8, 2/9/26, 3/27/38, 4/39/56, 5/57/68}{
        \filldraw [draw = teal, fill = teal!50, line width = 1.2pt]
          (-2, \b + .2) rectangle (2, \t - .2);
        \node [white, scale = .9] at (0, {(\b + \t) / 2}) {$v_{\i}$};
    }
    \filldraw [draw = blue, fill = cyan!50, line width = 1.2pt] (11, 2.3) rectangle (14, 64.7);
    \node [blue, scale = .8] at (12.5, 33.5) {$c_1^+$};
    \filldraw [draw = blue, fill = cyan!50, line width = 1.2pt] (6, 20.3) rectangle (9, 46.7);
    \node [blue, scale = .8] at (7.5, 33.5) {$c_2^+$};
    \filldraw [draw = red, fill = red!50, line width = 1.2pt] (-9, 2.3) rectangle (-6, 22.7);
    \node [white, scale = .8] at (-7.5, 12.5) {$c_1^-$};
    \filldraw [draw = red, fill = red!50, line width = 1.2pt] (-9, 32.3) rectangle (-6, 64.7);
    \node [white, scale = .8] at (-7.5, 48.5) {$c_2^-$};
  \end{tikzpicture}
  \caption{The arrangement.}
  \label{fig:mrsat}
\end{subfigure}
\hspace{0.02\linewidth}
\begin{subfigure}[t]{0.62\linewidth}
  \centering
  \begin{tikzpicture}[x = .16cm, y = .16cm]
      \pgfmathsetmacro{\xmin}{- 15}
      \pgfmathsetmacro{\xmax}{  20}
      \pgfmathsetmacro{\ymin}{0 - 4}
\pgfmathsetmacro{\ymax}{6 * (2 + 3 + 2 + 3 + 2) - 4}

      \begin{scope}
\whitebg{\xmin}{\xmax + 3}{\ymin}{\ymax}
          \clip (\xmin, \ymin) rectangle (\xmax, \ymax);

\def\wireregions{orange/orange/6/13/2/5, orange/orange/6/13/14/17,
            orange/orange/6/13/50/53, orange/orange/6/13/62/65,
            orange/orange/-7/-6/2/5, orange/orange/-7/-6/20/23, orange/orange/-7/-6/32/35,
            orange/orange/-7/-6/44/47, orange/orange/-7/-6/62/65}
          \def\gadgetregions{teal/teal/-5/5/-3/8, teal/teal/-5/5/9/26, teal/teal/-5/5/27/38,
            teal/teal/-5/5/39/56, teal/teal/-5/5/57/68,
            cyan/blue/14/19/2/65, cyan/blue/6/11/20/47,
            red/red/-13/-8/2/23, red/red/-13/-8/32/65}
          \foreach \c/\d/\xa/\xb/\ya/\yb in \wireregions {
              \fill [\c, opacity = .15] (\xa - .5, \ya - .5) rectangle (\xb + .5, \yb + .5);
          }
          \foreach \c/\d/\xa/\xb/\ya/\yb in \gadgetregions {
              \fill [\c, opacity = .15] (\xa - .5, \ya - .5) rectangle (\xb + .5, \yb + .5);
          }

\filler{-15}{67}
\filler{-14}{67}
\filler{-13}{67}
\filler{-12}{67}
\filler{-11}{67}
\filler{-10}{67}
\filler{-9}{67}
\filler{-8}{67}
\filler{-7}{67}
\filler{-6}{67}
\filler{-5}{67}
\filler{-4}{67}
\filler{4}{67}
\filler{5}{67}
\filler{6}{67}
\filler{7}{67}
\filler{8}{67}
\filler{9}{67}
\filler{10}{67}
\filler{11}{67}
\filler{12}{67}
\filler{13}{67}
\filler{14}{67}
\filler{15}{67}
\filler{16}{67}
\filler{17}{67}
\filler{18}{67}
\filler{19}{67}
\filler{20}{67}
\filler{21}{67}
\filler{22}{67}
\filler{-15}{66}
\filler{-14}{66}
\filler{-13}{66}
\filler{-12}{66}
\filler{-11}{66}
\filler{-10}{66}
\filler{-9}{66}
\filler{-8}{66}
\filler{-7}{66}
\filler{-6}{66}
\filler{-5}{66}
\filler{-4}{66}
\filler{4}{66}
\filler{5}{66}
\filler{6}{66}
\filler{7}{66}
\filler{8}{66}
\filler{9}{66}
\filler{10}{66}
\filler{11}{66}
\filler{12}{66}
\filler{13}{66}
\filler{14}{66}
\filler{15}{66}
\filler{16}{66}
\filler{17}{66}
\filler{18}{66}
\filler{19}{66}
\filler{20}{66}
\filler{21}{66}
\filler{22}{66}
\filler{-15}{65}
\filler{-14}{65}
\filler{-13}{65}
\filler{-12}{65}
\filler{-11}{65}
\filler{17}{65}
\filler{18}{65}
\filler{19}{65}
\filler{20}{65}
\filler{21}{65}
\filler{22}{65}
\filler{-15}{64}
\filler{-14}{64}
\filler{-13}{64}
\filler{-12}{64}
\filler{18}{64}
\filler{19}{64}
\filler{20}{64}
\filler{21}{64}
\filler{22}{64}
\filler{-15}{63}
\filler{-14}{63}
\filler{-13}{63}
\filler{19}{63}
\filler{20}{63}
\filler{21}{63}
\filler{22}{63}
\filler{-15}{62}
\filler{-14}{62}
\filler{-13}{62}
\filler{19}{62}
\filler{20}{62}
\filler{21}{62}
\filler{22}{62}
\filler{-15}{61}
\filler{-14}{61}
\filler{-13}{61}
\filler{-12}{61}
\filler{-7}{61}
\filler{-6}{61}
\filler{-5}{61}
\filler{5}{61}
\filler{6}{61}
\filler{7}{61}
\filler{8}{61}
\filler{9}{61}
\filler{10}{61}
\filler{11}{61}
\filler{12}{61}
\filler{13}{61}
\filler{18}{61}
\filler{19}{61}
\filler{20}{61}
\filler{21}{61}
\filler{22}{61}
\filler{-15}{60}
\filler{-14}{60}
\filler{-13}{60}
\filler{-12}{60}
\filler{-7}{60}
\filler{-6}{60}
\filler{-5}{60}
\filler{-4}{60}
\filler{4}{60}
\filler{5}{60}
\filler{6}{60}
\filler{7}{60}
\filler{8}{60}
\filler{9}{60}
\filler{10}{60}
\filler{11}{60}
\filler{12}{60}
\filler{13}{60}
\filler{18}{60}
\filler{19}{60}
\filler{20}{60}
\filler{21}{60}
\filler{22}{60}
\filler{-15}{59}
\filler{-14}{59}
\filler{-13}{59}
\filler{-12}{59}
\filler{-7}{59}
\filler{-6}{59}
\filler{-5}{59}
\filler{-4}{59}
\filler{4}{59}
\filler{5}{59}
\filler{6}{59}
\filler{7}{59}
\filler{8}{59}
\filler{9}{59}
\filler{10}{59}
\filler{11}{59}
\filler{12}{59}
\filler{13}{59}
\filler{18}{59}
\filler{19}{59}
\filler{20}{59}
\filler{21}{59}
\filler{22}{59}
\filler{-15}{58}
\filler{-14}{58}
\filler{-13}{58}
\filler{-12}{58}
\filler{-7}{58}
\filler{-6}{58}
\filler{-5}{58}
\filler{-4}{58}
\filler{4}{58}
\filler{5}{58}
\filler{6}{58}
\filler{7}{58}
\filler{8}{58}
\filler{9}{58}
\filler{10}{58}
\filler{11}{58}
\filler{12}{58}
\filler{13}{58}
\filler{19}{58}
\filler{20}{58}
\filler{21}{58}
\filler{22}{58}
\filler{-15}{57}
\filler{-14}{57}
\filler{-13}{57}
\filler{-12}{57}
\filler{-7}{57}
\filler{-6}{57}
\filler{-5}{57}
\filler{-4}{57}
\filler{-3}{57}
\filler{3}{57}
\filler{4}{57}
\filler{5}{57}
\filler{6}{57}
\filler{7}{57}
\filler{8}{57}
\filler{9}{57}
\filler{10}{57}
\filler{11}{57}
\filler{12}{57}
\filler{13}{57}
\filler{20}{57}
\filler{21}{57}
\filler{22}{57}
\filler{-15}{56}
\filler{-14}{56}
\filler{-13}{56}
\filler{-12}{56}
\filler{-7}{56}
\filler{-6}{56}
\filler{-5}{56}
\filler{-4}{56}
\filler{-3}{56}
\filler{3}{56}
\filler{4}{56}
\filler{5}{56}
\filler{6}{56}
\filler{7}{56}
\filler{8}{56}
\filler{9}{56}
\filler{10}{56}
\filler{11}{56}
\filler{12}{56}
\filler{13}{56}
\filler{14}{56}
\filler{21}{56}
\filler{22}{56}
\filler{-15}{55}
\filler{-14}{55}
\filler{-13}{55}
\filler{-12}{55}
\filler{-7}{55}
\filler{-6}{55}
\filler{-5}{55}
\filler{-4}{55}
\filler{4}{55}
\filler{5}{55}
\filler{6}{55}
\filler{7}{55}
\filler{8}{55}
\filler{9}{55}
\filler{10}{55}
\filler{11}{55}
\filler{12}{55}
\filler{13}{55}
\filler{14}{55}
\filler{21}{55}
\filler{22}{55}
\filler{-15}{54}
\filler{-14}{54}
\filler{-13}{54}
\filler{-12}{54}
\filler{-7}{54}
\filler{-6}{54}
\filler{-5}{54}
\filler{-4}{54}
\filler{4}{54}
\filler{5}{54}
\filler{6}{54}
\filler{7}{54}
\filler{8}{54}
\filler{9}{54}
\filler{10}{54}
\filler{11}{54}
\filler{12}{54}
\filler{13}{54}
\filler{14}{54}
\filler{21}{54}
\filler{22}{54}
\filler{-15}{53}
\filler{-14}{53}
\filler{-13}{53}
\filler{-12}{53}
\filler{-7}{53}
\filler{-6}{53}
\filler{-5}{53}
\filler{-4}{53}
\filler{20}{53}
\filler{21}{53}
\filler{22}{53}
\filler{-15}{52}
\filler{-14}{52}
\filler{-13}{52}
\filler{-7}{52}
\filler{-6}{52}
\filler{-5}{52}
\filler{20}{52}
\filler{21}{52}
\filler{22}{52}
\filler{-15}{51}
\filler{-14}{51}
\filler{-7}{51}
\filler{-6}{51}
\filler{-5}{51}
\filler{19}{51}
\filler{20}{51}
\filler{21}{51}
\filler{22}{51}
\filler{-15}{50}
\filler{-8}{50}
\filler{-7}{50}
\filler{-6}{50}
\filler{-5}{50}
\filler{19}{50}
\filler{20}{50}
\filler{21}{50}
\filler{22}{50}
\filler{-15}{49}
\filler{-8}{49}
\filler{-7}{49}
\filler{-6}{49}
\filler{-5}{49}
\filler{5}{49}
\filler{6}{49}
\filler{7}{49}
\filler{8}{49}
\filler{9}{49}
\filler{10}{49}
\filler{11}{49}
\filler{12}{49}
\filler{13}{49}
\filler{18}{49}
\filler{19}{49}
\filler{20}{49}
\filler{21}{49}
\filler{22}{49}
\filler{-15}{48}
\filler{-8}{48}
\filler{-7}{48}
\filler{-6}{48}
\filler{-5}{48}
\filler{-4}{48}
\filler{4}{48}
\filler{5}{48}
\filler{6}{48}
\filler{7}{48}
\filler{8}{48}
\filler{9}{48}
\filler{10}{48}
\filler{11}{48}
\filler{12}{48}
\filler{13}{48}
\filler{18}{48}
\filler{19}{48}
\filler{20}{48}
\filler{21}{48}
\filler{22}{48}
\filler{-15}{47}
\filler{-14}{47}
\filler{9}{47}
\filler{10}{47}
\filler{11}{47}
\filler{12}{47}
\filler{13}{47}
\filler{18}{47}
\filler{19}{47}
\filler{20}{47}
\filler{21}{47}
\filler{22}{47}
\filler{-15}{46}
\filler{-14}{46}
\filler{10}{46}
\filler{11}{46}
\filler{12}{46}
\filler{13}{46}
\filler{18}{46}
\filler{19}{46}
\filler{20}{46}
\filler{21}{46}
\filler{22}{46}
\filler{-15}{45}
\filler{-14}{45}
\filler{-13}{45}
\filler{11}{45}
\filler{12}{45}
\filler{13}{45}
\filler{18}{45}
\filler{19}{45}
\filler{20}{45}
\filler{21}{45}
\filler{22}{45}
\filler{-15}{44}
\filler{-14}{44}
\filler{-13}{44}
\filler{11}{44}
\filler{12}{44}
\filler{13}{44}
\filler{18}{44}
\filler{19}{44}
\filler{20}{44}
\filler{21}{44}
\filler{22}{44}
\filler{-15}{43}
\filler{-14}{43}
\filler{-13}{43}
\filler{-12}{43}
\filler{-7}{43}
\filler{-6}{43}
\filler{-5}{43}
\filler{5}{43}
\filler{10}{43}
\filler{11}{43}
\filler{12}{43}
\filler{13}{43}
\filler{18}{43}
\filler{19}{43}
\filler{20}{43}
\filler{21}{43}
\filler{22}{43}
\filler{-15}{42}
\filler{-14}{42}
\filler{-13}{42}
\filler{-12}{42}
\filler{-7}{42}
\filler{-6}{42}
\filler{-5}{42}
\filler{-4}{42}
\filler{4}{42}
\filler{5}{42}
\filler{10}{42}
\filler{11}{42}
\filler{12}{42}
\filler{13}{42}
\filler{18}{42}
\filler{19}{42}
\filler{20}{42}
\filler{21}{42}
\filler{22}{42}
\filler{-15}{41}
\filler{-14}{41}
\filler{-13}{41}
\filler{-12}{41}
\filler{-7}{41}
\filler{-6}{41}
\filler{-5}{41}
\filler{-4}{41}
\filler{4}{41}
\filler{5}{41}
\filler{10}{41}
\filler{11}{41}
\filler{12}{41}
\filler{13}{41}
\filler{18}{41}
\filler{19}{41}
\filler{20}{41}
\filler{21}{41}
\filler{22}{41}
\filler{-15}{40}
\filler{-14}{40}
\filler{-13}{40}
\filler{-7}{40}
\filler{-6}{40}
\filler{-5}{40}
\filler{-4}{40}
\filler{4}{40}
\filler{5}{40}
\filler{11}{40}
\filler{12}{40}
\filler{13}{40}
\filler{18}{40}
\filler{19}{40}
\filler{20}{40}
\filler{21}{40}
\filler{22}{40}
\filler{-15}{39}
\filler{-14}{39}
\filler{-7}{39}
\filler{-6}{39}
\filler{-5}{39}
\filler{-4}{39}
\filler{-3}{39}
\filler{3}{39}
\filler{4}{39}
\filler{5}{39}
\filler{12}{39}
\filler{13}{39}
\filler{18}{39}
\filler{19}{39}
\filler{20}{39}
\filler{21}{39}
\filler{22}{39}
\filler{-15}{38}
\filler{-8}{38}
\filler{-7}{38}
\filler{-6}{38}
\filler{-5}{38}
\filler{-4}{38}
\filler{-3}{38}
\filler{3}{38}
\filler{4}{38}
\filler{5}{38}
\filler{6}{38}
\filler{13}{38}
\filler{18}{38}
\filler{19}{38}
\filler{20}{38}
\filler{21}{38}
\filler{22}{38}
\filler{-15}{37}
\filler{-8}{37}
\filler{-7}{37}
\filler{-6}{37}
\filler{-5}{37}
\filler{-4}{37}
\filler{4}{37}
\filler{5}{37}
\filler{6}{37}
\filler{13}{37}
\filler{18}{37}
\filler{19}{37}
\filler{20}{37}
\filler{21}{37}
\filler{22}{37}
\filler{-15}{36}
\filler{-8}{36}
\filler{-7}{36}
\filler{-6}{36}
\filler{-5}{36}
\filler{-4}{36}
\filler{4}{36}
\filler{5}{36}
\filler{6}{36}
\filler{13}{36}
\filler{18}{36}
\filler{19}{36}
\filler{20}{36}
\filler{21}{36}
\filler{22}{36}
\filler{-15}{35}
\filler{-14}{35}
\filler{12}{35}
\filler{13}{35}
\filler{18}{35}
\filler{19}{35}
\filler{20}{35}
\filler{21}{35}
\filler{22}{35}
\filler{-15}{34}
\filler{-14}{34}
\filler{12}{34}
\filler{13}{34}
\filler{18}{34}
\filler{19}{34}
\filler{20}{34}
\filler{21}{34}
\filler{22}{34}
\filler{-15}{33}
\filler{-14}{33}
\filler{-13}{33}
\filler{11}{33}
\filler{12}{33}
\filler{13}{33}
\filler{18}{33}
\filler{19}{33}
\filler{20}{33}
\filler{21}{33}
\filler{22}{33}
\filler{-15}{32}
\filler{-14}{32}
\filler{-13}{32}
\filler{-12}{32}
\filler{11}{32}
\filler{12}{32}
\filler{13}{32}
\filler{18}{32}
\filler{19}{32}
\filler{20}{32}
\filler{21}{32}
\filler{22}{32}
\filler{-15}{31}
\filler{-14}{31}
\filler{-13}{31}
\filler{-12}{31}
\filler{-11}{31}
\filler{-10}{31}
\filler{-9}{31}
\filler{-8}{31}
\filler{-7}{31}
\filler{-6}{31}
\filler{-5}{31}
\filler{5}{31}
\filler{10}{31}
\filler{11}{31}
\filler{12}{31}
\filler{13}{31}
\filler{18}{31}
\filler{19}{31}
\filler{20}{31}
\filler{21}{31}
\filler{22}{31}
\filler{-15}{30}
\filler{-14}{30}
\filler{-13}{30}
\filler{-12}{30}
\filler{-11}{30}
\filler{-10}{30}
\filler{-9}{30}
\filler{-8}{30}
\filler{-7}{30}
\filler{-6}{30}
\filler{-5}{30}
\filler{-4}{30}
\filler{4}{30}
\filler{5}{30}
\filler{10}{30}
\filler{11}{30}
\filler{12}{30}
\filler{13}{30}
\filler{18}{30}
\filler{19}{30}
\filler{20}{30}
\filler{21}{30}
\filler{22}{30}
\filler{-15}{29}
\filler{-14}{29}
\filler{-13}{29}
\filler{-12}{29}
\filler{-11}{29}
\filler{-10}{29}
\filler{-9}{29}
\filler{-8}{29}
\filler{-7}{29}
\filler{-6}{29}
\filler{-5}{29}
\filler{-4}{29}
\filler{4}{29}
\filler{5}{29}
\filler{10}{29}
\filler{11}{29}
\filler{12}{29}
\filler{13}{29}
\filler{18}{29}
\filler{19}{29}
\filler{20}{29}
\filler{21}{29}
\filler{22}{29}
\filler{-15}{28}
\filler{-14}{28}
\filler{-13}{28}
\filler{-12}{28}
\filler{-11}{28}
\filler{-10}{28}
\filler{-9}{28}
\filler{-8}{28}
\filler{-7}{28}
\filler{-6}{28}
\filler{-5}{28}
\filler{-4}{28}
\filler{4}{28}
\filler{5}{28}
\filler{11}{28}
\filler{12}{28}
\filler{13}{28}
\filler{18}{28}
\filler{19}{28}
\filler{20}{28}
\filler{21}{28}
\filler{22}{28}
\filler{-15}{27}
\filler{-14}{27}
\filler{-13}{27}
\filler{-12}{27}
\filler{-11}{27}
\filler{-10}{27}
\filler{-9}{27}
\filler{-8}{27}
\filler{-7}{27}
\filler{-6}{27}
\filler{-5}{27}
\filler{-4}{27}
\filler{-3}{27}
\filler{3}{27}
\filler{4}{27}
\filler{5}{27}
\filler{12}{27}
\filler{13}{27}
\filler{18}{27}
\filler{19}{27}
\filler{20}{27}
\filler{21}{27}
\filler{22}{27}
\filler{-15}{26}
\filler{-14}{26}
\filler{-13}{26}
\filler{-12}{26}
\filler{-11}{26}
\filler{-10}{26}
\filler{-9}{26}
\filler{-8}{26}
\filler{-7}{26}
\filler{-6}{26}
\filler{-5}{26}
\filler{-4}{26}
\filler{-3}{26}
\filler{3}{26}
\filler{4}{26}
\filler{5}{26}
\filler{6}{26}
\filler{13}{26}
\filler{18}{26}
\filler{19}{26}
\filler{20}{26}
\filler{21}{26}
\filler{22}{26}
\filler{-15}{25}
\filler{-14}{25}
\filler{-13}{25}
\filler{-12}{25}
\filler{-11}{25}
\filler{-10}{25}
\filler{-9}{25}
\filler{-8}{25}
\filler{-7}{25}
\filler{-6}{25}
\filler{-5}{25}
\filler{-4}{25}
\filler{4}{25}
\filler{5}{25}
\filler{6}{25}
\filler{13}{25}
\filler{18}{25}
\filler{19}{25}
\filler{20}{25}
\filler{21}{25}
\filler{22}{25}
\filler{-15}{24}
\filler{-14}{24}
\filler{-13}{24}
\filler{-12}{24}
\filler{-11}{24}
\filler{-10}{24}
\filler{-9}{24}
\filler{-8}{24}
\filler{-7}{24}
\filler{-6}{24}
\filler{-5}{24}
\filler{-4}{24}
\filler{4}{24}
\filler{5}{24}
\filler{6}{24}
\filler{13}{24}
\filler{18}{24}
\filler{19}{24}
\filler{20}{24}
\filler{21}{24}
\filler{22}{24}
\filler{-15}{23}
\filler{-14}{23}
\filler{-13}{23}
\filler{-12}{23}
\filler{-11}{23}
\filler{12}{23}
\filler{13}{23}
\filler{18}{23}
\filler{19}{23}
\filler{20}{23}
\filler{21}{23}
\filler{22}{23}
\filler{-15}{22}
\filler{-14}{22}
\filler{-13}{22}
\filler{-12}{22}
\filler{12}{22}
\filler{13}{22}
\filler{19}{22}
\filler{20}{22}
\filler{21}{22}
\filler{22}{22}
\filler{-15}{21}
\filler{-14}{21}
\filler{-13}{21}
\filler{11}{21}
\filler{12}{21}
\filler{13}{21}
\filler{20}{21}
\filler{21}{21}
\filler{22}{21}
\filler{-15}{20}
\filler{-14}{20}
\filler{-13}{20}
\filler{10}{20}
\filler{11}{20}
\filler{12}{20}
\filler{13}{20}
\filler{14}{20}
\filler{21}{20}
\filler{22}{20}
\filler{-15}{19}
\filler{-14}{19}
\filler{-13}{19}
\filler{-12}{19}
\filler{-7}{19}
\filler{-6}{19}
\filler{-5}{19}
\filler{5}{19}
\filler{6}{19}
\filler{7}{19}
\filler{8}{19}
\filler{9}{19}
\filler{10}{19}
\filler{11}{19}
\filler{12}{19}
\filler{13}{19}
\filler{14}{19}
\filler{21}{19}
\filler{22}{19}
\filler{-15}{18}
\filler{-14}{18}
\filler{-13}{18}
\filler{-12}{18}
\filler{-7}{18}
\filler{-6}{18}
\filler{-5}{18}
\filler{-4}{18}
\filler{4}{18}
\filler{5}{18}
\filler{6}{18}
\filler{7}{18}
\filler{8}{18}
\filler{9}{18}
\filler{10}{18}
\filler{11}{18}
\filler{12}{18}
\filler{13}{18}
\filler{14}{18}
\filler{21}{18}
\filler{22}{18}
\filler{-15}{17}
\filler{-14}{17}
\filler{-13}{17}
\filler{-12}{17}
\filler{-7}{17}
\filler{-6}{17}
\filler{-5}{17}
\filler{-4}{17}
\filler{20}{17}
\filler{21}{17}
\filler{22}{17}
\filler{-15}{16}
\filler{-14}{16}
\filler{-13}{16}
\filler{-12}{16}
\filler{-7}{16}
\filler{-6}{16}
\filler{-5}{16}
\filler{20}{16}
\filler{21}{16}
\filler{22}{16}
\filler{-15}{15}
\filler{-14}{15}
\filler{-13}{15}
\filler{-12}{15}
\filler{-7}{15}
\filler{-6}{15}
\filler{-5}{15}
\filler{19}{15}
\filler{20}{15}
\filler{21}{15}
\filler{22}{15}
\filler{-15}{14}
\filler{-14}{14}
\filler{-13}{14}
\filler{-12}{14}
\filler{-7}{14}
\filler{-6}{14}
\filler{-5}{14}
\filler{19}{14}
\filler{20}{14}
\filler{21}{14}
\filler{22}{14}
\filler{-15}{13}
\filler{-14}{13}
\filler{-13}{13}
\filler{-12}{13}
\filler{-7}{13}
\filler{-6}{13}
\filler{-5}{13}
\filler{5}{13}
\filler{6}{13}
\filler{7}{13}
\filler{8}{13}
\filler{9}{13}
\filler{10}{13}
\filler{11}{13}
\filler{12}{13}
\filler{13}{13}
\filler{18}{13}
\filler{19}{13}
\filler{20}{13}
\filler{21}{13}
\filler{22}{13}
\filler{-15}{12}
\filler{-14}{12}
\filler{-13}{12}
\filler{-12}{12}
\filler{-7}{12}
\filler{-6}{12}
\filler{-5}{12}
\filler{-4}{12}
\filler{4}{12}
\filler{5}{12}
\filler{6}{12}
\filler{7}{12}
\filler{8}{12}
\filler{9}{12}
\filler{10}{12}
\filler{11}{12}
\filler{12}{12}
\filler{13}{12}
\filler{18}{12}
\filler{19}{12}
\filler{20}{12}
\filler{21}{12}
\filler{22}{12}
\filler{-15}{11}
\filler{-14}{11}
\filler{-13}{11}
\filler{-12}{11}
\filler{-7}{11}
\filler{-6}{11}
\filler{-5}{11}
\filler{-4}{11}
\filler{4}{11}
\filler{5}{11}
\filler{6}{11}
\filler{7}{11}
\filler{8}{11}
\filler{9}{11}
\filler{10}{11}
\filler{11}{11}
\filler{12}{11}
\filler{13}{11}
\filler{18}{11}
\filler{19}{11}
\filler{20}{11}
\filler{21}{11}
\filler{22}{11}
\filler{-15}{10}
\filler{-14}{10}
\filler{-13}{10}
\filler{-7}{10}
\filler{-6}{10}
\filler{-5}{10}
\filler{-4}{10}
\filler{4}{10}
\filler{5}{10}
\filler{6}{10}
\filler{7}{10}
\filler{8}{10}
\filler{9}{10}
\filler{10}{10}
\filler{11}{10}
\filler{12}{10}
\filler{13}{10}
\filler{19}{10}
\filler{20}{10}
\filler{21}{10}
\filler{22}{10}
\filler{-15}{9}
\filler{-14}{9}
\filler{-7}{9}
\filler{-6}{9}
\filler{-5}{9}
\filler{-4}{9}
\filler{-3}{9}
\filler{3}{9}
\filler{4}{9}
\filler{5}{9}
\filler{6}{9}
\filler{7}{9}
\filler{8}{9}
\filler{9}{9}
\filler{10}{9}
\filler{11}{9}
\filler{12}{9}
\filler{13}{9}
\filler{20}{9}
\filler{21}{9}
\filler{22}{9}
\filler{-15}{8}
\filler{-8}{8}
\filler{-7}{8}
\filler{-6}{8}
\filler{-5}{8}
\filler{-4}{8}
\filler{-3}{8}
\filler{3}{8}
\filler{4}{8}
\filler{5}{8}
\filler{6}{8}
\filler{7}{8}
\filler{8}{8}
\filler{9}{8}
\filler{10}{8}
\filler{11}{8}
\filler{12}{8}
\filler{13}{8}
\filler{14}{8}
\filler{21}{8}
\filler{22}{8}
\filler{-15}{7}
\filler{-8}{7}
\filler{-7}{7}
\filler{-6}{7}
\filler{-5}{7}
\filler{-4}{7}
\filler{4}{7}
\filler{5}{7}
\filler{6}{7}
\filler{7}{7}
\filler{8}{7}
\filler{9}{7}
\filler{10}{7}
\filler{11}{7}
\filler{12}{7}
\filler{13}{7}
\filler{14}{7}
\filler{21}{7}
\filler{22}{7}
\filler{-15}{6}
\filler{-8}{6}
\filler{-7}{6}
\filler{-6}{6}
\filler{-5}{6}
\filler{-4}{6}
\filler{4}{6}
\filler{5}{6}
\filler{6}{6}
\filler{7}{6}
\filler{8}{6}
\filler{9}{6}
\filler{10}{6}
\filler{11}{6}
\filler{12}{6}
\filler{13}{6}
\filler{14}{6}
\filler{21}{6}
\filler{22}{6}
\filler{-15}{5}
\filler{-14}{5}
\filler{20}{5}
\filler{21}{5}
\filler{22}{5}
\filler{-15}{4}
\filler{-14}{4}
\filler{20}{4}
\filler{21}{4}
\filler{22}{4}
\filler{-15}{3}
\filler{-14}{3}
\filler{-13}{3}
\filler{19}{3}
\filler{20}{3}
\filler{21}{3}
\filler{22}{3}
\filler{-15}{2}
\filler{-14}{2}
\filler{-13}{2}
\filler{-12}{2}
\filler{18}{2}
\filler{19}{2}
\filler{20}{2}
\filler{21}{2}
\filler{22}{2}
\filler{-15}{1}
\filler{-14}{1}
\filler{-13}{1}
\filler{-12}{1}
\filler{-11}{1}
\filler{-10}{1}
\filler{-9}{1}
\filler{-8}{1}
\filler{-7}{1}
\filler{-6}{1}
\filler{-5}{1}
\filler{5}{1}
\filler{6}{1}
\filler{7}{1}
\filler{8}{1}
\filler{9}{1}
\filler{10}{1}
\filler{11}{1}
\filler{12}{1}
\filler{13}{1}
\filler{14}{1}
\filler{15}{1}
\filler{16}{1}
\filler{17}{1}
\filler{18}{1}
\filler{19}{1}
\filler{20}{1}
\filler{21}{1}
\filler{22}{1}
\filler{-15}{0}
\filler{-14}{0}
\filler{-13}{0}
\filler{-12}{0}
\filler{-11}{0}
\filler{-10}{0}
\filler{-9}{0}
\filler{-8}{0}
\filler{-7}{0}
\filler{-6}{0}
\filler{-5}{0}
\filler{-4}{0}
\filler{4}{0}
\filler{5}{0}
\filler{6}{0}
\filler{7}{0}
\filler{8}{0}
\filler{9}{0}
\filler{10}{0}
\filler{11}{0}
\filler{12}{0}
\filler{13}{0}
\filler{14}{0}
\filler{15}{0}
\filler{16}{0}
\filler{17}{0}
\filler{18}{0}
\filler{19}{0}
\filler{20}{0}
\filler{21}{0}
\filler{22}{0}
\filler{-15}{-1}
\filler{-14}{-1}
\filler{-13}{-1}
\filler{-12}{-1}
\filler{-11}{-1}
\filler{-10}{-1}
\filler{-9}{-1}
\filler{-8}{-1}
\filler{-7}{-1}
\filler{-6}{-1}
\filler{-5}{-1}
\filler{-4}{-1}
\filler{4}{-1}
\filler{5}{-1}
\filler{6}{-1}
\filler{7}{-1}
\filler{8}{-1}
\filler{9}{-1}
\filler{10}{-1}
\filler{11}{-1}
\filler{12}{-1}
\filler{13}{-1}
\filler{14}{-1}
\filler{15}{-1}
\filler{16}{-1}
\filler{17}{-1}
\filler{18}{-1}
\filler{19}{-1}
\filler{20}{-1}
\filler{21}{-1}
\filler{22}{-1}
\filler{-15}{-2}
\filler{-14}{-2}
\filler{-13}{-2}
\filler{-12}{-2}
\filler{-11}{-2}
\filler{-10}{-2}
\filler{-9}{-2}
\filler{-8}{-2}
\filler{-7}{-2}
\filler{-6}{-2}
\filler{-5}{-2}
\filler{-4}{-2}
\filler{4}{-2}
\filler{5}{-2}
\filler{6}{-2}
\filler{7}{-2}
\filler{8}{-2}
\filler{9}{-2}
\filler{10}{-2}
\filler{11}{-2}
\filler{12}{-2}
\filler{13}{-2}
\filler{14}{-2}
\filler{15}{-2}
\filler{16}{-2}
\filler{17}{-2}
\filler{18}{-2}
\filler{19}{-2}
\filler{20}{-2}
\filler{21}{-2}
\filler{22}{-2}
\filler{-15}{-3}
\filler{-14}{-3}
\filler{-13}{-3}
\filler{-12}{-3}
\filler{-11}{-3}
\filler{-10}{-3}
\filler{-9}{-3}
\filler{-8}{-3}
\filler{-7}{-3}
\filler{-6}{-3}
\filler{-5}{-3}
\filler{-4}{-3}
\filler{-3}{-3}
\filler{3}{-3}
\filler{4}{-3}
\filler{5}{-3}
\filler{6}{-3}
\filler{7}{-3}
\filler{8}{-3}
\filler{9}{-3}
\filler{10}{-3}
\filler{11}{-3}
\filler{12}{-3}
\filler{13}{-3}
\filler{14}{-3}
\filler{15}{-3}
\filler{16}{-3}
\filler{17}{-3}
\filler{18}{-3}
\filler{19}{-3}
\filler{20}{-3}
\filler{21}{-3}
\filler{22}{-3}

          \vars{4}{2/{2/5}, 2/{0/5, 2/1}, 2/{2/1}, 2/{2/1, 0/5}, 2/{2/5}}

\clauseblock{0}{0}{5}{0}{0}{0}{4}
          \clauseblock{0}{2}{5}{1}{0}{0}{4}
          \clauseblock{0}{8}{5}{1}{0}{2}{4}
          \clauseblock{0}{10}{5}{2}{0}{8}{4}

\clauseblock{0}{3}{1}{0}{0}{0}{4}
          \clauseblock{0}{5}{1}{1}{0}{3}{4}
          \clauseblock{0}{7}{1}{2}{0}{5}{4}

\clauseblock{0}{0}{-2}{0}{0}{0}{4}
          \clauseblock{0}{3}{-2}{2}{0}{0}{4}

\clauseblock{0}{5}{-2}{0}{0}{0}{4}
          \clauseblock{0}{7}{-2}{1}{0}{5}{4}
          \clauseblock{0}{10}{-2}{2}{0}{7}{4}

\pgfmathsetmacro{\floorright}{\xmax - 2}
          \foreach \x in {\xmin, ..., \floorright}{
              \filler{\x + .5}{-4}
          }
\foreach \x in {\xmin, ..., -2}{
              \filler{\x + .5}{\ymax - .5}
          }
          \pgfmathsetmacro{\lidright}{\xmax - 2}
          \foreach \x in {\lidright, ..., 1}{
              \filler{\x + .5}{\ymax - .5}
          }

\foreach \c/\d/\xa/\xb/\ya/\yb in \gadgetregions {
              \draw [\d, uwidth = .3] (\xa - .5, \ya - .5) rectangle (\xb + .5, \yb + .5);
          }

\tikzset{gadgetlabel/.style = {text = #1!70!black, inner sep = 0pt, font = \small}}
          \foreach \i/\y in {1/2.5, 2/20.5, 3/32.5, 4/50.5, 5/62.5}{
              \node [gadgetlabel = teal] at (0, \y) {$v_{\i}$};
          }
          \node [gadgetlabel = blue] at (17.5, 55) {$c_1^+$};
          \node [gadgetlabel = blue] at (9.5, 37) {$c_2^+$};
          \node [gadgetlabel = red] at (-11.5, 7) {$c_1^-$};
          \node [gadgetlabel = red] at (-11.5, 49) {$c_2^-$};

\node [star, star points = 5, star point ratio = 2.2, fill = black, inner sep = 0pt, minimum size = .9 * .16cm] at (0, -3) {};
      \end{scope}
      \draw [draw = black, uwidth = 1] (\xmin - 1, \ymin - 1) rectangle (\xmax + 4, \ymax + 1);

      \node [draw = black, uwidth = 1] at (\xmax + 14, \ymax) [below] {
          \begin{tikzpicture}[x = .28cm, y = .28cm, inner xsep = 4pt]
              \node (omino) at (11.5, 15) {
                  \begin{tikzpicture}
                      \omino{0}{0}{1}
                  \end{tikzpicture}
              };
              \node (times) at (omino.south) [below] {$\times$};
              \node [scale = 1.25] (num) at (times.south) [below] {$204$};
          \end{tikzpicture}
      };
  \end{tikzpicture}
  \caption{The board.}
  \label{fig:global}
\end{subfigure}
\caption{\subref{fig:mrsat}~A rectilinear arrangement of the formula~\eqref{eq:running}.
\subref{fig:global}~The board built from it, each gadget coloured as its counterpart in~\subref{fig:mrsat}; grey cells are filled and all others are empty, and the star marks the coordinate origin $(0,0)$.
Row $H_*$ and the rows above it, where pieces spawn, are not drawn.}
\end{figure}
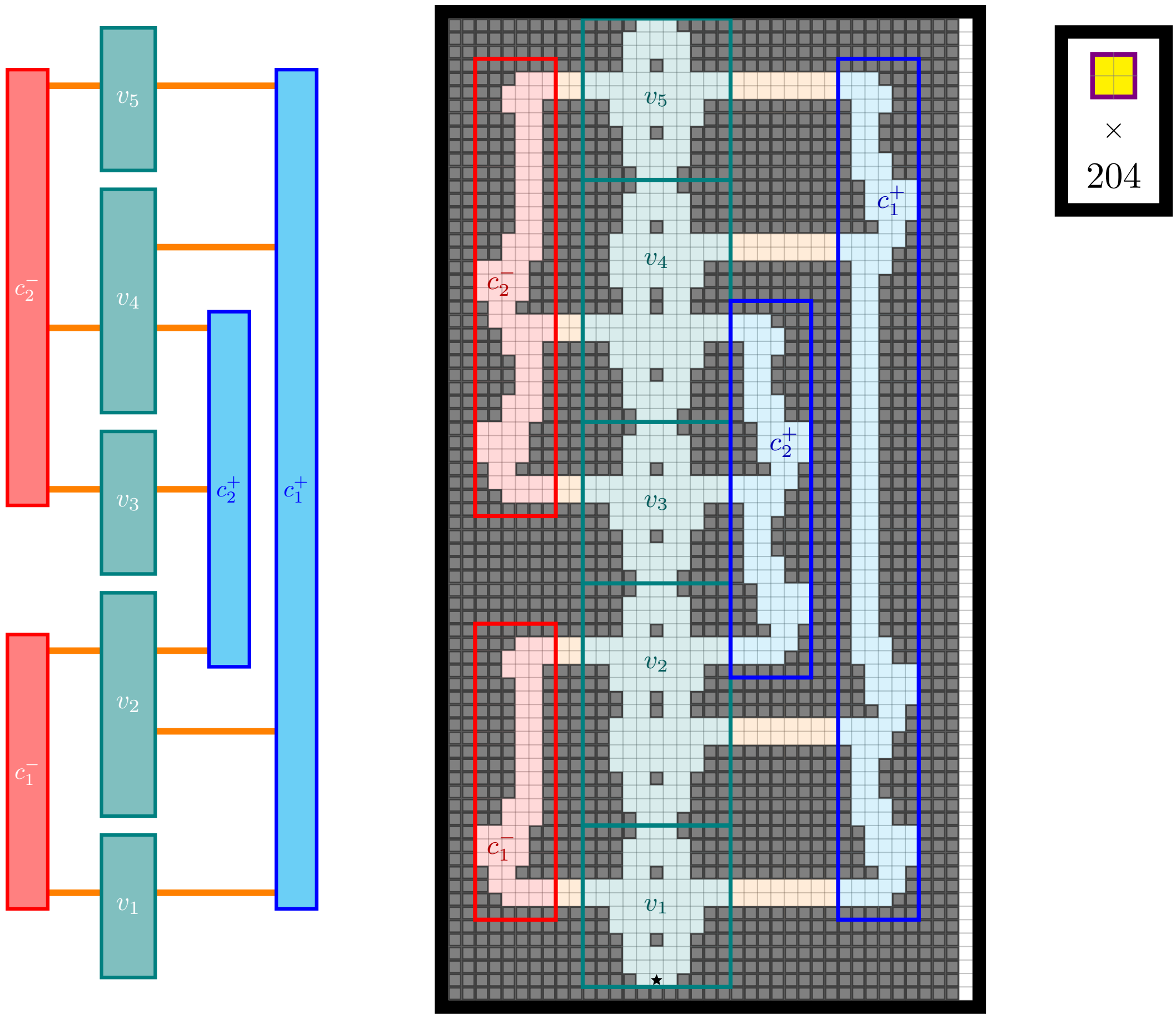

Given an instance $\phi$ of Monotone Rectilinear SAT together with its rectilinear arrangement, we build the board shown in Figure~\ref{fig:global}.
We introduce a coordinate system, where $(x,y)$ denotes the cell in column $x$ and row $y$, with columns counted to the right and rows upward.

\emph{Variable gadgets} are piled on the $y$-axis from the origin $(0,0)$ (marked by a star in Figure~\ref{fig:global}), one gadget per variable, in the order of the arrangement of $\phi$.
We call the pile the \emph{variable pile}.

\emph{Clause gadgets} for the positive clauses are placed to the right of the variable pile and clause gadgets for the negative clauses are placed to its left, in accordance with the arrangement.

\emph{Wires}, horizontal corridors of height two and even length, connect each variable gadget to the clause gadgets of the clauses containing the corresponding literal.

In addition, an \emph{empty column} runs up the right-hand edge of the board through all of its rows.
No \ilo~piece fits into it, so no row is ever cleared.
For technical convenience, the empty column is placed at an odd $x$.

Every remaining cell of the board is filled.
The construction is closed at the top except for an opening above the variable pile, where the spawned pieces enter the construction.
The losing row $H_*$ lies just above the construction, and pieces spawn above it.

Next we determine the number $N$ of pieces given to the player.
We take it to be a trivial upper bound on the number of pieces that can be packed.
A cell $(x,y)$ is a \emph{pin} if both $x$ and $y$ are even, $y < H_*$, and it is empty in the initial board.
The empty column contains no pin.
Obviously, an \ilo~piece in any position covers exactly one pin.
Pins play the role of the reference centres of Abrahamsen and Stade~\cite{AS26}.

A \emph{lossy slot} consists of five consecutive empty cells in a single row, bounded on both sides by filled cells, and is positioned so that its first, third and fifth cells are pins.
In the picture below, the dots are pins and the grey cells are filled.
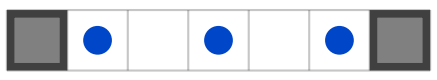
\begin{center}
\begin{tikzpicture}[x = .4cm, y = .4cm]
  \whitebg{-3}{3}{0}{0}
  \filler{-3}{0}\filler{3}{0}
  \foreach \x in {-2, 0, 2}{ \legendpin{\x}{0} }
\end{tikzpicture}
\end{center}
Obviously, at most two of the three pins of a lossy slot are covered by any packing.
The construction contains exactly two lossy slots per variable gadget.

Let $P$ be the total number of pins, and $Q = 2n$ the number of lossy slots in the construction, where $n$ is the number of variables of $\phi$.
Then, $P-Q$ is a trivial upper bound on the number of pieces that can be packed; we give the player exactly $N = P-Q$ pieces.
Therefore, to survive, the player must cover every pin on the board other than one in each of the lossy slots.
A \emph{winning packing} is a packing of $N$ pieces into the empty cells below row $H_*$.
We will show that the player can place all $N$ pieces without losing if and only if $\phi$ is satisfiable.
Note that finding a winning packing is only a necessary condition for survival, not a sufficient one: the player has to \emph{place} all $N$ pieces in accordance with the Tetris rules.

For the formula in~\eqref{eq:running} the count gives $P = 214$ and $Q = 10$, so $N = 204$.

\subsection{Variable gadget}
\label{sec:vargadget}

\begin{figure}[t]
\centering
\begin{subfigure}[t]{0.39\linewidth}
  \centering
\begin{tikzpicture}[x = .28cm, y = .28cm, baseline = (current bounding box.north),
    plabel/.style = {align = center, font = \small},
    leader/.style = {draw = orange, uwidth = .2}]
    \whitebg{-5}{5}{-3}{20}
    \filler{-5}{-3}
\filler{-5}{-2}
\filler{-5}{-1}
\filler{-5}{0}
\filler{-5}{1}
\filler{-5}{2}
\filler{-5}{3}
\filler{-5}{4}
\filler{-5}{5}
\filler{-5}{6}
\filler{-5}{7}
\filler{-5}{12}
\filler{-5}{13}
\filler{-5}{18}
\filler{-5}{19}
\filler{-5}{20}
\filler{-4}{-3}
\filler{-4}{-2}
\filler{-4}{-1}
\filler{-4}{0}
\filler{-4}{5}
\filler{-4}{6}
\filler{-4}{12}
\filler{-4}{18}
\filler{-4}{19}
\filler{-4}{20}
\filler{-3}{-3}
\filler{-3}{20}
\filler{3}{-3}
\filler{3}{20}
\filler{4}{-3}
\filler{4}{-2}
\filler{4}{-1}
\filler{4}{0}
\filler{4}{6}
\filler{4}{11}
\filler{4}{12}
\filler{4}{18}
\filler{4}{19}
\filler{4}{20}
\filler{5}{-3}
\filler{5}{-2}
\filler{5}{-1}
\filler{5}{0}
\filler{5}{1}
\filler{5}{6}
\filler{5}{7}
\filler{5}{8}
\filler{5}{9}
\filler{5}{10}
\filler{5}{11}
\filler{5}{12}
\filler{5}{13}
\filler{5}{18}
\filler{5}{19}
\filler{5}{20}
     \foreach \y in {-3, -1, ..., 19}{
        \foreach \x in {-2, 0, 2}{ \legendpin{\x}{\y} }
    }
    \foreach \y in {3, 9, 15}{
        \foreach \x in {-4, 4}{ \legendpin{\x}{\y} }
    }
    \varbot{0}{0}{2}
    \varblock{0}{0}{0}{1}{2}
    \varblock{0}{6}{1}{0}{2}
    \varblock{0}{12}{1}{1}{2}
    \vartop{0}{18}{2}
    \begin{scope}
        \clip (-3.5, -.5) -- ++ (0, -2) -- ++ (1, 0) -- ++ (0, -1) -- ++ (5, 0) -- ++ (0, 1) -- ++ (1, 0) -- ++ (0, 2) -- cycle;
        \draw [draw = orange, uwidth = .5] (-3.5, -.5) -- ++ (0, -2) -- ++ (1, 0) -- ++ (0, -1) -- ++ (5, 0) -- ++ (0, 1) -- ++ (1, 0) -- ++ (0, 2) -- cycle;
    \end{scope}
    \begin{scope}
        \clip (-3.5, 17.5) -- ++ (0, 2) -- ++ (1, 0) -- ++ (0, 1) -- ++ (5, 0) -- ++ (0, -1) -- ++ (1, 0) -- ++ (0, -2) -- cycle;
        \draw [draw = orange, uwidth = .5] (-3.5, 17.5) -- ++ (0, 2) -- ++ (1, 0) -- ++ (0, 1) -- ++ (5, 0) -- ++ (0, -1) -- ++ (1, 0) -- ++ (0, -2) -- cycle;
    \end{scope}
    \begin{scope}
        \clip (3.5, -.5) -- ++ (-7, 0) -- ++ (0, 1) -- ++ (-1, 0) -- ++ (0, 4) -- ++ (1, 0) -- ++ (0, 1) -- ++ (7, 0) -- ++ (2, 0) -- ++ (0, -4) -- ++ (-1, 0) -- ++ (0, -1) -- ++ (-1, 0) -- ++ (0, -1) -- cycle;
        \draw [draw = orange, uwidth = .6] (3.5, -.5) -- ++ (-7, 0) -- ++ (0, 1) -- ++ (-1, 0) -- ++ (0, 4) -- ++ (1, 0) -- ++ (0, 1) -- ++ (7, 0) -- ++ (2, 0) -- ++ (0, -4) -- ++ (-1, 0) -- ++ (0, -1) -- ++ (-1, 0) -- ++ (0, -1) -- cycle;
    \end{scope}
    \begin{scope}
        \clip (-3.5, 5.5) -- ++ (7, 0) -- ++ (0, 1) -- ++ (1, 0) -- ++ (0, 4) -- ++ (-1, 0) -- ++ (0, 1) -- ++ (-7, 0) -- ++ (-2, 0) -- ++ (0, -4) -- ++ (1, 0) -- ++ (0, -1) -- ++ (1, 0) -- ++ (0, -1) -- cycle;
        \draw [draw = orange, uwidth = .5] (-3.5, 5.5) -- ++ (7, 0) -- ++ (0, 1) -- ++ (1, 0) -- ++ (0, 4) -- ++ (-1, 0) -- ++ (0, 1) -- ++ (-7, 0) -- ++ (-2, 0) -- ++ (0, -4) -- ++ (1, 0) -- ++ (0, -1) -- ++ (1, 0) -- ++ (0, -1) -- cycle;
    \end{scope}
    \begin{scope}
        \clip (-3.5, 11.5) -- ++ (7, 0) -- ++ (0, 1) -- ++ (1, 0) -- ++ (0, 1) -- ++ (1, 0) -- ++ (0, 4) -- ++ (-2, 0) -- ++ (-7, 0) -- ++ (-2, 0) -- ++ (0, -4) -- ++ (1, 0) -- ++ (0, -1) -- ++ (1, 0) -- ++ (0, -1) -- cycle;
        \draw [draw = orange, uwidth = .5] (-3.5, 11.5) -- ++ (7, 0) -- ++ (0, 1) -- ++ (1, 0) -- ++ (0, 1) -- ++ (1, 0) -- ++ (0, 4) -- ++ (-2, 0) -- ++ (-7, 0) -- ++ (-2, 0) -- ++ (0, -4) -- ++ (1, 0) -- ++ (0, -1) -- ++ (1, 0) -- ++ (0, -1) -- cycle;
    \end{scope}
    \node [plabel] (foot)  at (-11, -2)  {foot};
    \node [plabel] (head)  at (-11, 19)  {head};
    \node [plabel] (bodyr) at (-11, 2.5) {body\\ (w/\,positive port)};
    \node [plabel] (bodyl) at (-11, 9)   {body\\ (w/\,negative port)};
    \node [plabel] (bodyb) at (-11, 15)  {body\\ (w/\,both ports)};
    \draw [leader] (foot.east)  -- (-3.5, -1.5);
    \draw [leader] (head.east)  -- (-3.5, 18.5);
    \draw [leader] (bodyr)      -- (-4.5, 2.5);
    \draw [leader] (bodyl)      -- (-5.5, 9);
    \draw [leader] (bodyb)      -- (-5.5, 15);
\filldraw [draw = framecolor, fill = framecolor, fill opacity = .3, uwidth = .1]
      (-13, -5) rectangle ++ (2, -1);
    \node [anchor = west, font = \small] at (-10.6, -5.5) {lossy slot};
\filldraw [draw = violet!60, fill = violet, fill opacity = .15, uwidth = .1]
      (-2.8, -5) rectangle ++ (1.2, -1);
    \node [anchor = west, font = \small] at (-1.5, -5.5) {output port};
\foreach \x/\y in {3.5/2.5, -5.5/8.5, 3.5/14.5, -5.5/14.5}{
        \filldraw [draw = violet!60, fill = violet, fill opacity = .15, uwidth = .1]
          (\x, \y) rectangle ++ (2, 2);
    }
    \foreach \x/\y in {4/3, -4/9, 4/15, -4/15}{ \legendpin{\x}{\y} }
    \filldraw [draw = framecolor, fill = framecolor, fill opacity = .3, uwidth = .1]
      (-2.5, -.5) rectangle ++ (5, -1);
    \filldraw [draw = framecolor, fill = framecolor, fill opacity = .3, uwidth = .1]
      (-2.5, 19.5) rectangle ++ (5, -1);
    \foreach \y in {-1, 19}{
        \foreach \x in {-2, 0, 2}{ \legendpin{\x}{\y} }
    }
\end{tikzpicture}
  \caption{The gadget and its constituents.}
  \label{fig:var-parts}
\end{subfigure}
\hfill
\begin{subfigure}[t]{0.28\linewidth}
  \centering
\begin{tikzpicture}[x = .28cm, y = .28cm, baseline = (current bounding box.north)]
    \whitebg{-5}{5}{-3}{20}
    \filler{-5}{-3}
\filler{-5}{-2}
\filler{-5}{-1}
\filler{-5}{0}
\filler{-5}{1}
\filler{-5}{2}
\filler{-5}{3}
\filler{-5}{4}
\filler{-5}{5}
\filler{-5}{6}
\filler{-5}{7}
\filler{-5}{12}
\filler{-5}{13}
\filler{-5}{18}
\filler{-5}{19}
\filler{-5}{20}
\filler{-4}{-3}
\filler{-4}{-2}
\filler{-4}{-1}
\filler{-4}{0}
\filler{-4}{5}
\filler{-4}{6}
\filler{-4}{12}
\filler{-4}{18}
\filler{-4}{19}
\filler{-4}{20}
\filler{-3}{-3}
\filler{-3}{20}
\filler{3}{-3}
\filler{3}{20}
\filler{4}{-3}
\filler{4}{-2}
\filler{4}{-1}
\filler{4}{0}
\filler{4}{6}
\filler{4}{11}
\filler{4}{12}
\filler{4}{18}
\filler{4}{19}
\filler{4}{20}
\filler{5}{-3}
\filler{5}{-2}
\filler{5}{-1}
\filler{5}{0}
\filler{5}{1}
\filler{5}{6}
\filler{5}{7}
\filler{5}{8}
\filler{5}{9}
\filler{5}{10}
\filler{5}{11}
\filler{5}{12}
\filler{5}{13}
\filler{5}{18}
\filler{5}{19}
\filler{5}{20}
     \varbot{0}{0}{1}
    \varblock{0}{0}{0}{1}{1}
    \varblock{0}{6}{1}{0}{1}
    \varblock{0}{12}{1}{1}{1}
    \vartop{0}{18}{1}

    \foreach \y in {-3, -1, ..., 19}{
        \foreach \x in {-2, 0, 2}{ \legendpin{\x}{\y} }
    }
    \foreach \y in {3, 9, 15}{
        \foreach \x in {-4, 4}{ \legendpin{\x}{\y} }
    }

    \varbot{0}{0}{2}
    \varblock{0}{0}{0}{1}{2}
    \varblock{0}{6}{1}{0}{2}
    \varblock{0}{12}{1}{1}{2}
    \vartop{0}{18}{2}

\foreach \x/\y/\l in {-1/-3/A, -2/-1/B, -2/17/C, 1/18/D, 1/10/E,
                          -1/7/F, -1/9/G, -3/9/H, -5/9/I, -1/19/J}{
        \node [font = \scriptsize\sffamily, inner sep = 0pt] at (\x + .5, \y + .5) {\l};
    }

\draw[->, >=stealth, black, line width = 1] (5.2, 3)  -- ++ (2, 0);
    \draw[->, >=stealth, black, line width = 1] (-5.2, 9) -- ++ (-2, 0);
    \draw[->, >=stealth, black, line width = 1] (5.2, 15) -- ++ (2, 0);
    \draw[->, >=stealth, black, line width = 1] (-5.2, 15) -- ++ (-2, 0);
\node [anchor = west, font = \small, opacity = 0] at (-1.5, -5.5) {output port};
  \end{tikzpicture}
  \caption{\True.}
  \label{fig:var-true}
\end{subfigure}
\begin{subfigure}[t]{0.28\linewidth}
  \centering
  \begin{tikzpicture}[x = .28cm, y = .28cm, baseline = (current bounding box.north)]
    \whitebg{-5}{5}{-3}{20}
    \filler{-5}{-3}
\filler{-5}{-2}
\filler{-5}{-1}
\filler{-5}{0}
\filler{-5}{1}
\filler{-5}{2}
\filler{-5}{3}
\filler{-5}{4}
\filler{-5}{5}
\filler{-5}{6}
\filler{-5}{7}
\filler{-5}{12}
\filler{-5}{13}
\filler{-5}{18}
\filler{-5}{19}
\filler{-5}{20}
\filler{-4}{-3}
\filler{-4}{-2}
\filler{-4}{-1}
\filler{-4}{0}
\filler{-4}{5}
\filler{-4}{6}
\filler{-4}{12}
\filler{-4}{18}
\filler{-4}{19}
\filler{-4}{20}
\filler{-3}{-3}
\filler{-3}{20}
\filler{3}{-3}
\filler{3}{20}
\filler{4}{-3}
\filler{4}{-2}
\filler{4}{-1}
\filler{4}{0}
\filler{4}{6}
\filler{4}{11}
\filler{4}{12}
\filler{4}{18}
\filler{4}{19}
\filler{4}{20}
\filler{5}{-3}
\filler{5}{-2}
\filler{5}{-1}
\filler{5}{0}
\filler{5}{1}
\filler{5}{6}
\filler{5}{7}
\filler{5}{8}
\filler{5}{9}
\filler{5}{10}
\filler{5}{11}
\filler{5}{12}
\filler{5}{13}
\filler{5}{18}
\filler{5}{19}
\filler{5}{20}
     \varbot{0}{0}{0}
    \varblock{0}{0}{0}{1}{0}
    \varblock{0}{6}{1}{0}{0}
    \varblock{0}{12}{1}{1}{0}
    \vartop{0}{18}{0}

    \foreach \y in {-3, -1, ..., 19}{
        \foreach \x in {-2, 0, 2}{ \legendpin{\x}{\y} }
    }
    \foreach \y in {3, 9, 15}{
        \foreach \x in {-4, 4}{ \legendpin{\x}{\y} }
    }

    \varbot{0}{0}{2}
    \varblock{0}{0}{0}{1}{2}
    \varblock{0}{6}{1}{0}{2}
    \varblock{0}{12}{1}{1}{2}
    \vartop{0}{18}{2}

\draw[->, >=stealth, black, line width = 1] (5.2, 3)  -- ++ (2, 0);
    \draw[->, >=stealth, black, line width = 1] (-5.2, 9) -- ++ (-2, 0);
    \draw[->, >=stealth, black, line width = 1] (5.2, 15) -- ++ (2, 0);
    \draw[->, >=stealth, black, line width = 1] (-5.2, 15) -- ++ (-2, 0);
\node [anchor = west, font = \small, opacity = 0] at (-1.5, -5.5) {output port};
  \end{tikzpicture}
  \caption{\False.}
  \label{fig:var-false}
\end{subfigure}
\caption{A variable gadget is a stack of body modules between a foot and a head.
The three body modules drawn are equipped with a positive port, a negative port, and both.
The gadget has one lossy slot in its head and one in its foot, shaded green in~\subref{fig:var-parts}.
The arrows in~\subref{fig:var-true} and~\subref{fig:var-false} mark the output ports and the direction in which each sends its signal.
The two fillings are symmetric: in~\subref{fig:var-true} every positive port sends \True and every negative port sends \False, and in~\subref{fig:var-false} it is the other way round.
The letters in~\subref{fig:var-true} merely label pieces for the proof of Lemma~\ref{lem:nodouble}.}
\label{fig:var}
\end{figure}
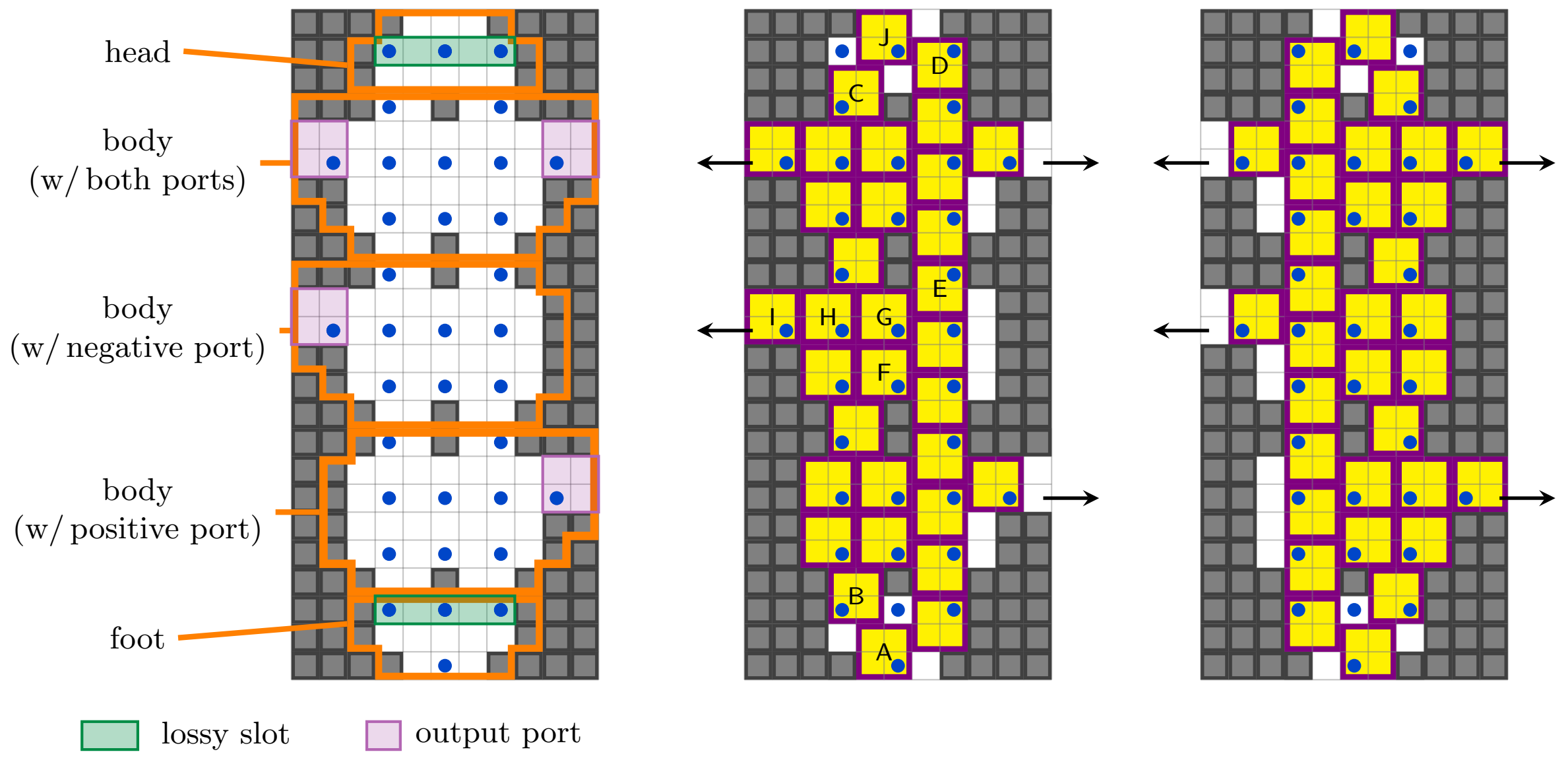

Figure~\ref{fig:var}\subref{fig:var-parts} shows the gadget for a variable $v_i$.
It is a vertical stack of \emph{body modules}, capped above by a \emph{head module} and below by a \emph{foot module} that join it to its neighbours in the variable pile.
A body module carries an output port of $2 \times 2$ size on its right, on its left, or on both.
A port on the right, a \emph{positive port}, serves a positive occurrence of $v_i$, and a port on the left, a \emph{negative port}, a negative one, so a positive and a negative occurrence can share one module.
The gadget therefore needs
\(
  \max (d_i^{+},\, d_i^{-} )
\)
body modules, where $d_i^{+}$ and $d_i^{-}$ are the numbers of positive and negative occurrences of $v_i$.
A body module is six rows tall and the head and the foot are three rows each, so the gadget occupies $6\max(d_i^{+}, d_i^{-}) + 6$ rows.
With this many body modules, all the modules with a single port have the same kind of port, so no gadget in the construction has both a positive-port and a negative-port module as Figure~\ref{fig:var}\subref{fig:var-parts} does; the figure shows all three forms of body module only for illustration.
The height of a variable gadget is even, so every gadget starts at an even row and has its pins where Figure~\ref{fig:var} shows them;
in particular, every output port has a pin in its lower row.

A \emph{filling} of a gadget or a module is a set of non-overlapping pieces on empty cells, each covering one of its pins; it is \emph{pin-covering} if it covers all of its pins.
Figure~\ref{fig:var}\subref{fig:var-true} and \subref{fig:var-false} give two fillings of the variable gadget, which are mirror images.
Let us say that a positive port sends \True if the piece covering the pin $(x,y)$ in the port also covers $(x-1,y)$, and \False if it covers $(x+1,y)$.
Symmetrically, a negative port with a pin $(x,y)$ sends \True if the piece covering the pin $(x,y)$ in the port also covers $(x+1,y)$, and \False if it covers $(x-1,y)$.
In the \emph{\True filling} of Figure~\ref{fig:var}\subref{fig:var-true}, every positive port sends \True and every negative port sends \False; in the \emph{\False filling} of Figure~\ref{fig:var}\subref{fig:var-false}, the ports send the opposite values.
Apart from their ports, all body modules have the same shape, and the fillings repeat from one module to the next, so they exist for every sequence of body modules.
Clearly, the filling chosen for $v_i$ places no constraint on the fillings of the other variable gadgets.

\begin{lemma}
\label{lem:nodouble}
In a winning packing, the gadget of $v_i$ does not send \True along both a positive and a negative port.
Moreover, no piece can occupy cells in different variable gadgets.
\end{lemma}
\begin{proof}
We renumber the rows so that the bottom row of the gadget is row $0$.
Let $h = 6\max(d_i^+,d_i^-)+6$ be the height of the variable gadget; the gadget occupies rows $0$ to $h-1$.
Consider the piece covering the bottom pin $(0,0)$ of the foot (piece~\textsf{A} in Figure~\ref{fig:var}\subref{fig:var-true}).
The second claim of the lemma is shown by induction on $i$;
so by the induction hypothesis, the piece also covers $(0,1)$.
Suppose first that the piece also covers $(-1,0)$, as in Figure~\ref{fig:var}\subref{fig:var-true};
the other case is the mirror image.

The centre pin $(0,2)$ of the lossy slot in the foot cannot be covered, so $(-2,2)$ must be covered, necessarily by a piece in rows $2$ and $3$ (piece~\textsf{B}).
Going up, each pin $(-2,y)$ for $y \in \{2,4,\dots,h-4\}$ is then covered in rows $y$ and $y+1$, because the piece covering $(-2,y-2)$ takes $(-2,y-1)$; at the top, piece~\textsf{C} leaves the pin $(-2,h-2)$ uncovered, and it is the uncovered pin of the lossy slot in the head.
Every other pin of the gadget is covered.
Going down from piece~\textsf{D}, the pins $(2,y)$ for $y \in \{2,4,\dots,h-2\}$ are covered in rows $y$ and $y-1$, since $(2,h-1)$ is filled.

Consider a body module with a negative port, occupying rows $y$ to $y+5$ (Figure~\ref{fig:var}\subref{fig:var-true} shows one, containing pieces~\textsf{E}--\textsf{I}).
The piece covering the pin $(2,y+5)$ covers $(2,y+4)$, so it also covers $(1,y+4)$ (piece~\textsf{E}).
Since $(0,y)$ is filled, the pin $(0,y+1)$ is covered in rows $y+1$ and $y+2$ (piece~\textsf{F}), and so the pin $(0,y+3)$ in rows $y+3$ and $y+4$, hence in columns $0$ and $-1$ (piece~\textsf{G}).
Then the pins $(-2,y+3)$ and $(-4,y+3)$ are covered in columns $-3,-2$ and $-5,-4$ (pieces~\textsf{H} and~\textsf{I}), meaning that the negative port sends \False.

Finally, we show that no piece occupies cells in both this gadget and the next.
Since piece~\textsf{C} takes $(-1,h-3)$ and piece~\textsf{D} takes $(1,h-3)$ and $(1,h-2)$, the only way to cover the topmost pin $(0,h-2)$ is to place a piece in rows $h-2$ and $h-1$ (piece~\textsf{J}).
\end{proof}

A gadget may, however, send \False along ports on both sides, since the piece at a port that could send \True may instead lie one cell further out.
This is harmless, as a \False sent where \True was possible only weakens what reaches the clause gadgets.

\subsection{Wire}
\label{sec:wire}

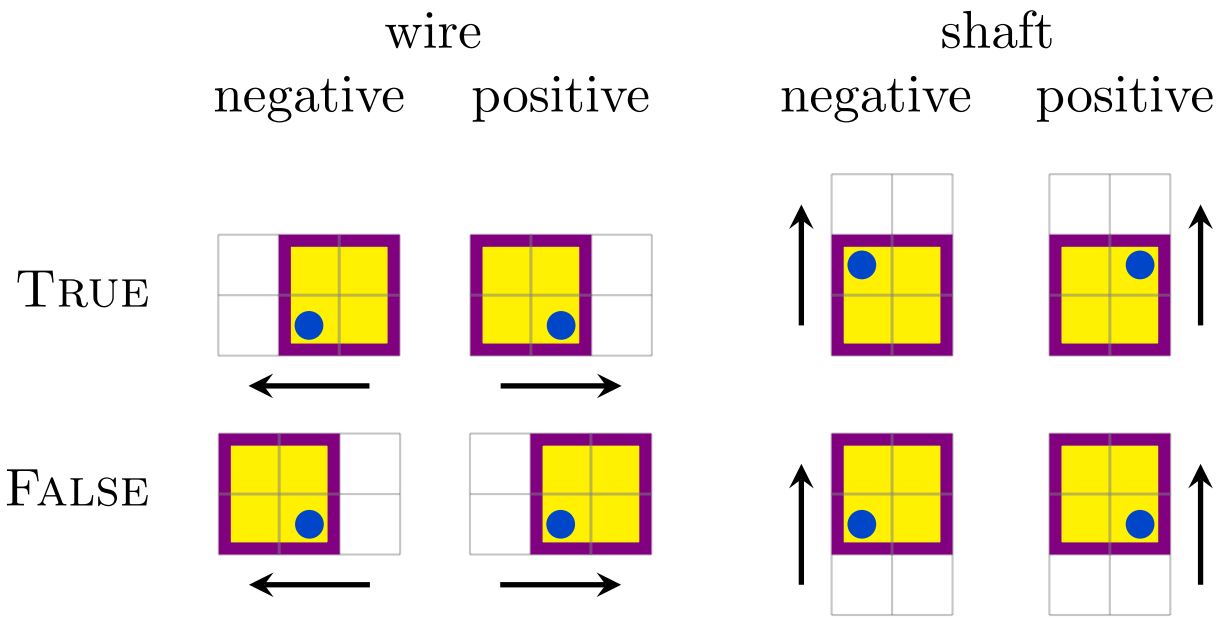
\begin{figure}[t]
\centering
\tikzset{legendpic/.style = {x = .4cm, y = .4cm,
  baseline = {([yshift = -.5ex] 0pt, {#1 * .4cm})}}}
\begin{tabular}{@{}r@{\hspace{1.2em}}c@{\hspace{1.2em}}c@{\hspace{2.4em}}c@{\hspace{1.2em}}c@{}}
& \multicolumn{2}{@{}c@{\hspace{2.4em}}}{wire} & \multicolumn{2}{@{}c@{}}{shaft} \\
  & negative & positive & negative & positive \\[8pt]
  \True
    & \begin{tikzpicture}[legendpic = 0.5]
        \whitebg{0}{2}{0}{1}
        \omino{1}{0}{1}
        \legendpin{1}{0}
        \draw[->, >=stealth, black, line width = 1] (2, -1) -- ++ (-2, 0);
      \end{tikzpicture}
    & \begin{tikzpicture}[legendpic = 0.5]
        \whitebg{0}{2}{0}{1}
        \omino{0}{0}{1}
        \legendpin{1}{0}
        \draw[->, >=stealth, black, line width = 1] (0, -1) -- ++ (2, 0);
      \end{tikzpicture}
    & \begin{tikzpicture}[legendpic = 0.5]
        \whitebg{0}{1}{0}{2}
        \omino{0}{0}{1}
        \legendpin{0}{1}
        \draw[->, >=stealth, black, line width = 1] (-1, 0) -- ++ (0, 2);
      \end{tikzpicture}
    & \begin{tikzpicture}[legendpic = 0.5]
        \whitebg{0}{1}{0}{2}
        \omino{0}{0}{1}
        \legendpin{1}{1}
        \draw[->, >=stealth, black, line width = 1] (2, 0) -- ++ (0, 2);
      \end{tikzpicture} \\[20pt]
  \False
    & \begin{tikzpicture}[legendpic = 0.5]
        \whitebg{0}{2}{0}{1}
        \omino{0}{0}{1}
        \legendpin{1}{0}
        \draw[->, >=stealth, black, line width = 1] (2, -1) -- ++ (-2, 0);
      \end{tikzpicture}
    & \begin{tikzpicture}[legendpic = 0.5]
        \whitebg{0}{2}{0}{1}
        \omino{1}{0}{1}
        \legendpin{1}{0}
        \draw[->, >=stealth, black, line width = 1] (0, -1) -- ++ (2, 0);
      \end{tikzpicture}
    & \begin{tikzpicture}[legendpic = 1.5]
        \whitebg{0}{1}{0}{2}
        \omino{0}{1}{1}
        \legendpin{0}{1}
        \draw[->, >=stealth, black, line width = 1] (-1, 0) -- ++ (0, 2);
      \end{tikzpicture}
    & \begin{tikzpicture}[legendpic = 1.5]
        \whitebg{0}{1}{0}{2}
        \omino{0}{1}{1}
        \legendpin{1}{1}
        \draw[->, >=stealth, black, line width = 1] (2, 0) -- ++ (0, 2);
      \end{tikzpicture} \\
\end{tabular}
\caption{How to read a truth value off a piece.
A piece reads \True if the pin it covers lies in the half of it further along the direction of the signal, shown by the arrow, and \False otherwise.}
\label{fig:legend}
\end{figure}

A wire is a horizontal corridor of height two and even length, running from an output port of a variable gadget to a clause gadget.
Recall that the variable gadget sends \True via a positive port if the pin $(x,y)$ and the neighbouring cell $(x-1,y)$ are covered by the same piece, and sends \False otherwise.
The signal is sent via a negative port symmetrically.
The same rule applies to every piece in a wire (Figure~\ref{fig:legend}).
If the variable gadget sends \False, the wire delivers \False to the clause gadget, as long as all of its pins are covered.
If the variable gadget sends \True, the wire may still deliver \False, but it is harmless.
Wires may be lengthened by any even number of cells, which is what lets the construction match the arrangement.
A wire may also have length zero, in which case the clause gadget is attached directly to the variable gadget, as $c_2^+$ is in Figure~\ref{fig:global}.

\subsection{Clause gadget}
\label{sec:clause}

\begin{figure}[t]
\centering
\begin{tikzpicture}[x = .28cm, y = .28cm,
    plabel/.style = {align = center, font = \small},
    leader/.style = {draw = orange, uwidth = .2}]
    \whitebg{6}{12}{-1}{26}
    \begin{scope}
        \clip (5.5, -1.5) rectangle (12.5, 26.5);
        \filler{0}{-1}
\filler{1}{-1}
\filler{2}{-1}
\filler{3}{-1}
\filler{4}{-1}
\filler{5}{-1}
\filler{6}{-1}
\filler{7}{-1}
\filler{8}{-1}
\filler{9}{-1}
\filler{10}{-1}
\filler{11}{-1}
\filler{12}{-1}
\filler{10}{0}
\filler{11}{0}
\filler{12}{0}
\filler{11}{1}
\filler{12}{1}
\filler{0}{2}
\filler{1}{2}
\filler{2}{2}
\filler{3}{2}
\filler{4}{2}
\filler{5}{2}
\filler{6}{2}
\filler{7}{2}
\filler{8}{2}
\filler{11}{2}
\filler{12}{2}
\filler{0}{3}
\filler{1}{3}
\filler{2}{3}
\filler{3}{3}
\filler{4}{3}
\filler{5}{3}
\filler{6}{3}
\filler{7}{3}
\filler{12}{3}
\filler{0}{4}
\filler{1}{4}
\filler{2}{4}
\filler{3}{4}
\filler{4}{4}
\filler{5}{4}
\filler{6}{4}
\filler{7}{4}
\filler{12}{4}
\filler{0}{5}
\filler{1}{5}
\filler{2}{5}
\filler{3}{5}
\filler{4}{5}
\filler{5}{5}
\filler{6}{5}
\filler{7}{5}
\filler{12}{5}
\filler{0}{6}
\filler{1}{6}
\filler{2}{6}
\filler{3}{6}
\filler{4}{6}
\filler{5}{6}
\filler{6}{6}
\filler{10}{6}
\filler{11}{6}
\filler{12}{6}
\filler{0}{7}
\filler{1}{7}
\filler{2}{7}
\filler{3}{7}
\filler{4}{7}
\filler{5}{7}
\filler{6}{7}
\filler{10}{7}
\filler{11}{7}
\filler{12}{7}
\filler{0}{8}
\filler{1}{8}
\filler{2}{8}
\filler{3}{8}
\filler{4}{8}
\filler{5}{8}
\filler{6}{8}
\filler{9}{8}
\filler{10}{8}
\filler{11}{8}
\filler{12}{8}
\filler{0}{9}
\filler{1}{9}
\filler{2}{9}
\filler{3}{9}
\filler{4}{9}
\filler{5}{9}
\filler{6}{9}
\filler{9}{9}
\filler{10}{9}
\filler{11}{9}
\filler{12}{9}
\filler{0}{10}
\filler{1}{10}
\filler{2}{10}
\filler{3}{10}
\filler{4}{10}
\filler{5}{10}
\filler{6}{10}
\filler{9}{10}
\filler{10}{10}
\filler{11}{10}
\filler{12}{10}
\filler{0}{11}
\filler{1}{11}
\filler{2}{11}
\filler{3}{11}
\filler{4}{11}
\filler{5}{11}
\filler{6}{11}
\filler{10}{11}
\filler{11}{11}
\filler{12}{11}
\filler{10}{12}
\filler{11}{12}
\filler{12}{12}
\filler{11}{13}
\filler{12}{13}
\filler{0}{14}
\filler{1}{14}
\filler{2}{14}
\filler{3}{14}
\filler{4}{14}
\filler{5}{14}
\filler{6}{14}
\filler{7}{14}
\filler{8}{14}
\filler{11}{14}
\filler{12}{14}
\filler{0}{15}
\filler{1}{15}
\filler{2}{15}
\filler{3}{15}
\filler{4}{15}
\filler{5}{15}
\filler{6}{15}
\filler{7}{15}
\filler{12}{15}
\filler{0}{16}
\filler{1}{16}
\filler{2}{16}
\filler{3}{16}
\filler{4}{16}
\filler{5}{16}
\filler{6}{16}
\filler{7}{16}
\filler{12}{16}
\filler{0}{17}
\filler{1}{17}
\filler{2}{17}
\filler{3}{17}
\filler{4}{17}
\filler{5}{17}
\filler{6}{17}
\filler{7}{17}
\filler{12}{17}
\filler{0}{18}
\filler{1}{18}
\filler{2}{18}
\filler{3}{18}
\filler{4}{18}
\filler{5}{18}
\filler{6}{18}
\filler{10}{18}
\filler{11}{18}
\filler{12}{18}
\filler{0}{19}
\filler{1}{19}
\filler{2}{19}
\filler{3}{19}
\filler{4}{19}
\filler{5}{19}
\filler{6}{19}
\filler{10}{19}
\filler{11}{19}
\filler{12}{19}
\filler{0}{20}
\filler{1}{20}
\filler{2}{20}
\filler{3}{20}
\filler{4}{20}
\filler{5}{20}
\filler{6}{20}
\filler{9}{20}
\filler{10}{20}
\filler{11}{20}
\filler{12}{20}
\filler{0}{21}
\filler{1}{21}
\filler{2}{21}
\filler{3}{21}
\filler{4}{21}
\filler{5}{21}
\filler{6}{21}
\filler{9}{21}
\filler{10}{21}
\filler{11}{21}
\filler{12}{21}
\filler{0}{22}
\filler{1}{22}
\filler{2}{22}
\filler{3}{22}
\filler{4}{22}
\filler{5}{22}
\filler{6}{22}
\filler{9}{22}
\filler{10}{22}
\filler{11}{22}
\filler{12}{22}
\filler{0}{23}
\filler{1}{23}
\filler{2}{23}
\filler{3}{23}
\filler{4}{23}
\filler{5}{23}
\filler{6}{23}
\filler{10}{23}
\filler{11}{23}
\filler{12}{23}
\filler{10}{24}
\filler{11}{24}
\filler{12}{24}
\filler{9}{25}
\filler{10}{25}
\filler{11}{25}
\filler{12}{25}
\filler{0}{26}
\filler{1}{26}
\filler{2}{26}
\filler{3}{26}
\filler{4}{26}
\filler{5}{26}
\filler{6}{26}
\filler{7}{26}
\filler{8}{26}
\filler{9}{26}
\filler{10}{26}
\filler{11}{26}
\filler{12}{26}
     \end{scope}
    \foreach \x/\y in {6/0, 8/0, 10/2, 8/4, 10/4, 8/6, 8/8, 8/10,
                       6/12, 8/12, 10/14, 8/16, 10/16, 8/18, 8/20, 8/22,
                       6/24, 8/24}{
        \legendpin{\x}{\y}
    }
\foreach \y in {0, 12, 24}{
        \draw [->, >=stealth, black, line width = 1] (5.1, \y + .5) -- ++ (-2, 0)
            node [anchor = east, font = \small, inner sep = 2pt] {wire};
    }
\foreach \ya/\yb in {0/7, 10/19, 22/25}{
        \draw [draw = orange, uwidth = .25] (5.5 + .125, \ya - .5 + .125) rectangle (11.5 - .125, \yb + .5 - .125);
    }
    \node [plabel] (top)    at (17, 23.5) {top};
    \node [plabel] (or)     at (17, 14.5) {middle};
    \node [plabel] (bottom) at (17, 3.5)  {bottom};
    \node [plabel] (corra)  at (17, 8.5)  {shaft};
    \node [plabel] (corrb)  at (17, 20.5) {shaft};
    \draw [leader] (top)    -- (11.5, 23.5);
    \draw [leader] (or)     -- (11.5, 14.5);
    \draw [leader] (bottom) -- (11.5, 3.5);
\draw [leader] (corra)  -- (8.5, 8.5);
    \draw [leader] (corrb)  -- (8.5, 20.5);
\end{tikzpicture}
\caption{A gadget for a positive clause with three literals. The length of a shaft is two more than a multiple of six, since the ports of different variable gadgets lie a multiple of six rows apart.}
\label{fig:clause-parts}
\end{figure}
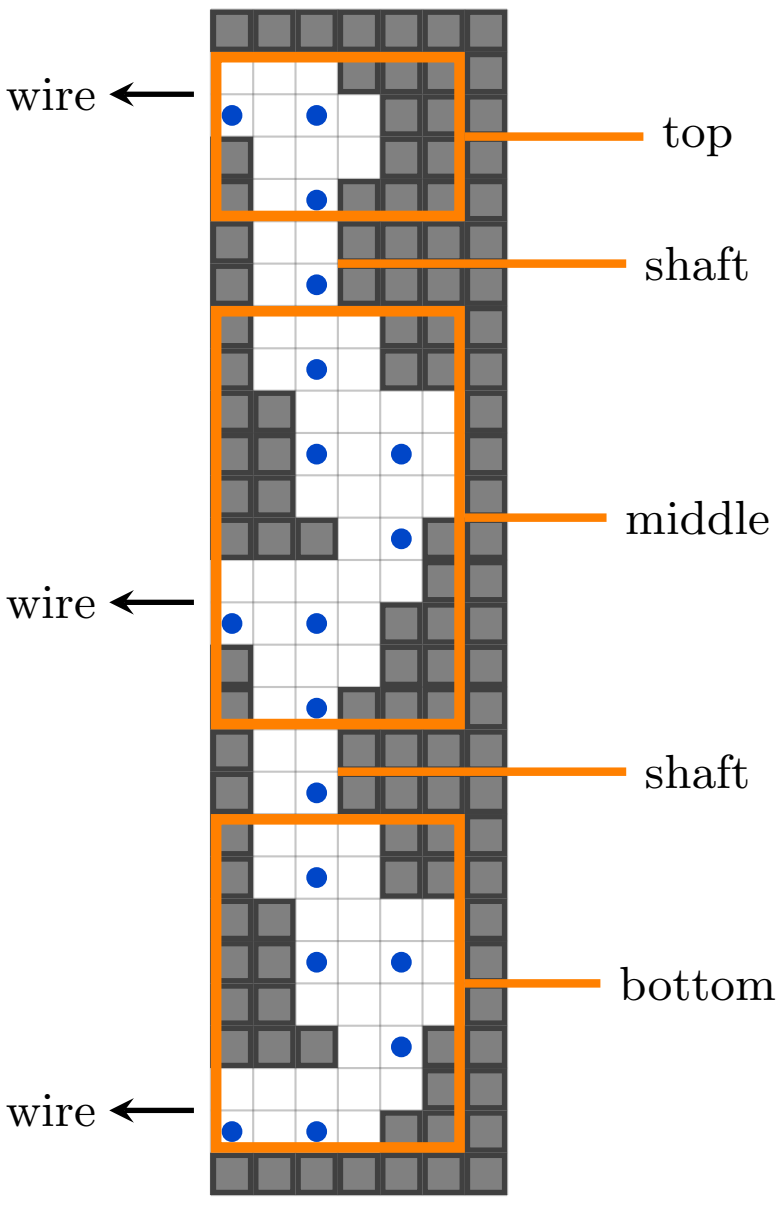

Let $c = \ell_{i_1} \vee \cdots \vee \ell_{i_k}$ with $i_1 < \cdots < i_k$, where $\ell_{i_j}$ is a literal of the variable $v_{i_j}$.
Its gadget consists of $k$ modules, one per literal, placed one above another and joined by \emph{shafts}, vertical strips of empty cells two cells wide (Figure~\ref{fig:clause-parts}).
The module for the first literal $\ell_{i_1}$, called the \emph{bottom module}, converts the truth value from its wire into a vertical output to the module above.
The module for each of $\ell_{i_2},\dots,\ell_{i_{k-1}}$, called a \emph{middle module}, takes a horizontal input from the wire of its variable and a vertical input from the module below, and emits their disjunction as a vertical output to the module above.
The module for the last literal $\ell_{i_k}$, called the \emph{top module}, can have all of its pins covered if and only if the disjunction of its horizontal and vertical inputs is \True.
The gadget therefore evaluates $c$ from the bottom up, carrying a running \True upward as soon as one literal is satisfied, so that all of its pins can be covered if and only if $c$ is satisfied by the values it receives.

Figure~\ref{fig:clause} draws the three modules of a positive clause gadget.
Negative clauses use the mirror images of the same modules; every statement about a module carries over to its mirror image unchanged.
The truth value is transmitted in a clause gadget by the following rule.
If a pin $(x,y)$ and the cell below $(x,y-1)$ are covered by the same piece, that piece carries \True; if a pin $(x,y)$ and the cell above $(x,y+1)$ are covered by the same piece, that piece carries \False (Figure~\ref{fig:legend}).
A shaft passes the truth value from the module below to the module above unchanged.

\begin{figure}[tp]
\centering
\begin{subfigure}[t]{\linewidth}
  \centering
\begin{tabular}{@{}cc@{}}
      \begin{tikzpicture}[x = .28cm, y = .28cm]
\clip (3.5, 24.5) rectangle ++ (12, 14);
        \whitebg{4}{16}{22}{38}
        \filler{4}{22}
\filler{4}{23}
\filler{4}{24}
\filler{4}{25}
\filler{4}{30}
\filler{4}{31}
\filler{4}{32}
\filler{4}{33}
\filler{4}{34}
\filler{4}{35}
\filler{4}{36}
\filler{4}{37}
\filler{4}{38}
\filler{5}{22}
\filler{5}{23}
\filler{5}{24}
\filler{5}{25}
\filler{5}{30}
\filler{5}{31}
\filler{5}{32}
\filler{5}{33}
\filler{5}{34}
\filler{5}{35}
\filler{5}{36}
\filler{5}{37}
\filler{5}{38}
\filler{6}{22}
\filler{6}{23}
\filler{6}{24}
\filler{6}{25}
\filler{6}{30}
\filler{6}{31}
\filler{6}{32}
\filler{6}{33}
\filler{6}{34}
\filler{6}{35}
\filler{6}{36}
\filler{6}{37}
\filler{6}{38}
\filler{7}{22}
\filler{7}{23}
\filler{7}{24}
\filler{7}{25}
\filler{7}{30}
\filler{7}{31}
\filler{7}{32}
\filler{7}{33}
\filler{7}{34}
\filler{7}{35}
\filler{7}{36}
\filler{7}{37}
\filler{7}{38}
\filler{8}{22}
\filler{8}{23}
\filler{8}{24}
\filler{8}{25}
\filler{8}{30}
\filler{8}{31}
\filler{8}{32}
\filler{9}{22}
\filler{9}{23}
\filler{9}{24}
\filler{9}{25}
\filler{10}{22}
\filler{10}{23}
\filler{10}{24}
\filler{10}{25}
\filler{11}{22}
\filler{11}{23}
\filler{11}{24}
\filler{11}{25}
\filler{12}{22}
\filler{12}{23}
\filler{12}{24}
\filler{12}{25}
\filler{12}{26}
\filler{12}{35}
\filler{12}{36}
\filler{12}{37}
\filler{12}{38}
\filler{13}{22}
\filler{13}{23}
\filler{13}{24}
\filler{13}{25}
\filler{13}{26}
\filler{13}{27}
\filler{13}{34}
\filler{13}{35}
\filler{13}{36}
\filler{13}{37}
\filler{13}{38}
\filler{14}{22}
\filler{14}{23}
\filler{14}{24}
\filler{14}{25}
\filler{14}{26}
\filler{14}{27}
\filler{14}{28}
\filler{14}{29}
\filler{14}{33}
\filler{14}{34}
\filler{14}{35}
\filler{14}{36}
\filler{14}{37}
\filler{14}{38}
\filler{15}{22}
\filler{15}{23}
\filler{15}{24}
\filler{15}{25}
\filler{15}{26}
\filler{15}{27}
\filler{15}{28}
\filler{15}{29}
\filler{15}{30}
\filler{15}{31}
\filler{15}{32}
\filler{15}{33}
\filler{15}{34}
\filler{15}{35}
\filler{15}{36}
\filler{15}{37}
\filler{15}{38}
\filler{16}{22}
\filler{16}{23}
\filler{16}{24}
\filler{16}{25}
\filler{16}{26}
\filler{16}{27}
\filler{16}{28}
\filler{16}{29}
\filler{16}{30}
\filler{16}{31}
\filler{16}{32}
\filler{16}{33}
\filler{16}{34}
\filler{16}{35}
\filler{16}{36}
\filler{16}{37}
\filler{16}{38}
         \clauseblock{1}{0}{2}{0}{0}{0}{2}

        \filler{4}{26}
        \filler{5}{26}
        \filler{6}{26}
        \filler{7}{26}
        \filler{4}{29}
        \filler{5}{29}
        \filler{6}{29}
        \filler{7}{29}
        \filler{13}{28}
        \filler{13}{29}
        \filler{14}{30}
        \filler{9}{30}
        \filler{14}{31}
        \filler{9}{31}
        \filler{14}{32}
        \filler{9}{32}
        \filler{13}{33}
        \filler{12}{33}
        \filler{8}{33}
        \filler{12}{34}
        \filler{8}{34}
        \filler{11}{35}
        \filler{8}{35}
        \filler{11}{36}
        \filler{8}{36}
        \filler{11}{37}
        \filler{8}{37}
        \filler{11}{38}
        \filler{8}{38}

        \omino{3}{27}{1}
        \omino{5}{27}{1}

        \omino{11}{28}{1}
        \omino{12}{30}{1}
        \omino{10}{30}{1}
        \omino{10}{32}{1}
        \omino{9}{34}{1}
        \omino{9}{36}{1}
\omino{9}{38}{1}

        \foreach \x in {4, 6, ..., 10}{ \legendpin{\x}{27} }
        \foreach \y in {31, 33, ..., 37}{ \legendpin{10}{\y} }
        \foreach \y in {27}{ \legendpin{10}{\y} }
        \legendpin{12}{29}
        \legendpin{12}{31}
\draw [draw = orange, uwidth = .25] (7.625, 26.625) rectangle (13.375, 34.375);
\draw [draw = cyan, uwidth = .3] (2.5, 26.5) rectangle ++ (2, 2);
        \draw [draw = cyan, uwidth = .3] (4.5, 26.5) rectangle ++ (2, 2);
        \draw [draw = cyan, uwidth = .3] (8.5, 33.5) rectangle ++ (2, 2);
        \draw [draw = cyan, uwidth = .3] (8.5, 35.5) rectangle ++ (2, 2);
        \draw [draw = cyan, uwidth = .3] (8.5, 37.5) rectangle ++ (2, 2);
      \end{tikzpicture} &
      \begin{tikzpicture}[x = .28cm, y = .28cm]
\clip (3.5, 24.5) rectangle ++ (12, 14);
        \whitebg{4}{16}{22}{38}
        \filler{4}{22}
\filler{4}{23}
\filler{4}{24}
\filler{4}{25}
\filler{4}{30}
\filler{4}{31}
\filler{4}{32}
\filler{4}{33}
\filler{4}{34}
\filler{4}{35}
\filler{4}{36}
\filler{4}{37}
\filler{4}{38}
\filler{5}{22}
\filler{5}{23}
\filler{5}{24}
\filler{5}{25}
\filler{5}{30}
\filler{5}{31}
\filler{5}{32}
\filler{5}{33}
\filler{5}{34}
\filler{5}{35}
\filler{5}{36}
\filler{5}{37}
\filler{5}{38}
\filler{6}{22}
\filler{6}{23}
\filler{6}{24}
\filler{6}{25}
\filler{6}{30}
\filler{6}{31}
\filler{6}{32}
\filler{6}{33}
\filler{6}{34}
\filler{6}{35}
\filler{6}{36}
\filler{6}{37}
\filler{6}{38}
\filler{7}{22}
\filler{7}{23}
\filler{7}{24}
\filler{7}{25}
\filler{7}{30}
\filler{7}{31}
\filler{7}{32}
\filler{7}{33}
\filler{7}{34}
\filler{7}{35}
\filler{7}{36}
\filler{7}{37}
\filler{7}{38}
\filler{8}{22}
\filler{8}{23}
\filler{8}{24}
\filler{8}{25}
\filler{8}{30}
\filler{8}{31}
\filler{8}{32}
\filler{9}{22}
\filler{9}{23}
\filler{9}{24}
\filler{9}{25}
\filler{10}{22}
\filler{10}{23}
\filler{10}{24}
\filler{10}{25}
\filler{11}{22}
\filler{11}{23}
\filler{11}{24}
\filler{11}{25}
\filler{12}{22}
\filler{12}{23}
\filler{12}{24}
\filler{12}{25}
\filler{12}{26}
\filler{12}{35}
\filler{12}{36}
\filler{12}{37}
\filler{12}{38}
\filler{13}{22}
\filler{13}{23}
\filler{13}{24}
\filler{13}{25}
\filler{13}{26}
\filler{13}{27}
\filler{13}{34}
\filler{13}{35}
\filler{13}{36}
\filler{13}{37}
\filler{13}{38}
\filler{14}{22}
\filler{14}{23}
\filler{14}{24}
\filler{14}{25}
\filler{14}{26}
\filler{14}{27}
\filler{14}{28}
\filler{14}{29}
\filler{14}{33}
\filler{14}{34}
\filler{14}{35}
\filler{14}{36}
\filler{14}{37}
\filler{14}{38}
\filler{15}{22}
\filler{15}{23}
\filler{15}{24}
\filler{15}{25}
\filler{15}{26}
\filler{15}{27}
\filler{15}{28}
\filler{15}{29}
\filler{15}{30}
\filler{15}{31}
\filler{15}{32}
\filler{15}{33}
\filler{15}{34}
\filler{15}{35}
\filler{15}{36}
\filler{15}{37}
\filler{15}{38}
\filler{16}{22}
\filler{16}{23}
\filler{16}{24}
\filler{16}{25}
\filler{16}{26}
\filler{16}{27}
\filler{16}{28}
\filler{16}{29}
\filler{16}{30}
\filler{16}{31}
\filler{16}{32}
\filler{16}{33}
\filler{16}{34}
\filler{16}{35}
\filler{16}{36}
\filler{16}{37}
\filler{16}{38}
         \clauseblock{1}{0}{2}{0}{0}{0}{3}

        \filler{4}{26}
        \filler{5}{26}
        \filler{6}{26}
        \filler{7}{26}
        \filler{4}{29}
        \filler{5}{29}
        \filler{6}{29}
        \filler{7}{29}
        \filler{13}{28}
        \filler{13}{29}
        \filler{14}{30}
        \filler{9}{30}
        \filler{14}{31}
        \filler{9}{31}
        \filler{14}{32}
        \filler{9}{32}
        \filler{13}{33}
        \filler{12}{33}
        \filler{8}{33}
        \filler{12}{34}
        \filler{8}{34}
        \filler{11}{35}
        \filler{8}{35}
        \filler{11}{36}
        \filler{8}{36}
        \filler{11}{37}
        \filler{8}{37}
        \filler{11}{38}
        \filler{8}{38}

        \omino{4}{27}{1}
        \omino{6}{27}{1}

        \omino{11}{29}{1}
        \omino{12}{31}{1}
        \omino{10}{31}{1}
        \omino{10}{33}{1}
        \omino{9}{35}{1}
        \omino{9}{37}{1}

        \foreach \x in {4, 6, ..., 10}{ \legendpin{\x}{27} }
        \foreach \y in {31, 33, ..., 37}{ \legendpin{10}{\y} }
        \foreach \y in {27}{ \legendpin{10}{\y} }
        \legendpin{12}{29}
        \legendpin{12}{31}
\draw [draw = orange, uwidth = .25] (7.625, 26.625) rectangle (13.375, 34.375);
\draw [draw = red, uwidth = .3] (3.5, 26.5) rectangle ++ (2, 2);
        \draw [draw = red, uwidth = .3] (5.5, 26.5) rectangle ++ (2, 2);
        \draw [draw = red, uwidth = .3] (8.5, 34.5) rectangle ++ (2, 2);
        \draw [draw = red, uwidth = .3] (8.5, 36.5) rectangle ++ (2, 2);
      \end{tikzpicture} \\[2pt]
      \scriptsize \True & \scriptsize \False
  \end{tabular}
  \caption{The bottom module, whose vertical input is filled.}
  \label{fig:clause-bottom}
\end{subfigure}

\vspace{1.2em}
\begin{subfigure}[t]{\linewidth}
  \centering
\begin{tabular}{@{}cccc@{}}
      \begin{tikzpicture}[x = .28cm, y = .28cm]
\clip (3.5, 21.5) rectangle ++ (12, 17);
        \whitebg{4}{16}{22}{38}
        \filler{4}{22}
\filler{4}{23}
\filler{4}{24}
\filler{4}{25}
\filler{4}{30}
\filler{4}{31}
\filler{4}{32}
\filler{4}{33}
\filler{4}{34}
\filler{4}{35}
\filler{4}{36}
\filler{4}{37}
\filler{4}{38}
\filler{5}{22}
\filler{5}{23}
\filler{5}{24}
\filler{5}{25}
\filler{5}{30}
\filler{5}{31}
\filler{5}{32}
\filler{5}{33}
\filler{5}{34}
\filler{5}{35}
\filler{5}{36}
\filler{5}{37}
\filler{5}{38}
\filler{6}{22}
\filler{6}{23}
\filler{6}{24}
\filler{6}{25}
\filler{6}{30}
\filler{6}{31}
\filler{6}{32}
\filler{6}{33}
\filler{6}{34}
\filler{6}{35}
\filler{6}{36}
\filler{6}{37}
\filler{6}{38}
\filler{7}{22}
\filler{7}{23}
\filler{7}{24}
\filler{7}{25}
\filler{7}{30}
\filler{7}{31}
\filler{7}{32}
\filler{7}{33}
\filler{7}{34}
\filler{7}{35}
\filler{7}{36}
\filler{7}{37}
\filler{7}{38}
\filler{8}{30}
\filler{8}{31}
\filler{8}{32}
\filler{12}{22}
\filler{12}{23}
\filler{12}{24}
\filler{12}{25}
\filler{12}{35}
\filler{12}{36}
\filler{12}{37}
\filler{12}{38}
\filler{13}{22}
\filler{13}{23}
\filler{13}{24}
\filler{13}{25}
\filler{13}{26}
\filler{13}{27}
\filler{13}{34}
\filler{13}{35}
\filler{13}{36}
\filler{13}{37}
\filler{13}{38}
\filler{14}{22}
\filler{14}{23}
\filler{14}{24}
\filler{14}{25}
\filler{14}{26}
\filler{14}{27}
\filler{14}{28}
\filler{14}{29}
\filler{14}{33}
\filler{14}{34}
\filler{14}{35}
\filler{14}{36}
\filler{14}{37}
\filler{14}{38}
\filler{15}{22}
\filler{15}{23}
\filler{15}{24}
\filler{15}{25}
\filler{15}{26}
\filler{15}{27}
\filler{15}{28}
\filler{15}{29}
\filler{15}{30}
\filler{15}{31}
\filler{15}{32}
\filler{15}{33}
\filler{15}{34}
\filler{15}{35}
\filler{15}{36}
\filler{15}{37}
\filler{15}{38}
\filler{16}{22}
\filler{16}{23}
\filler{16}{24}
\filler{16}{25}
\filler{16}{26}
\filler{16}{27}
\filler{16}{28}
\filler{16}{29}
\filler{16}{30}
\filler{16}{31}
\filler{16}{32}
\filler{16}{33}
\filler{16}{34}
\filler{16}{35}
\filler{16}{36}
\filler{16}{37}
\filler{16}{38}
         \clauseblock{1}{0}{2}{1}{0}{0}{0}

        \filler{4}{26}
        \filler{5}{26}
        \filler{6}{26}
        \filler{7}{26}
        \filler{4}{29}
        \filler{5}{29}
        \filler{6}{29}
        \filler{7}{29}

        \filler{13}{28}
        \filler{13}{29}
        \filler{14}{30}
        \filler{9}{30}
        \filler{14}{31}
        \filler{9}{31}
        \filler{14}{32}
        \filler{9}{32}
        \filler{13}{33}
        \filler{12}{33}
        \filler{8}{33}
        \filler{12}{34}
        \filler{8}{34}
        \filler{11}{35}
        \filler{8}{35}
        \filler{11}{36}
        \filler{8}{36}
        \filler{11}{37}
        \filler{8}{37}
        \filler{11}{38}
        \filler{8}{38}

        \omino{3}{27}{1}
        \omino{5}{27}{1}

        \omino{9}{24}{1}
        \omino{9}{22}{1}

        \omino{11}{28}{1}
        \omino{12}{30}{1}
        \omino{10}{30}{1}
        \omino{10}{32}{1}
        \omino{9}{34}{1}
        \omino{9}{36}{1}
\omino{9}{38}{1}

        \foreach \x in {4, 6, ..., 10}{ \legendpin{\x}{27} }
        \foreach \y in {31, 33, ..., 37}{ \legendpin{10}{\y} }
        \foreach \y in {27, 25, 23}{ \legendpin{10}{\y} }
        \legendpin{12}{29}
        \legendpin{12}{31}
\draw [draw = orange, uwidth = .25] (7.625, 24.625) rectangle (13.375, 34.375);
\draw [draw = cyan, uwidth = .3] (10.5, 3.5) rectangle ++ (2, 2);
        \draw [draw = cyan, uwidth = .3] (9.5, 5.5) rectangle ++ (2, 2);
        \draw [draw = cyan, uwidth = .3] (11.5, 5.5) rectangle ++ (2, 2);
        \draw [draw = cyan, uwidth = .3] (9.5, 7.5) rectangle ++ (2, 2);
        \draw [draw = cyan, uwidth = .3] (8.5, 9.5) rectangle ++ (2, 2);
        \draw [draw = cyan, uwidth = .3] (8.5, 11.5) rectangle ++ (2, 2);
        \draw [draw = cyan, uwidth = .3] (8.5, 13.5) rectangle ++ (2, 2);
        \draw [draw = cyan, uwidth = .3] (8.5, 15.5) rectangle ++ (2, 2);
        \draw [draw = cyan, uwidth = .3] (8.5, 17.5) rectangle ++ (2, 2);
        \draw [draw = cyan, uwidth = .3] (8.5, 19.5) rectangle ++ (2, 2);
        \draw [draw = cyan, uwidth = .3] (8.5, 21.5) rectangle ++ (2, 2);
        \draw [draw = cyan, uwidth = .3] (2.5, 26.5) rectangle ++ (2, 2);
        \draw [draw = cyan, uwidth = .3] (4.5, 26.5) rectangle ++ (2, 2);
        \draw [draw = cyan, uwidth = .3] (8.5, 33.5) rectangle ++ (2, 2);
        \draw [draw = cyan, uwidth = .3] (8.5, 35.5) rectangle ++ (2, 2);
        \draw [draw = cyan, uwidth = .3] (8.5, 37.5) rectangle ++ (2, 2);
      \end{tikzpicture} &
      \begin{tikzpicture}[x = .28cm, y = .28cm]
\clip (3.5, 21.5) rectangle ++ (12, 17);
        \whitebg{4}{16}{22}{38}
        \filler{4}{22}
\filler{4}{23}
\filler{4}{24}
\filler{4}{25}
\filler{4}{30}
\filler{4}{31}
\filler{4}{32}
\filler{4}{33}
\filler{4}{34}
\filler{4}{35}
\filler{4}{36}
\filler{4}{37}
\filler{4}{38}
\filler{5}{22}
\filler{5}{23}
\filler{5}{24}
\filler{5}{25}
\filler{5}{30}
\filler{5}{31}
\filler{5}{32}
\filler{5}{33}
\filler{5}{34}
\filler{5}{35}
\filler{5}{36}
\filler{5}{37}
\filler{5}{38}
\filler{6}{22}
\filler{6}{23}
\filler{6}{24}
\filler{6}{25}
\filler{6}{30}
\filler{6}{31}
\filler{6}{32}
\filler{6}{33}
\filler{6}{34}
\filler{6}{35}
\filler{6}{36}
\filler{6}{37}
\filler{6}{38}
\filler{7}{22}
\filler{7}{23}
\filler{7}{24}
\filler{7}{25}
\filler{7}{30}
\filler{7}{31}
\filler{7}{32}
\filler{7}{33}
\filler{7}{34}
\filler{7}{35}
\filler{7}{36}
\filler{7}{37}
\filler{7}{38}
\filler{8}{30}
\filler{8}{31}
\filler{8}{32}
\filler{12}{22}
\filler{12}{23}
\filler{12}{24}
\filler{12}{25}
\filler{12}{35}
\filler{12}{36}
\filler{12}{37}
\filler{12}{38}
\filler{13}{22}
\filler{13}{23}
\filler{13}{24}
\filler{13}{25}
\filler{13}{26}
\filler{13}{27}
\filler{13}{34}
\filler{13}{35}
\filler{13}{36}
\filler{13}{37}
\filler{13}{38}
\filler{14}{22}
\filler{14}{23}
\filler{14}{24}
\filler{14}{25}
\filler{14}{26}
\filler{14}{27}
\filler{14}{28}
\filler{14}{29}
\filler{14}{33}
\filler{14}{34}
\filler{14}{35}
\filler{14}{36}
\filler{14}{37}
\filler{14}{38}
\filler{15}{22}
\filler{15}{23}
\filler{15}{24}
\filler{15}{25}
\filler{15}{26}
\filler{15}{27}
\filler{15}{28}
\filler{15}{29}
\filler{15}{30}
\filler{15}{31}
\filler{15}{32}
\filler{15}{33}
\filler{15}{34}
\filler{15}{35}
\filler{15}{36}
\filler{15}{37}
\filler{15}{38}
\filler{16}{22}
\filler{16}{23}
\filler{16}{24}
\filler{16}{25}
\filler{16}{26}
\filler{16}{27}
\filler{16}{28}
\filler{16}{29}
\filler{16}{30}
\filler{16}{31}
\filler{16}{32}
\filler{16}{33}
\filler{16}{34}
\filler{16}{35}
\filler{16}{36}
\filler{16}{37}
\filler{16}{38}
         \clauseblock{1}{0}{2}{1}{0}{0}{2}

        \filler{4}{26}
        \filler{5}{26}
        \filler{6}{26}
        \filler{7}{26}
        \filler{4}{29}
        \filler{5}{29}
        \filler{6}{29}
        \filler{7}{29}

        \filler{13}{28}
        \filler{13}{29}
        \filler{14}{30}
        \filler{9}{30}
        \filler{14}{31}
        \filler{9}{31}
        \filler{14}{32}
        \filler{9}{32}
        \filler{13}{33}
        \filler{12}{33}
        \filler{8}{33}
        \filler{12}{34}
        \filler{8}{34}
        \filler{11}{35}
        \filler{8}{35}
        \filler{11}{36}
        \filler{8}{36}
        \filler{11}{37}
        \filler{8}{37}
        \filler{11}{38}
        \filler{8}{38}

        \omino{3}{27}{1}
        \omino{5}{27}{1}

        \omino{9}{25}{1}
        \omino{9}{23}{1}
        \omino{9}{21}{1}

        \omino{11}{28}{1}
        \omino{12}{30}{1}
        \omino{10}{30}{1}
        \omino{10}{32}{1}
        \omino{9}{34}{1}
        \omino{9}{36}{1}
\omino{9}{38}{1}

        \foreach \x in {4, 6, ..., 10}{ \legendpin{\x}{27} }
        \foreach \y in {31, 33, ..., 37}{ \legendpin{10}{\y} }
        \foreach \y in {27, 25, 23}{ \legendpin{10}{\y} }
        \legendpin{12}{29}
        \legendpin{12}{31}
\draw [draw = orange, uwidth = .25] (7.625, 24.625) rectangle (13.375, 34.375);
\draw [draw = red, uwidth = .3] (10.5, 4.5) rectangle ++ (2, 2);
        \draw [draw = red, uwidth = .3] (9.5, 6.5) rectangle ++ (2, 2);
        \draw [draw = red, uwidth = .3] (11.5, 6.5) rectangle ++ (2, 2);
        \draw [draw = red, uwidth = .3] (9.5, 8.5) rectangle ++ (2, 2);
        \draw [draw = red, uwidth = .3] (8.5, 10.5) rectangle ++ (2, 2);
        \draw [draw = red, uwidth = .3] (8.5, 12.5) rectangle ++ (2, 2);
        \draw [draw = red, uwidth = .3] (8.5, 14.5) rectangle ++ (2, 2);
        \draw [draw = red, uwidth = .3] (8.5, 16.5) rectangle ++ (2, 2);
        \draw [draw = red, uwidth = .3] (8.5, 18.5) rectangle ++ (2, 2);
        \draw [draw = red, uwidth = .3] (8.5, 20.5) rectangle ++ (2, 2);
        \draw [draw = red, uwidth = .3] (8.5, 22.5) rectangle ++ (2, 2);
        \draw [draw = cyan, uwidth = .3] (2.5, 26.5) rectangle ++ (2, 2);
        \draw [draw = cyan, uwidth = .3] (4.5, 26.5) rectangle ++ (2, 2);
        \draw [draw = cyan, uwidth = .3] (8.5, 33.5) rectangle ++ (2, 2);
        \draw [draw = cyan, uwidth = .3] (8.5, 35.5) rectangle ++ (2, 2);
        \draw [draw = cyan, uwidth = .3] (8.5, 37.5) rectangle ++ (2, 2);
      \end{tikzpicture} &
      \begin{tikzpicture}[x = .28cm, y = .28cm]
\clip (3.5, 21.5) rectangle ++ (12, 17);
        \whitebg{4}{16}{22}{38}
        \filler{4}{22}
\filler{4}{23}
\filler{4}{24}
\filler{4}{25}
\filler{4}{30}
\filler{4}{31}
\filler{4}{32}
\filler{4}{33}
\filler{4}{34}
\filler{4}{35}
\filler{4}{36}
\filler{4}{37}
\filler{4}{38}
\filler{5}{22}
\filler{5}{23}
\filler{5}{24}
\filler{5}{25}
\filler{5}{30}
\filler{5}{31}
\filler{5}{32}
\filler{5}{33}
\filler{5}{34}
\filler{5}{35}
\filler{5}{36}
\filler{5}{37}
\filler{5}{38}
\filler{6}{22}
\filler{6}{23}
\filler{6}{24}
\filler{6}{25}
\filler{6}{30}
\filler{6}{31}
\filler{6}{32}
\filler{6}{33}
\filler{6}{34}
\filler{6}{35}
\filler{6}{36}
\filler{6}{37}
\filler{6}{38}
\filler{7}{22}
\filler{7}{23}
\filler{7}{24}
\filler{7}{25}
\filler{7}{30}
\filler{7}{31}
\filler{7}{32}
\filler{7}{33}
\filler{7}{34}
\filler{7}{35}
\filler{7}{36}
\filler{7}{37}
\filler{7}{38}
\filler{8}{30}
\filler{8}{31}
\filler{8}{32}
\filler{12}{22}
\filler{12}{23}
\filler{12}{24}
\filler{12}{25}
\filler{12}{35}
\filler{12}{36}
\filler{12}{37}
\filler{12}{38}
\filler{13}{22}
\filler{13}{23}
\filler{13}{24}
\filler{13}{25}
\filler{13}{26}
\filler{13}{27}
\filler{13}{34}
\filler{13}{35}
\filler{13}{36}
\filler{13}{37}
\filler{13}{38}
\filler{14}{22}
\filler{14}{23}
\filler{14}{24}
\filler{14}{25}
\filler{14}{26}
\filler{14}{27}
\filler{14}{28}
\filler{14}{29}
\filler{14}{33}
\filler{14}{34}
\filler{14}{35}
\filler{14}{36}
\filler{14}{37}
\filler{14}{38}
\filler{15}{22}
\filler{15}{23}
\filler{15}{24}
\filler{15}{25}
\filler{15}{26}
\filler{15}{27}
\filler{15}{28}
\filler{15}{29}
\filler{15}{30}
\filler{15}{31}
\filler{15}{32}
\filler{15}{33}
\filler{15}{34}
\filler{15}{35}
\filler{15}{36}
\filler{15}{37}
\filler{15}{38}
\filler{16}{22}
\filler{16}{23}
\filler{16}{24}
\filler{16}{25}
\filler{16}{26}
\filler{16}{27}
\filler{16}{28}
\filler{16}{29}
\filler{16}{30}
\filler{16}{31}
\filler{16}{32}
\filler{16}{33}
\filler{16}{34}
\filler{16}{35}
\filler{16}{36}
\filler{16}{37}
\filler{16}{38}
         \clauseblock{1}{0}{2}{1}{0}{0}{1}

        \filler{4}{26}
        \filler{5}{26}
        \filler{6}{26}
        \filler{7}{26}
        \filler{4}{29}
        \filler{5}{29}
        \filler{6}{29}
        \filler{7}{29}

        \filler{13}{28}
        \filler{13}{29}
        \filler{14}{30}
        \filler{9}{30}
        \filler{14}{31}
        \filler{9}{31}
        \filler{14}{32}
        \filler{9}{32}
        \filler{13}{33}
        \filler{12}{33}
        \filler{8}{33}
        \filler{12}{34}
        \filler{8}{34}
        \filler{11}{35}
        \filler{8}{35}
        \filler{11}{36}
        \filler{8}{36}
        \filler{11}{37}
        \filler{8}{37}
        \filler{11}{38}
        \filler{8}{38}

        \omino{4}{27}{1}
        \omino{6}{27}{1}

        \omino{9}{24}{1}
        \omino{9}{22}{1}

        \omino{11}{28}{1}
        \omino{12}{30}{1}
        \omino{10}{30}{1}
        \omino{10}{32}{1}
        \omino{9}{34}{1}
        \omino{9}{36}{1}
\omino{9}{38}{1}

        \foreach \x in {4, 6, ..., 10}{ \legendpin{\x}{27} }
        \foreach \y in {31, 33, ..., 37}{ \legendpin{10}{\y} }
        \foreach \y in {27, 25, 23}{ \legendpin{10}{\y} }
        \legendpin{12}{29}
        \legendpin{12}{31}
\draw [draw = orange, uwidth = .25] (7.625, 24.625) rectangle (13.375, 34.375);
\draw [draw = cyan, uwidth = .3] (10.5, 3.5) rectangle ++ (2, 2);
        \draw [draw = cyan, uwidth = .3] (9.5, 5.5) rectangle ++ (2, 2);
        \draw [draw = cyan, uwidth = .3] (11.5, 5.5) rectangle ++ (2, 2);
        \draw [draw = cyan, uwidth = .3] (9.5, 7.5) rectangle ++ (2, 2);
        \draw [draw = cyan, uwidth = .3] (8.5, 9.5) rectangle ++ (2, 2);
        \draw [draw = cyan, uwidth = .3] (8.5, 11.5) rectangle ++ (2, 2);
        \draw [draw = cyan, uwidth = .3] (8.5, 13.5) rectangle ++ (2, 2);
        \draw [draw = cyan, uwidth = .3] (8.5, 15.5) rectangle ++ (2, 2);
        \draw [draw = cyan, uwidth = .3] (8.5, 17.5) rectangle ++ (2, 2);
        \draw [draw = cyan, uwidth = .3] (8.5, 19.5) rectangle ++ (2, 2);
        \draw [draw = cyan, uwidth = .3] (8.5, 21.5) rectangle ++ (2, 2);
        \draw [draw = red, uwidth = .3] (3.5, 26.5) rectangle ++ (2, 2);
        \draw [draw = red, uwidth = .3] (5.5, 26.5) rectangle ++ (2, 2);
        \draw [draw = cyan, uwidth = .3] (8.5, 33.5) rectangle ++ (2, 2);
        \draw [draw = cyan, uwidth = .3] (8.5, 35.5) rectangle ++ (2, 2);
        \draw [draw = cyan, uwidth = .3] (8.5, 37.5) rectangle ++ (2, 2);
      \end{tikzpicture} &
      \begin{tikzpicture}[x = .28cm, y = .28cm]
\clip (3.5, 21.5) rectangle ++ (12, 17);
        \whitebg{4}{16}{22}{38}
        \filler{4}{22}
\filler{4}{23}
\filler{4}{24}
\filler{4}{25}
\filler{4}{30}
\filler{4}{31}
\filler{4}{32}
\filler{4}{33}
\filler{4}{34}
\filler{4}{35}
\filler{4}{36}
\filler{4}{37}
\filler{4}{38}
\filler{5}{22}
\filler{5}{23}
\filler{5}{24}
\filler{5}{25}
\filler{5}{30}
\filler{5}{31}
\filler{5}{32}
\filler{5}{33}
\filler{5}{34}
\filler{5}{35}
\filler{5}{36}
\filler{5}{37}
\filler{5}{38}
\filler{6}{22}
\filler{6}{23}
\filler{6}{24}
\filler{6}{25}
\filler{6}{30}
\filler{6}{31}
\filler{6}{32}
\filler{6}{33}
\filler{6}{34}
\filler{6}{35}
\filler{6}{36}
\filler{6}{37}
\filler{6}{38}
\filler{7}{22}
\filler{7}{23}
\filler{7}{24}
\filler{7}{25}
\filler{7}{30}
\filler{7}{31}
\filler{7}{32}
\filler{7}{33}
\filler{7}{34}
\filler{7}{35}
\filler{7}{36}
\filler{7}{37}
\filler{7}{38}
\filler{8}{30}
\filler{8}{31}
\filler{8}{32}
\filler{12}{22}
\filler{12}{23}
\filler{12}{24}
\filler{12}{25}
\filler{12}{35}
\filler{12}{36}
\filler{12}{37}
\filler{12}{38}
\filler{13}{22}
\filler{13}{23}
\filler{13}{24}
\filler{13}{25}
\filler{13}{26}
\filler{13}{27}
\filler{13}{34}
\filler{13}{35}
\filler{13}{36}
\filler{13}{37}
\filler{13}{38}
\filler{14}{22}
\filler{14}{23}
\filler{14}{24}
\filler{14}{25}
\filler{14}{26}
\filler{14}{27}
\filler{14}{28}
\filler{14}{29}
\filler{14}{33}
\filler{14}{34}
\filler{14}{35}
\filler{14}{36}
\filler{14}{37}
\filler{14}{38}
\filler{15}{22}
\filler{15}{23}
\filler{15}{24}
\filler{15}{25}
\filler{15}{26}
\filler{15}{27}
\filler{15}{28}
\filler{15}{29}
\filler{15}{30}
\filler{15}{31}
\filler{15}{32}
\filler{15}{33}
\filler{15}{34}
\filler{15}{35}
\filler{15}{36}
\filler{15}{37}
\filler{15}{38}
\filler{16}{22}
\filler{16}{23}
\filler{16}{24}
\filler{16}{25}
\filler{16}{26}
\filler{16}{27}
\filler{16}{28}
\filler{16}{29}
\filler{16}{30}
\filler{16}{31}
\filler{16}{32}
\filler{16}{33}
\filler{16}{34}
\filler{16}{35}
\filler{16}{36}
\filler{16}{37}
\filler{16}{38}
         \clauseblock{1}{0}{2}{1}{0}{0}{3}

        \filler{4}{26}
        \filler{5}{26}
        \filler{6}{26}
        \filler{7}{26}
        \filler{4}{29}
        \filler{5}{29}
        \filler{6}{29}
        \filler{7}{29}

        \filler{13}{28}
        \filler{13}{29}
        \filler{14}{30}
        \filler{9}{30}
        \filler{14}{31}
        \filler{9}{31}
        \filler{14}{32}
        \filler{9}{32}
        \filler{13}{33}
        \filler{12}{33}
        \filler{8}{33}
        \filler{12}{34}
        \filler{8}{34}
        \filler{11}{35}
        \filler{8}{35}
        \filler{11}{36}
        \filler{8}{36}
        \filler{11}{37}
        \filler{8}{37}
        \filler{11}{38}
        \filler{8}{38}

        \omino{4}{27}{1}
        \omino{6}{27}{1}

        \omino{9}{25}{1}
        \omino{9}{23}{1}
        \omino{9}{21}{1}

        \omino{11}{29}{1}
        \omino{12}{31}{1}
        \omino{10}{31}{1}
        \omino{10}{33}{1}
        \omino{9}{35}{1}
        \omino{9}{37}{1}

        \foreach \x in {4, 6, ..., 10}{ \legendpin{\x}{27} }
        \foreach \y in {31, 33, ..., 37}{ \legendpin{10}{\y} }
        \foreach \y in {27, 25, 23}{ \legendpin{10}{\y} }
        \legendpin{12}{29}
        \legendpin{12}{31}
\draw [draw = orange, uwidth = .25] (7.625, 24.625) rectangle (13.375, 34.375);
\draw [draw = red, uwidth = .3] (10.5, 4.5) rectangle ++ (2, 2);
        \draw [draw = red, uwidth = .3] (9.5, 6.5) rectangle ++ (2, 2);
        \draw [draw = red, uwidth = .3] (11.5, 6.5) rectangle ++ (2, 2);
        \draw [draw = red, uwidth = .3] (9.5, 8.5) rectangle ++ (2, 2);
        \draw [draw = red, uwidth = .3] (8.5, 10.5) rectangle ++ (2, 2);
        \draw [draw = red, uwidth = .3] (8.5, 12.5) rectangle ++ (2, 2);
        \draw [draw = red, uwidth = .3] (8.5, 14.5) rectangle ++ (2, 2);
        \draw [draw = red, uwidth = .3] (8.5, 16.5) rectangle ++ (2, 2);
        \draw [draw = red, uwidth = .3] (8.5, 18.5) rectangle ++ (2, 2);
        \draw [draw = red, uwidth = .3] (8.5, 20.5) rectangle ++ (2, 2);
        \draw [draw = red, uwidth = .3] (8.5, 22.5) rectangle ++ (2, 2);
        \draw [draw = red, uwidth = .3] (3.5, 26.5) rectangle ++ (2, 2);
        \draw [draw = red, uwidth = .3] (5.5, 26.5) rectangle ++ (2, 2);
        \draw [draw = red, uwidth = .3] (8.5, 34.5) rectangle ++ (2, 2);
        \draw [draw = red, uwidth = .3] (8.5, 36.5) rectangle ++ (2, 2);
      \end{tikzpicture} \\[2pt]
      \scriptsize \True, \True & \scriptsize \True, \False & \scriptsize \False, \True & \scriptsize \False, \False
  \end{tabular}
  \caption{A middle module, in the four combinations of its two inputs.}
  \label{fig:clause-or}
\end{subfigure}

\vspace{1.2em}
\begin{subfigure}[t]{\linewidth}
  \centering
\begin{tabular}{@{}cccc@{}}
      \begin{tikzpicture}[x = .28cm, y = .28cm]
\clip (3.5, 21.5) rectangle ++ (12, 9);
        \whitebg{4}{16}{22}{38}
        \filler{4}{22}
\filler{4}{23}
\filler{4}{24}
\filler{4}{25}
\filler{4}{30}
\filler{4}{31}
\filler{4}{32}
\filler{4}{33}
\filler{4}{34}
\filler{4}{35}
\filler{4}{36}
\filler{4}{37}
\filler{4}{38}
\filler{5}{22}
\filler{5}{23}
\filler{5}{24}
\filler{5}{25}
\filler{5}{30}
\filler{5}{31}
\filler{5}{32}
\filler{5}{33}
\filler{5}{34}
\filler{5}{35}
\filler{5}{36}
\filler{5}{37}
\filler{5}{38}
\filler{6}{22}
\filler{6}{23}
\filler{6}{24}
\filler{6}{25}
\filler{6}{30}
\filler{6}{31}
\filler{6}{32}
\filler{6}{33}
\filler{6}{34}
\filler{6}{35}
\filler{6}{36}
\filler{6}{37}
\filler{6}{38}
\filler{7}{22}
\filler{7}{23}
\filler{7}{24}
\filler{7}{25}
\filler{7}{30}
\filler{7}{31}
\filler{7}{32}
\filler{7}{33}
\filler{7}{34}
\filler{7}{35}
\filler{7}{36}
\filler{7}{37}
\filler{7}{38}
\filler{8}{30}
\filler{8}{31}
\filler{8}{32}
\filler{8}{33}
\filler{8}{34}
\filler{8}{35}
\filler{8}{36}
\filler{8}{37}
\filler{8}{38}
\filler{9}{30}
\filler{9}{31}
\filler{9}{32}
\filler{9}{33}
\filler{9}{34}
\filler{9}{35}
\filler{9}{36}
\filler{9}{37}
\filler{9}{38}
\filler{10}{30}
\filler{10}{31}
\filler{10}{32}
\filler{10}{33}
\filler{10}{34}
\filler{10}{35}
\filler{10}{36}
\filler{10}{37}
\filler{10}{38}
\filler{11}{29}
\filler{11}{30}
\filler{11}{31}
\filler{11}{32}
\filler{11}{33}
\filler{11}{34}
\filler{11}{35}
\filler{11}{36}
\filler{11}{37}
\filler{11}{38}
\filler{12}{22}
\filler{12}{23}
\filler{12}{24}
\filler{12}{25}
\filler{12}{28}
\filler{12}{29}
\filler{12}{30}
\filler{12}{31}
\filler{12}{32}
\filler{12}{33}
\filler{12}{34}
\filler{12}{35}
\filler{12}{36}
\filler{12}{37}
\filler{12}{38}
\filler{13}{22}
\filler{13}{23}
\filler{13}{24}
\filler{13}{25}
\filler{13}{26}
\filler{13}{27}
\filler{13}{28}
\filler{13}{29}
\filler{13}{30}
\filler{13}{31}
\filler{13}{32}
\filler{13}{33}
\filler{13}{34}
\filler{13}{35}
\filler{13}{36}
\filler{13}{37}
\filler{13}{38}
\filler{14}{22}
\filler{14}{23}
\filler{14}{24}
\filler{14}{25}
\filler{14}{26}
\filler{14}{27}
\filler{14}{28}
\filler{14}{29}
\filler{14}{30}
\filler{14}{31}
\filler{14}{32}
\filler{14}{33}
\filler{14}{34}
\filler{14}{35}
\filler{14}{36}
\filler{14}{37}
\filler{14}{38}
\filler{15}{22}
\filler{15}{23}
\filler{15}{24}
\filler{15}{25}
\filler{15}{26}
\filler{15}{27}
\filler{15}{28}
\filler{15}{29}
\filler{15}{30}
\filler{15}{31}
\filler{15}{32}
\filler{15}{33}
\filler{15}{34}
\filler{15}{35}
\filler{15}{36}
\filler{15}{37}
\filler{15}{38}
\filler{16}{22}
\filler{16}{23}
\filler{16}{24}
\filler{16}{25}
\filler{16}{26}
\filler{16}{27}
\filler{16}{28}
\filler{16}{29}
\filler{16}{30}
\filler{16}{31}
\filler{16}{32}
\filler{16}{33}
\filler{16}{34}
\filler{16}{35}
\filler{16}{36}
\filler{16}{37}
\filler{16}{38}
         \clauseblock{1}{0}{2}{2}{0}{0}{0}

        \filler{4}{26}
        \filler{5}{26}
        \filler{6}{26}
        \filler{7}{26}
        \filler{4}{29}
        \filler{5}{29}
        \filler{6}{29}
        \filler{7}{29}

        \omino{3}{27}{1}
        \omino{5}{27}{1}

        \omino{9}{24}{1}
        \omino{9}{22}{1}

        \foreach \x in {4, 6, ..., 10}{ \legendpin{\x}{27} }
        \foreach \y in {27, 25, 23}{ \legendpin{10}{\y} }
\draw [draw = orange, uwidth = .25] (7.625, 24.625) rectangle (13.375, 28.375);
\draw [draw = cyan, uwidth = .3] (10.5, 3.5) rectangle ++ (2, 2);
        \draw [draw = cyan, uwidth = .3] (9.5, 5.5) rectangle ++ (2, 2);
        \draw [draw = cyan, uwidth = .3] (11.5, 5.5) rectangle ++ (2, 2);
        \draw [draw = cyan, uwidth = .3] (9.5, 7.5) rectangle ++ (2, 2);
        \draw [draw = cyan, uwidth = .3] (8.5, 9.5) rectangle ++ (2, 2);
        \draw [draw = cyan, uwidth = .3] (8.5, 11.5) rectangle ++ (2, 2);
        \draw [draw = cyan, uwidth = .3] (8.5, 13.5) rectangle ++ (2, 2);
        \draw [draw = cyan, uwidth = .3] (8.5, 15.5) rectangle ++ (2, 2);
        \draw [draw = cyan, uwidth = .3] (8.5, 17.5) rectangle ++ (2, 2);
        \draw [draw = cyan, uwidth = .3] (8.5, 19.5) rectangle ++ (2, 2);
        \draw [draw = cyan, uwidth = .3] (8.5, 21.5) rectangle ++ (2, 2);
        \draw [draw = cyan, uwidth = .3] (2.5, 26.5) rectangle ++ (2, 2);
        \draw [draw = cyan, uwidth = .3] (4.5, 26.5) rectangle ++ (2, 2);
      \end{tikzpicture} &
      \begin{tikzpicture}[x = .28cm, y = .28cm]
\clip (3.5, 21.5) rectangle ++ (12, 9);
        \whitebg{4}{16}{22}{38}
        \filler{4}{22}
\filler{4}{23}
\filler{4}{24}
\filler{4}{25}
\filler{4}{30}
\filler{4}{31}
\filler{4}{32}
\filler{4}{33}
\filler{4}{34}
\filler{4}{35}
\filler{4}{36}
\filler{4}{37}
\filler{4}{38}
\filler{5}{22}
\filler{5}{23}
\filler{5}{24}
\filler{5}{25}
\filler{5}{30}
\filler{5}{31}
\filler{5}{32}
\filler{5}{33}
\filler{5}{34}
\filler{5}{35}
\filler{5}{36}
\filler{5}{37}
\filler{5}{38}
\filler{6}{22}
\filler{6}{23}
\filler{6}{24}
\filler{6}{25}
\filler{6}{30}
\filler{6}{31}
\filler{6}{32}
\filler{6}{33}
\filler{6}{34}
\filler{6}{35}
\filler{6}{36}
\filler{6}{37}
\filler{6}{38}
\filler{7}{22}
\filler{7}{23}
\filler{7}{24}
\filler{7}{25}
\filler{7}{30}
\filler{7}{31}
\filler{7}{32}
\filler{7}{33}
\filler{7}{34}
\filler{7}{35}
\filler{7}{36}
\filler{7}{37}
\filler{7}{38}
\filler{8}{30}
\filler{8}{31}
\filler{8}{32}
\filler{8}{33}
\filler{8}{34}
\filler{8}{35}
\filler{8}{36}
\filler{8}{37}
\filler{8}{38}
\filler{9}{30}
\filler{9}{31}
\filler{9}{32}
\filler{9}{33}
\filler{9}{34}
\filler{9}{35}
\filler{9}{36}
\filler{9}{37}
\filler{9}{38}
\filler{10}{30}
\filler{10}{31}
\filler{10}{32}
\filler{10}{33}
\filler{10}{34}
\filler{10}{35}
\filler{10}{36}
\filler{10}{37}
\filler{10}{38}
\filler{11}{29}
\filler{11}{30}
\filler{11}{31}
\filler{11}{32}
\filler{11}{33}
\filler{11}{34}
\filler{11}{35}
\filler{11}{36}
\filler{11}{37}
\filler{11}{38}
\filler{12}{22}
\filler{12}{23}
\filler{12}{24}
\filler{12}{25}
\filler{12}{28}
\filler{12}{29}
\filler{12}{30}
\filler{12}{31}
\filler{12}{32}
\filler{12}{33}
\filler{12}{34}
\filler{12}{35}
\filler{12}{36}
\filler{12}{37}
\filler{12}{38}
\filler{13}{22}
\filler{13}{23}
\filler{13}{24}
\filler{13}{25}
\filler{13}{26}
\filler{13}{27}
\filler{13}{28}
\filler{13}{29}
\filler{13}{30}
\filler{13}{31}
\filler{13}{32}
\filler{13}{33}
\filler{13}{34}
\filler{13}{35}
\filler{13}{36}
\filler{13}{37}
\filler{13}{38}
\filler{14}{22}
\filler{14}{23}
\filler{14}{24}
\filler{14}{25}
\filler{14}{26}
\filler{14}{27}
\filler{14}{28}
\filler{14}{29}
\filler{14}{30}
\filler{14}{31}
\filler{14}{32}
\filler{14}{33}
\filler{14}{34}
\filler{14}{35}
\filler{14}{36}
\filler{14}{37}
\filler{14}{38}
\filler{15}{22}
\filler{15}{23}
\filler{15}{24}
\filler{15}{25}
\filler{15}{26}
\filler{15}{27}
\filler{15}{28}
\filler{15}{29}
\filler{15}{30}
\filler{15}{31}
\filler{15}{32}
\filler{15}{33}
\filler{15}{34}
\filler{15}{35}
\filler{15}{36}
\filler{15}{37}
\filler{15}{38}
\filler{16}{22}
\filler{16}{23}
\filler{16}{24}
\filler{16}{25}
\filler{16}{26}
\filler{16}{27}
\filler{16}{28}
\filler{16}{29}
\filler{16}{30}
\filler{16}{31}
\filler{16}{32}
\filler{16}{33}
\filler{16}{34}
\filler{16}{35}
\filler{16}{36}
\filler{16}{37}
\filler{16}{38}
         \clauseblock{1}{0}{2}{2}{0}{0}{2}

        \filler{4}{26}
        \filler{5}{26}
        \filler{6}{26}
        \filler{7}{26}
        \filler{4}{29}
        \filler{5}{29}
        \filler{6}{29}
        \filler{7}{29}

        \omino{3}{27}{1}
        \omino{5}{27}{1}

        \omino{9}{25}{1}
        \omino{9}{23}{1}
        \omino{9}{21}{1}

        \foreach \x in {4, 6, ..., 10}{ \legendpin{\x}{27} }
        \foreach \y in {27, 25, 23}{ \legendpin{10}{\y} }
\draw [draw = orange, uwidth = .25] (7.625, 24.625) rectangle (13.375, 28.375);
\draw [draw = red, uwidth = .3] (10.5, 4.5) rectangle ++ (2, 2);
        \draw [draw = red, uwidth = .3] (9.5, 6.5) rectangle ++ (2, 2);
        \draw [draw = red, uwidth = .3] (11.5, 6.5) rectangle ++ (2, 2);
        \draw [draw = red, uwidth = .3] (9.5, 8.5) rectangle ++ (2, 2);
        \draw [draw = red, uwidth = .3] (8.5, 10.5) rectangle ++ (2, 2);
        \draw [draw = red, uwidth = .3] (8.5, 12.5) rectangle ++ (2, 2);
        \draw [draw = red, uwidth = .3] (8.5, 14.5) rectangle ++ (2, 2);
        \draw [draw = red, uwidth = .3] (8.5, 16.5) rectangle ++ (2, 2);
        \draw [draw = red, uwidth = .3] (8.5, 18.5) rectangle ++ (2, 2);
        \draw [draw = red, uwidth = .3] (8.5, 20.5) rectangle ++ (2, 2);
        \draw [draw = red, uwidth = .3] (8.5, 22.5) rectangle ++ (2, 2);
        \draw [draw = cyan, uwidth = .3] (2.5, 26.5) rectangle ++ (2, 2);
        \draw [draw = cyan, uwidth = .3] (4.5, 26.5) rectangle ++ (2, 2);
      \end{tikzpicture} &
      \begin{tikzpicture}[x = .28cm, y = .28cm]
\clip (3.5, 21.5) rectangle ++ (12, 9);
        \whitebg{4}{16}{22}{38}
        \filler{4}{22}
\filler{4}{23}
\filler{4}{24}
\filler{4}{25}
\filler{4}{30}
\filler{4}{31}
\filler{4}{32}
\filler{4}{33}
\filler{4}{34}
\filler{4}{35}
\filler{4}{36}
\filler{4}{37}
\filler{4}{38}
\filler{5}{22}
\filler{5}{23}
\filler{5}{24}
\filler{5}{25}
\filler{5}{30}
\filler{5}{31}
\filler{5}{32}
\filler{5}{33}
\filler{5}{34}
\filler{5}{35}
\filler{5}{36}
\filler{5}{37}
\filler{5}{38}
\filler{6}{22}
\filler{6}{23}
\filler{6}{24}
\filler{6}{25}
\filler{6}{30}
\filler{6}{31}
\filler{6}{32}
\filler{6}{33}
\filler{6}{34}
\filler{6}{35}
\filler{6}{36}
\filler{6}{37}
\filler{6}{38}
\filler{7}{22}
\filler{7}{23}
\filler{7}{24}
\filler{7}{25}
\filler{7}{30}
\filler{7}{31}
\filler{7}{32}
\filler{7}{33}
\filler{7}{34}
\filler{7}{35}
\filler{7}{36}
\filler{7}{37}
\filler{7}{38}
\filler{8}{30}
\filler{8}{31}
\filler{8}{32}
\filler{8}{33}
\filler{8}{34}
\filler{8}{35}
\filler{8}{36}
\filler{8}{37}
\filler{8}{38}
\filler{9}{30}
\filler{9}{31}
\filler{9}{32}
\filler{9}{33}
\filler{9}{34}
\filler{9}{35}
\filler{9}{36}
\filler{9}{37}
\filler{9}{38}
\filler{10}{30}
\filler{10}{31}
\filler{10}{32}
\filler{10}{33}
\filler{10}{34}
\filler{10}{35}
\filler{10}{36}
\filler{10}{37}
\filler{10}{38}
\filler{11}{29}
\filler{11}{30}
\filler{11}{31}
\filler{11}{32}
\filler{11}{33}
\filler{11}{34}
\filler{11}{35}
\filler{11}{36}
\filler{11}{37}
\filler{11}{38}
\filler{12}{22}
\filler{12}{23}
\filler{12}{24}
\filler{12}{25}
\filler{12}{28}
\filler{12}{29}
\filler{12}{30}
\filler{12}{31}
\filler{12}{32}
\filler{12}{33}
\filler{12}{34}
\filler{12}{35}
\filler{12}{36}
\filler{12}{37}
\filler{12}{38}
\filler{13}{22}
\filler{13}{23}
\filler{13}{24}
\filler{13}{25}
\filler{13}{26}
\filler{13}{27}
\filler{13}{28}
\filler{13}{29}
\filler{13}{30}
\filler{13}{31}
\filler{13}{32}
\filler{13}{33}
\filler{13}{34}
\filler{13}{35}
\filler{13}{36}
\filler{13}{37}
\filler{13}{38}
\filler{14}{22}
\filler{14}{23}
\filler{14}{24}
\filler{14}{25}
\filler{14}{26}
\filler{14}{27}
\filler{14}{28}
\filler{14}{29}
\filler{14}{30}
\filler{14}{31}
\filler{14}{32}
\filler{14}{33}
\filler{14}{34}
\filler{14}{35}
\filler{14}{36}
\filler{14}{37}
\filler{14}{38}
\filler{15}{22}
\filler{15}{23}
\filler{15}{24}
\filler{15}{25}
\filler{15}{26}
\filler{15}{27}
\filler{15}{28}
\filler{15}{29}
\filler{15}{30}
\filler{15}{31}
\filler{15}{32}
\filler{15}{33}
\filler{15}{34}
\filler{15}{35}
\filler{15}{36}
\filler{15}{37}
\filler{15}{38}
\filler{16}{22}
\filler{16}{23}
\filler{16}{24}
\filler{16}{25}
\filler{16}{26}
\filler{16}{27}
\filler{16}{28}
\filler{16}{29}
\filler{16}{30}
\filler{16}{31}
\filler{16}{32}
\filler{16}{33}
\filler{16}{34}
\filler{16}{35}
\filler{16}{36}
\filler{16}{37}
\filler{16}{38}
         \clauseblock{1}{0}{2}{2}{0}{0}{1}

        \filler{4}{26}
        \filler{5}{26}
        \filler{6}{26}
        \filler{7}{26}
        \filler{4}{29}
        \filler{5}{29}
        \filler{6}{29}
        \filler{7}{29}

        \omino{4}{27}{1}
        \omino{6}{27}{1}

        \omino{9}{24}{1}
        \omino{9}{22}{1}

        \foreach \x in {4, 6, ..., 10}{ \legendpin{\x}{27} }
        \foreach \y in {27, 25, 23}{ \legendpin{10}{\y} }
\draw [draw = orange, uwidth = .25] (7.625, 24.625) rectangle (13.375, 28.375);
\draw [draw = cyan, uwidth = .3] (10.5, 3.5) rectangle ++ (2, 2);
        \draw [draw = cyan, uwidth = .3] (9.5, 5.5) rectangle ++ (2, 2);
        \draw [draw = cyan, uwidth = .3] (11.5, 5.5) rectangle ++ (2, 2);
        \draw [draw = cyan, uwidth = .3] (9.5, 7.5) rectangle ++ (2, 2);
        \draw [draw = cyan, uwidth = .3] (8.5, 9.5) rectangle ++ (2, 2);
        \draw [draw = cyan, uwidth = .3] (8.5, 11.5) rectangle ++ (2, 2);
        \draw [draw = cyan, uwidth = .3] (8.5, 13.5) rectangle ++ (2, 2);
        \draw [draw = cyan, uwidth = .3] (8.5, 15.5) rectangle ++ (2, 2);
        \draw [draw = cyan, uwidth = .3] (8.5, 17.5) rectangle ++ (2, 2);
        \draw [draw = cyan, uwidth = .3] (8.5, 19.5) rectangle ++ (2, 2);
        \draw [draw = cyan, uwidth = .3] (8.5, 21.5) rectangle ++ (2, 2);
        \draw [draw = red, uwidth = .3] (3.5, 26.5) rectangle ++ (2, 2);
        \draw [draw = red, uwidth = .3] (5.5, 26.5) rectangle ++ (2, 2);
      \end{tikzpicture} &
      \begin{tikzpicture}[x = .28cm, y = .28cm]
\clip (3.5, 21.5) rectangle ++ (12, 9);
        \whitebg{4}{16}{22}{38}
        \filler{4}{22}
\filler{4}{23}
\filler{4}{24}
\filler{4}{25}
\filler{4}{30}
\filler{4}{31}
\filler{4}{32}
\filler{4}{33}
\filler{4}{34}
\filler{4}{35}
\filler{4}{36}
\filler{4}{37}
\filler{4}{38}
\filler{5}{22}
\filler{5}{23}
\filler{5}{24}
\filler{5}{25}
\filler{5}{30}
\filler{5}{31}
\filler{5}{32}
\filler{5}{33}
\filler{5}{34}
\filler{5}{35}
\filler{5}{36}
\filler{5}{37}
\filler{5}{38}
\filler{6}{22}
\filler{6}{23}
\filler{6}{24}
\filler{6}{25}
\filler{6}{30}
\filler{6}{31}
\filler{6}{32}
\filler{6}{33}
\filler{6}{34}
\filler{6}{35}
\filler{6}{36}
\filler{6}{37}
\filler{6}{38}
\filler{7}{22}
\filler{7}{23}
\filler{7}{24}
\filler{7}{25}
\filler{7}{30}
\filler{7}{31}
\filler{7}{32}
\filler{7}{33}
\filler{7}{34}
\filler{7}{35}
\filler{7}{36}
\filler{7}{37}
\filler{7}{38}
\filler{8}{30}
\filler{8}{31}
\filler{8}{32}
\filler{8}{33}
\filler{8}{34}
\filler{8}{35}
\filler{8}{36}
\filler{8}{37}
\filler{8}{38}
\filler{9}{30}
\filler{9}{31}
\filler{9}{32}
\filler{9}{33}
\filler{9}{34}
\filler{9}{35}
\filler{9}{36}
\filler{9}{37}
\filler{9}{38}
\filler{10}{30}
\filler{10}{31}
\filler{10}{32}
\filler{10}{33}
\filler{10}{34}
\filler{10}{35}
\filler{10}{36}
\filler{10}{37}
\filler{10}{38}
\filler{11}{29}
\filler{11}{30}
\filler{11}{31}
\filler{11}{32}
\filler{11}{33}
\filler{11}{34}
\filler{11}{35}
\filler{11}{36}
\filler{11}{37}
\filler{11}{38}
\filler{12}{22}
\filler{12}{23}
\filler{12}{24}
\filler{12}{25}
\filler{12}{28}
\filler{12}{29}
\filler{12}{30}
\filler{12}{31}
\filler{12}{32}
\filler{12}{33}
\filler{12}{34}
\filler{12}{35}
\filler{12}{36}
\filler{12}{37}
\filler{12}{38}
\filler{13}{22}
\filler{13}{23}
\filler{13}{24}
\filler{13}{25}
\filler{13}{26}
\filler{13}{27}
\filler{13}{28}
\filler{13}{29}
\filler{13}{30}
\filler{13}{31}
\filler{13}{32}
\filler{13}{33}
\filler{13}{34}
\filler{13}{35}
\filler{13}{36}
\filler{13}{37}
\filler{13}{38}
\filler{14}{22}
\filler{14}{23}
\filler{14}{24}
\filler{14}{25}
\filler{14}{26}
\filler{14}{27}
\filler{14}{28}
\filler{14}{29}
\filler{14}{30}
\filler{14}{31}
\filler{14}{32}
\filler{14}{33}
\filler{14}{34}
\filler{14}{35}
\filler{14}{36}
\filler{14}{37}
\filler{14}{38}
\filler{15}{22}
\filler{15}{23}
\filler{15}{24}
\filler{15}{25}
\filler{15}{26}
\filler{15}{27}
\filler{15}{28}
\filler{15}{29}
\filler{15}{30}
\filler{15}{31}
\filler{15}{32}
\filler{15}{33}
\filler{15}{34}
\filler{15}{35}
\filler{15}{36}
\filler{15}{37}
\filler{15}{38}
\filler{16}{22}
\filler{16}{23}
\filler{16}{24}
\filler{16}{25}
\filler{16}{26}
\filler{16}{27}
\filler{16}{28}
\filler{16}{29}
\filler{16}{30}
\filler{16}{31}
\filler{16}{32}
\filler{16}{33}
\filler{16}{34}
\filler{16}{35}
\filler{16}{36}
\filler{16}{37}
\filler{16}{38}
         \clauseblock{1}{0}{2}{2}{0}{0}{3}

        \filler{4}{26}
        \filler{5}{26}
        \filler{6}{26}
        \filler{7}{26}
        \filler{4}{29}
        \filler{5}{29}
        \filler{6}{29}
        \filler{7}{29}

        \omino{4}{27}{1}
        \omino{6}{27}{1}

        \omino{9}{25}{1}
        \omino{9}{23}{1}
        \omino{9}{21}{1}

        \foreach \x in {4, 6, 8}{ \legendpin{\x}{27} }
        \foreach \y in {25, 23}{ \legendpin{10}{\y} }

\uncoveredpin{10}{27}
\draw [draw = orange, uwidth = .25] (7.625, 24.625) rectangle (13.375, 28.375);
\draw [draw = red, uwidth = .3] (10.5, 4.5) rectangle ++ (2, 2);
        \draw [draw = red, uwidth = .3] (9.5, 6.5) rectangle ++ (2, 2);
        \draw [draw = red, uwidth = .3] (11.5, 6.5) rectangle ++ (2, 2);
        \draw [draw = red, uwidth = .3] (9.5, 8.5) rectangle ++ (2, 2);
        \draw [draw = red, uwidth = .3] (8.5, 10.5) rectangle ++ (2, 2);
        \draw [draw = red, uwidth = .3] (8.5, 12.5) rectangle ++ (2, 2);
        \draw [draw = red, uwidth = .3] (8.5, 14.5) rectangle ++ (2, 2);
        \draw [draw = red, uwidth = .3] (8.5, 16.5) rectangle ++ (2, 2);
        \draw [draw = red, uwidth = .3] (8.5, 18.5) rectangle ++ (2, 2);
        \draw [draw = red, uwidth = .3] (8.5, 20.5) rectangle ++ (2, 2);
        \draw [draw = red, uwidth = .3] (8.5, 22.5) rectangle ++ (2, 2);
        \draw [draw = red, uwidth = .3] (3.5, 26.5) rectangle ++ (2, 2);
        \draw [draw = red, uwidth = .3] (5.5, 26.5) rectangle ++ (2, 2);
      \end{tikzpicture} \\[2pt]
      \scriptsize \True, \True & \scriptsize \True, \False & \scriptsize \False, \True & \scriptsize \False, \False
  \end{tabular}
  \caption{The top module, which ends at the height of its wire.}
  \label{fig:clause-top}
\end{subfigure}
\caption{The clause modules, each panel labelled with its inputs, the horizontal one first.
Each module is outlined in orange as in Figure~\ref{fig:clause-parts}, and the pieces with a pin in the wire and the shafts are framed, cyan where they carry \True and red where they carry \False.
In the all-\False case of~\subref{fig:clause-top}, the pin in magenta is left uncovered.}
\label{fig:clause}
\end{figure}

The bottom module takes only a horizontal input, since the cells below it are filled (Figure~\ref{fig:clause}\subref{fig:clause-bottom}).

\begin{lemma}[Bottom module]
\label{lem:clausefirst}
The bottom module has a pin-covering filling emitting the value of its horizontal input, and no pin-covering filling of it emits \True unless its horizontal input is \True.
\end{lemma}
\begin{proof}
  Figure~\ref{fig:clause}\subref{fig:clause-bottom} draws a pin-covering filling for each value of the horizontal input.
  If the horizontal input is \False, one can easily check, going upward, that every pin $(x,y)$ must be covered by a piece that also covers $(x,y+1)$; so the module emits \False.
\end{proof}

A middle module also takes a vertical input from the shaft below (Figure~\ref{fig:clause}\subref{fig:clause-or}).

\begin{lemma}[Middle module]
\label{lem:or}
A middle module has a pin-covering filling, and if at least one of its inputs is \True it has one emitting \True.
Conversely, if a pin-covering filling of the module emits \True, at least one of its inputs is \True.
\end{lemma}
\begin{proof}
  If at least one of the inputs is \True, the three left panels of Figure~\ref{fig:clause}\subref{fig:clause-or} draw fillings satisfying the lemma.
  Otherwise, the argument for the bottom module applies: going upward, every pin $(x,y)$ must be covered by a piece that also covers $(x,y+1)$.
\end{proof}
Bottom and middle modules may emit \False even when an input is \True; this is harmless.

\begin{lemma}[Top module]
\label{lem:clauselast}
The top module has a pin-covering filling if and only if at least one of its inputs is \True.
\end{lemma}
\begin{proof}
The three left panels of Figure~\ref{fig:clause}\subref{fig:clause-top} draw a pin-covering filling for each pair of inputs with a \True among them.
When both inputs are \False, the rightmost panel of Figure~\ref{fig:clause}\subref{fig:clause-top} shows the only way to cover the top-left and bottom-right pins, which leaves the top-right pin uncovered.
\end{proof}

\begin{corollary}
\label{cor:clause}
If at least one of the values the gadget for $c$ receives along its wires is \True, it admits a pin-covering filling.
If none of them is, every filling of it leaves a pin uncovered.
\end{corollary}

\subsection{Correctness}
\label{sec:correctness}

The board transcribes the arrangement of $\phi$ directly.
Its width $W$ is $O(m)$ and, by the heights of the variable gadgets, its height $H$ is $O(\sum_{i=1}^n (d_i^{+} + d_i^{-}))$, so the board and $N = P - Q$, which is smaller than $WH$, can be computed in polynomial time.
We show that the player can survive if and only if $\phi$ is satisfiable.
By the observations in Section~\ref{sec:structure}, the player survives if and only if all $N$ pieces can be placed inside the construction.

First suppose the player survives.
Then the placed pieces form a winning packing.
Let $\alpha(v_i)$ be \True if some positive port of the gadget of $v_i$ sends \True, and \False otherwise.
Then, if some negative port of $v_i$ sends \True, $\alpha(v_i)$ is \False by Lemma~\ref{lem:nodouble}.
Since every pin of every clause gadget is covered, by Corollary~\ref{cor:clause} at least one wire into each clause gadget delivers \True.
Therefore, each clause has a literal that is \True under $\alpha$, and so $\alpha$ satisfies $\phi$.

Conversely, suppose $\phi$ is satisfied by an assignment $\alpha$.
Then there is a winning packing in which each variable gadget has the \True or the \False filling of Figure~\ref{fig:var} in accordance with $\alpha$, every wire carries the value of its literal, and no pin of any clause gadget is left uncovered by Corollary~\ref{cor:clause}.
It remains to order the placements so that each piece can reach its position.
We place the pieces against the direction in which signals propagate, in three phases.

\begin{figure}[t]
\centering
\tikzset{order/.style = {font = \small\sffamily, inner sep = 0pt}}
\newcommand{\clauseorderpins}{\foreach \x/\y in {0/0, 2/0, 4/0, 6/0, 8/0, 10/2, 8/4, 10/4, 8/6, 8/8, 8/10,
                       0/12, 2/12, 4/12, 6/12, 8/12, 10/14, 8/16, 10/16, 8/18, 8/20, 8/22,
                       0/24, 2/24, 4/24, 6/24, 8/24}{
        \legendpin{\x}{\y}
    }}
\begin{subfigure}[t]{0.32\linewidth}
  \centering
  \begin{tikzpicture}[x = .28cm, y = .28cm]
    \whitebg{0}{12}{-1}{26}
    \clip (-.5, -1.5) rectangle (12.5, 26.5);
    \filler{0}{-1}
\filler{1}{-1}
\filler{2}{-1}
\filler{3}{-1}
\filler{4}{-1}
\filler{5}{-1}
\filler{6}{-1}
\filler{7}{-1}
\filler{8}{-1}
\filler{9}{-1}
\filler{10}{-1}
\filler{11}{-1}
\filler{12}{-1}
\filler{10}{0}
\filler{11}{0}
\filler{12}{0}
\filler{11}{1}
\filler{12}{1}
\filler{0}{2}
\filler{1}{2}
\filler{2}{2}
\filler{3}{2}
\filler{4}{2}
\filler{5}{2}
\filler{6}{2}
\filler{7}{2}
\filler{8}{2}
\filler{11}{2}
\filler{12}{2}
\filler{0}{3}
\filler{1}{3}
\filler{2}{3}
\filler{3}{3}
\filler{4}{3}
\filler{5}{3}
\filler{6}{3}
\filler{7}{3}
\filler{12}{3}
\filler{0}{4}
\filler{1}{4}
\filler{2}{4}
\filler{3}{4}
\filler{4}{4}
\filler{5}{4}
\filler{6}{4}
\filler{7}{4}
\filler{12}{4}
\filler{0}{5}
\filler{1}{5}
\filler{2}{5}
\filler{3}{5}
\filler{4}{5}
\filler{5}{5}
\filler{6}{5}
\filler{7}{5}
\filler{12}{5}
\filler{0}{6}
\filler{1}{6}
\filler{2}{6}
\filler{3}{6}
\filler{4}{6}
\filler{5}{6}
\filler{6}{6}
\filler{10}{6}
\filler{11}{6}
\filler{12}{6}
\filler{0}{7}
\filler{1}{7}
\filler{2}{7}
\filler{3}{7}
\filler{4}{7}
\filler{5}{7}
\filler{6}{7}
\filler{10}{7}
\filler{11}{7}
\filler{12}{7}
\filler{0}{8}
\filler{1}{8}
\filler{2}{8}
\filler{3}{8}
\filler{4}{8}
\filler{5}{8}
\filler{6}{8}
\filler{9}{8}
\filler{10}{8}
\filler{11}{8}
\filler{12}{8}
\filler{0}{9}
\filler{1}{9}
\filler{2}{9}
\filler{3}{9}
\filler{4}{9}
\filler{5}{9}
\filler{6}{9}
\filler{9}{9}
\filler{10}{9}
\filler{11}{9}
\filler{12}{9}
\filler{0}{10}
\filler{1}{10}
\filler{2}{10}
\filler{3}{10}
\filler{4}{10}
\filler{5}{10}
\filler{6}{10}
\filler{9}{10}
\filler{10}{10}
\filler{11}{10}
\filler{12}{10}
\filler{0}{11}
\filler{1}{11}
\filler{2}{11}
\filler{3}{11}
\filler{4}{11}
\filler{5}{11}
\filler{6}{11}
\filler{10}{11}
\filler{11}{11}
\filler{12}{11}
\filler{10}{12}
\filler{11}{12}
\filler{12}{12}
\filler{11}{13}
\filler{12}{13}
\filler{0}{14}
\filler{1}{14}
\filler{2}{14}
\filler{3}{14}
\filler{4}{14}
\filler{5}{14}
\filler{6}{14}
\filler{7}{14}
\filler{8}{14}
\filler{11}{14}
\filler{12}{14}
\filler{0}{15}
\filler{1}{15}
\filler{2}{15}
\filler{3}{15}
\filler{4}{15}
\filler{5}{15}
\filler{6}{15}
\filler{7}{15}
\filler{12}{15}
\filler{0}{16}
\filler{1}{16}
\filler{2}{16}
\filler{3}{16}
\filler{4}{16}
\filler{5}{16}
\filler{6}{16}
\filler{7}{16}
\filler{12}{16}
\filler{0}{17}
\filler{1}{17}
\filler{2}{17}
\filler{3}{17}
\filler{4}{17}
\filler{5}{17}
\filler{6}{17}
\filler{7}{17}
\filler{12}{17}
\filler{0}{18}
\filler{1}{18}
\filler{2}{18}
\filler{3}{18}
\filler{4}{18}
\filler{5}{18}
\filler{6}{18}
\filler{10}{18}
\filler{11}{18}
\filler{12}{18}
\filler{0}{19}
\filler{1}{19}
\filler{2}{19}
\filler{3}{19}
\filler{4}{19}
\filler{5}{19}
\filler{6}{19}
\filler{10}{19}
\filler{11}{19}
\filler{12}{19}
\filler{0}{20}
\filler{1}{20}
\filler{2}{20}
\filler{3}{20}
\filler{4}{20}
\filler{5}{20}
\filler{6}{20}
\filler{9}{20}
\filler{10}{20}
\filler{11}{20}
\filler{12}{20}
\filler{0}{21}
\filler{1}{21}
\filler{2}{21}
\filler{3}{21}
\filler{4}{21}
\filler{5}{21}
\filler{6}{21}
\filler{9}{21}
\filler{10}{21}
\filler{11}{21}
\filler{12}{21}
\filler{0}{22}
\filler{1}{22}
\filler{2}{22}
\filler{3}{22}
\filler{4}{22}
\filler{5}{22}
\filler{6}{22}
\filler{9}{22}
\filler{10}{22}
\filler{11}{22}
\filler{12}{22}
\filler{0}{23}
\filler{1}{23}
\filler{2}{23}
\filler{3}{23}
\filler{4}{23}
\filler{5}{23}
\filler{6}{23}
\filler{10}{23}
\filler{11}{23}
\filler{12}{23}
\filler{10}{24}
\filler{11}{24}
\filler{12}{24}
\filler{9}{25}
\filler{10}{25}
\filler{11}{25}
\filler{12}{25}
\filler{0}{26}
\filler{1}{26}
\filler{2}{26}
\filler{3}{26}
\filler{4}{26}
\filler{5}{26}
\filler{6}{26}
\filler{7}{26}
\filler{8}{26}
\filler{9}{26}
\filler{10}{26}
\filler{11}{26}
\filler{12}{26}
     \begin{scope}[shift = {(0, -21)}]
        \clauseblock{0}{3}{1}{0}{0}{0}{2}
        \clauseblock{0}{5}{1}{1}{0}{3}{1}
        \clauseblock{0}{7}{1}{2}{0}{5}{0}
    \end{scope}
    \draw [draw = orange, uwidth = .25] (5.5, -1.5) -- (5.5, 26.5);
    \omino{5}{0}{1}
    \omino{5}{24}{1}
    \clauseorderpins
\foreach \x/\y/\n in {7/0/1, 5/0/2, 9/1/3, 10/3/4, 8/3/5, 8/5/6, 7/7/7, 7/9/8,
                         8/11/9, 6/12/10, 9/13/11, 10/15/12, 8/15/13, 8/17/14, 7/19/15, 7/21/16,
                         7/23/17, 5/24/18}{
        \node [order] at (\x + .5, \y + .5) {\n};
    }
  \end{tikzpicture}
  \caption{Phase 1: a clause gadget.}
  \label{fig:order-1}
\end{subfigure}
\hfill
\begin{subfigure}[t]{0.32\linewidth}
  \centering
  \begin{tikzpicture}[x = .28cm, y = .28cm]
    \begin{scope}
      \clip (-4.5, -3.5) rectangle (12.5, 4.5);
      \whitebg{-4}{12}{-3}{4}
\begin{scope}[shift = {(-6, -3)}]
        \filler{-5}{-3}
\filler{-5}{-2}
\filler{-5}{-1}
\filler{-5}{0}
\filler{-5}{1}
\filler{-5}{2}
\filler{-5}{3}
\filler{-5}{4}
\filler{-5}{5}
\filler{-5}{6}
\filler{-5}{7}
\filler{-5}{12}
\filler{-5}{13}
\filler{-5}{18}
\filler{-5}{19}
\filler{-5}{20}
\filler{-4}{-3}
\filler{-4}{-2}
\filler{-4}{-1}
\filler{-4}{0}
\filler{-4}{5}
\filler{-4}{6}
\filler{-4}{12}
\filler{-4}{18}
\filler{-4}{19}
\filler{-4}{20}
\filler{-3}{-3}
\filler{-3}{20}
\filler{3}{-3}
\filler{3}{20}
\filler{4}{-3}
\filler{4}{-2}
\filler{4}{-1}
\filler{4}{0}
\filler{4}{6}
\filler{4}{11}
\filler{4}{12}
\filler{4}{18}
\filler{4}{19}
\filler{4}{20}
\filler{5}{-3}
\filler{5}{-2}
\filler{5}{-1}
\filler{5}{0}
\filler{5}{1}
\filler{5}{6}
\filler{5}{7}
\filler{5}{8}
\filler{5}{9}
\filler{5}{10}
\filler{5}{11}
\filler{5}{12}
\filler{5}{13}
\filler{5}{18}
\filler{5}{19}
\filler{5}{20}
         \foreach \y in {-3, -1, ..., 19}{ \foreach \x in {-2, 0, 2}{ \legendpin{\x}{\y} } }
        \foreach \y in {3, 9, 15}{ \foreach \x in {-4, 4}{ \legendpin{\x}{\y} } }
        \varbot{0}{0}{2}
        \varblock{0}{0}{0}{1}{2}
        \varblock{0}{6}{1}{0}{2}
        \varblock{0}{12}{1}{1}{2}
        \vartop{0}{18}{2}
      \end{scope}
      \filler{0}{-3} \filler{1}{-3} \filler{2}{-3} \filler{3}{-3} \filler{4}{-3} \filler{5}{-3} \filler{6}{-3} \filler{7}{-3}
      \filler{8}{-3} \filler{9}{-3} \filler{0}{-2} \filler{1}{-2} \filler{2}{-2} \filler{3}{-2} \filler{4}{-2} \filler{5}{-2}
      \filler{6}{-2} \filler{7}{-2} \filler{8}{-2} \filler{9}{-2} \filler{0}{-1} \filler{1}{-1} \filler{2}{-1} \filler{3}{-1}
      \filler{4}{-1} \filler{5}{-1} \filler{6}{-1} \filler{7}{-1} \filler{8}{-1} \filler{9}{-1} \filler{0}{2} \filler{1}{2}
      \filler{2}{2} \filler{3}{2} \filler{4}{2} \filler{5}{2} \filler{6}{2} \filler{7}{2} \filler{8}{2} \filler{9}{2}
      \filler{0}{3} \filler{1}{3} \filler{2}{3} \filler{3}{3} \filler{4}{3} \filler{5}{3} \filler{6}{3} \filler{7}{3}
      \filler{8}{3} \filler{9}{3} \filler{0}{4} \filler{1}{4} \filler{2}{4} \filler{3}{4} \filler{4}{4} \filler{5}{4}
      \filler{6}{4} \filler{7}{4} \filler{8}{4} \filler{9}{4}
\begin{scope}[shift = {(4, -12)}]
        \filler{0}{-1}
\filler{1}{-1}
\filler{2}{-1}
\filler{3}{-1}
\filler{4}{-1}
\filler{5}{-1}
\filler{6}{-1}
\filler{7}{-1}
\filler{8}{-1}
\filler{9}{-1}
\filler{10}{-1}
\filler{11}{-1}
\filler{12}{-1}
\filler{10}{0}
\filler{11}{0}
\filler{12}{0}
\filler{11}{1}
\filler{12}{1}
\filler{0}{2}
\filler{1}{2}
\filler{2}{2}
\filler{3}{2}
\filler{4}{2}
\filler{5}{2}
\filler{6}{2}
\filler{7}{2}
\filler{8}{2}
\filler{11}{2}
\filler{12}{2}
\filler{0}{3}
\filler{1}{3}
\filler{2}{3}
\filler{3}{3}
\filler{4}{3}
\filler{5}{3}
\filler{6}{3}
\filler{7}{3}
\filler{12}{3}
\filler{0}{4}
\filler{1}{4}
\filler{2}{4}
\filler{3}{4}
\filler{4}{4}
\filler{5}{4}
\filler{6}{4}
\filler{7}{4}
\filler{12}{4}
\filler{0}{5}
\filler{1}{5}
\filler{2}{5}
\filler{3}{5}
\filler{4}{5}
\filler{5}{5}
\filler{6}{5}
\filler{7}{5}
\filler{12}{5}
\filler{0}{6}
\filler{1}{6}
\filler{2}{6}
\filler{3}{6}
\filler{4}{6}
\filler{5}{6}
\filler{6}{6}
\filler{10}{6}
\filler{11}{6}
\filler{12}{6}
\filler{0}{7}
\filler{1}{7}
\filler{2}{7}
\filler{3}{7}
\filler{4}{7}
\filler{5}{7}
\filler{6}{7}
\filler{10}{7}
\filler{11}{7}
\filler{12}{7}
\filler{0}{8}
\filler{1}{8}
\filler{2}{8}
\filler{3}{8}
\filler{4}{8}
\filler{5}{8}
\filler{6}{8}
\filler{9}{8}
\filler{10}{8}
\filler{11}{8}
\filler{12}{8}
\filler{0}{9}
\filler{1}{9}
\filler{2}{9}
\filler{3}{9}
\filler{4}{9}
\filler{5}{9}
\filler{6}{9}
\filler{9}{9}
\filler{10}{9}
\filler{11}{9}
\filler{12}{9}
\filler{0}{10}
\filler{1}{10}
\filler{2}{10}
\filler{3}{10}
\filler{4}{10}
\filler{5}{10}
\filler{6}{10}
\filler{9}{10}
\filler{10}{10}
\filler{11}{10}
\filler{12}{10}
\filler{0}{11}
\filler{1}{11}
\filler{2}{11}
\filler{3}{11}
\filler{4}{11}
\filler{5}{11}
\filler{6}{11}
\filler{10}{11}
\filler{11}{11}
\filler{12}{11}
\filler{10}{12}
\filler{11}{12}
\filler{12}{12}
\filler{11}{13}
\filler{12}{13}
\filler{0}{14}
\filler{1}{14}
\filler{2}{14}
\filler{3}{14}
\filler{4}{14}
\filler{5}{14}
\filler{6}{14}
\filler{7}{14}
\filler{8}{14}
\filler{11}{14}
\filler{12}{14}
\filler{0}{15}
\filler{1}{15}
\filler{2}{15}
\filler{3}{15}
\filler{4}{15}
\filler{5}{15}
\filler{6}{15}
\filler{7}{15}
\filler{12}{15}
\filler{0}{16}
\filler{1}{16}
\filler{2}{16}
\filler{3}{16}
\filler{4}{16}
\filler{5}{16}
\filler{6}{16}
\filler{7}{16}
\filler{12}{16}
\filler{0}{17}
\filler{1}{17}
\filler{2}{17}
\filler{3}{17}
\filler{4}{17}
\filler{5}{17}
\filler{6}{17}
\filler{7}{17}
\filler{12}{17}
\filler{0}{18}
\filler{1}{18}
\filler{2}{18}
\filler{3}{18}
\filler{4}{18}
\filler{5}{18}
\filler{6}{18}
\filler{10}{18}
\filler{11}{18}
\filler{12}{18}
\filler{0}{19}
\filler{1}{19}
\filler{2}{19}
\filler{3}{19}
\filler{4}{19}
\filler{5}{19}
\filler{6}{19}
\filler{10}{19}
\filler{11}{19}
\filler{12}{19}
\filler{0}{20}
\filler{1}{20}
\filler{2}{20}
\filler{3}{20}
\filler{4}{20}
\filler{5}{20}
\filler{6}{20}
\filler{9}{20}
\filler{10}{20}
\filler{11}{20}
\filler{12}{20}
\filler{0}{21}
\filler{1}{21}
\filler{2}{21}
\filler{3}{21}
\filler{4}{21}
\filler{5}{21}
\filler{6}{21}
\filler{9}{21}
\filler{10}{21}
\filler{11}{21}
\filler{12}{21}
\filler{0}{22}
\filler{1}{22}
\filler{2}{22}
\filler{3}{22}
\filler{4}{22}
\filler{5}{22}
\filler{6}{22}
\filler{9}{22}
\filler{10}{22}
\filler{11}{22}
\filler{12}{22}
\filler{0}{23}
\filler{1}{23}
\filler{2}{23}
\filler{3}{23}
\filler{4}{23}
\filler{5}{23}
\filler{6}{23}
\filler{10}{23}
\filler{11}{23}
\filler{12}{23}
\filler{10}{24}
\filler{11}{24}
\filler{12}{24}
\filler{9}{25}
\filler{10}{25}
\filler{11}{25}
\filler{12}{25}
\filler{0}{26}
\filler{1}{26}
\filler{2}{26}
\filler{3}{26}
\filler{4}{26}
\filler{5}{26}
\filler{6}{26}
\filler{7}{26}
\filler{8}{26}
\filler{9}{26}
\filler{10}{26}
\filler{11}{26}
\filler{12}{26}
       \end{scope}
      \begin{scope}[shift = {(4, -33)}]
        \clauseblock{0}{5}{1}{1}{0}{3}{0}
        \clauseblock{0}{7}{1}{2}{0}{5}{0}
      \end{scope}
      \foreach \x in {-1, 1, ..., 7}{ \omino{\x}{0}{1} }
      \draw [draw = orange, uwidth = .25] (-.5, -3.5) -- (-.5, 4.5);
      \draw [draw = orange, uwidth = .25] (9.5, -3.5) -- (9.5, 4.5);
      \omino{-1}{0}{1}
      \omino{9}{0}{1}
      \foreach \x in {0, 2, ..., 12}{ \legendpin{\x}{0} }
      \legendpin{12}{-2}
      \legendpin{12}{4}
\foreach \x/\y/\n in {7/0/1, 5/0/2, 3/0/3, 1/0/4, -1/0/5}{
          \node [order] at (\x + .5, \y + .5) {\n};
      }
    \end{scope}
\end{tikzpicture}
  \caption{Phase 2: a wire.}
  \label{fig:order-2}
\end{subfigure}
\hfill
\begin{subfigure}[t]{0.32\linewidth}
  \centering
  \begin{tikzpicture}[x = .28cm, y = .28cm]
    \clip (-8.5, -3.5) rectangle (7.5, 20.5);
    \whitebg{-8}{7}{-3}{20}
    \filler{-5}{-3}
\filler{-5}{-2}
\filler{-5}{-1}
\filler{-5}{0}
\filler{-5}{1}
\filler{-5}{2}
\filler{-5}{3}
\filler{-5}{4}
\filler{-5}{5}
\filler{-5}{6}
\filler{-5}{7}
\filler{-5}{12}
\filler{-5}{13}
\filler{-5}{18}
\filler{-5}{19}
\filler{-5}{20}
\filler{-4}{-3}
\filler{-4}{-2}
\filler{-4}{-1}
\filler{-4}{0}
\filler{-4}{5}
\filler{-4}{6}
\filler{-4}{12}
\filler{-4}{18}
\filler{-4}{19}
\filler{-4}{20}
\filler{-3}{-3}
\filler{-3}{20}
\filler{3}{-3}
\filler{3}{20}
\filler{4}{-3}
\filler{4}{-2}
\filler{4}{-1}
\filler{4}{0}
\filler{4}{6}
\filler{4}{11}
\filler{4}{12}
\filler{4}{18}
\filler{4}{19}
\filler{4}{20}
\filler{5}{-3}
\filler{5}{-2}
\filler{5}{-1}
\filler{5}{0}
\filler{5}{1}
\filler{5}{6}
\filler{5}{7}
\filler{5}{8}
\filler{5}{9}
\filler{5}{10}
\filler{5}{11}
\filler{5}{12}
\filler{5}{13}
\filler{5}{18}
\filler{5}{19}
\filler{5}{20}
     \varbot{0}{0}{1}
    \varblock{0}{0}{0}{1}{1}
    \varblock{0}{6}{1}{0}{1}
    \varblock{0}{12}{1}{1}{1}
    \vartop{0}{18}{1}
\filler{-8}{-3} \filler{-7}{-3} \filler{-6}{-3} \filler{6}{-3} \filler{7}{-3} \filler{-8}{-2} \filler{-7}{-2} \filler{-6}{-2}
    \filler{6}{-2} \filler{7}{-2} \filler{-8}{-1} \filler{-7}{-1} \filler{-6}{-1} \filler{6}{-1} \filler{7}{-1} \filler{-8}{0}
    \filler{-7}{0} \filler{-6}{0} \filler{6}{0} \filler{7}{0} \filler{-8}{1} \filler{-7}{1} \filler{-6}{1} \filler{6}{1}
    \filler{7}{1} \filler{-8}{2} \filler{-7}{2} \filler{-6}{2} \filler{6}{2} \filler{7}{2} \filler{-8}{3} \filler{-7}{3}
    \filler{-6}{3} \filler{-8}{4} \filler{-7}{4} \filler{-6}{4} \filler{-8}{5} \filler{-7}{5} \filler{-6}{5} \filler{6}{5}
    \filler{7}{5} \filler{-8}{6} \filler{-7}{6} \filler{-6}{6} \filler{6}{6} \filler{7}{6} \filler{-8}{7} \filler{-7}{7}
    \filler{-6}{7} \filler{6}{7} \filler{7}{7} \filler{-8}{8} \filler{-7}{8} \filler{-6}{8} \filler{6}{8} \filler{7}{8}
    \filler{6}{9} \filler{7}{9} \filler{6}{10} \filler{7}{10} \filler{-8}{11} \filler{-7}{11} \filler{-6}{11} \filler{6}{11}
    \filler{7}{11} \filler{-8}{12} \filler{-7}{12} \filler{-6}{12} \filler{6}{12} \filler{7}{12} \filler{-8}{13} \filler{-7}{13}
    \filler{-6}{13} \filler{6}{13} \filler{7}{13} \filler{-8}{14} \filler{-7}{14} \filler{-6}{14} \filler{6}{14} \filler{7}{14}
    \filler{-8}{17} \filler{-7}{17} \filler{-6}{17} \filler{6}{17} \filler{7}{17} \filler{-8}{18} \filler{-7}{18} \filler{-6}{18}
    \filler{6}{18} \filler{7}{18} \filler{-8}{19} \filler{-7}{19} \filler{-6}{19} \filler{6}{19} \filler{7}{19} \filler{-8}{20}
    \filler{-7}{20} \filler{-6}{20} \filler{6}{20} \filler{7}{20}
    \foreach \y in {9, 15}{ \omino{-9}{\y}{1} \omino{-7}{\y}{1} }
    \foreach \y in {3, 15}{ \omino{5}{\y}{1} \omino{7}{\y}{1} }
    \foreach \y in {-3, -1, ..., 19}{
        \foreach \x in {-2, 0, 2}{ \legendpin{\x}{\y} }
    }
    \foreach \y in {3, 9, 15}{
        \foreach \x in {-4, 4}{ \legendpin{\x}{\y} }
    }
    \foreach \y in {9, 15}{ \legendpin{-8}{\y} \legendpin{-6}{\y} }
    \varbot{0}{0}{2}
    \varblock{0}{0}{0}{1}{2}
    \varblock{0}{6}{1}{0}{2}
    \varblock{0}{12}{1}{1}{2}
    \vartop{0}{18}{2}
    \draw [draw = orange, uwidth = .25] (-5.5, -3.5) -- (-5.5, 20.5);
    \draw [draw = orange, uwidth = .25] (5.5, -3.5) -- (5.5, 20.5);
    \foreach \y in {3, 15}{ \omino{5}{\y}{1} \legendpin{6}{\y} }
\foreach \x/\y/\n in {-1/-3/1, -2/-1/2, -3/1/3, -3/3/4, -2/5/5, -3/7/6, -5/9/7, -3/9/8,
                         -2/11/9, -3/13/10, -5/15/11, -3/15/12, -2/17/13, -1/1/14, -1/3/15, -1/7/16,
                         -1/9/17, -1/13/18, -1/15/19, 1/-2/20, 1/0/21, 3/3/22, 1/2/23, 1/4/24,
                         1/6/25, 1/8/26, 1/10/27, 1/12/28, 3/15/29, 1/14/30, 1/16/31, 1/18/32,
                         -1/19/33}{
        \node [order] at (\x + .5, \y + .5) {\n};
    }
  \end{tikzpicture}
  \caption{Phase 3: a variable gadget.}
  \label{fig:order-3}
\end{subfigure}
\caption{The order of placement, each panel showing the state at the end of its phase: \subref{fig:order-1} the clause gadget of Figure~\ref{fig:clause-parts}, \subref{fig:order-2} a wire, and \subref{fig:order-3} the variable gadget of Figure~\ref{fig:var} with its \True filling.
Beyond the orange lines, each panel shows parts of the neighbouring wires or gadgets; the pieces there without numbers were placed in an earlier phase.
The numbers give the order in which the pieces are placed within each panel; each piece reaches its position by moving left, right and down through the cells that are still empty.}
\label{fig:order}
\end{figure}
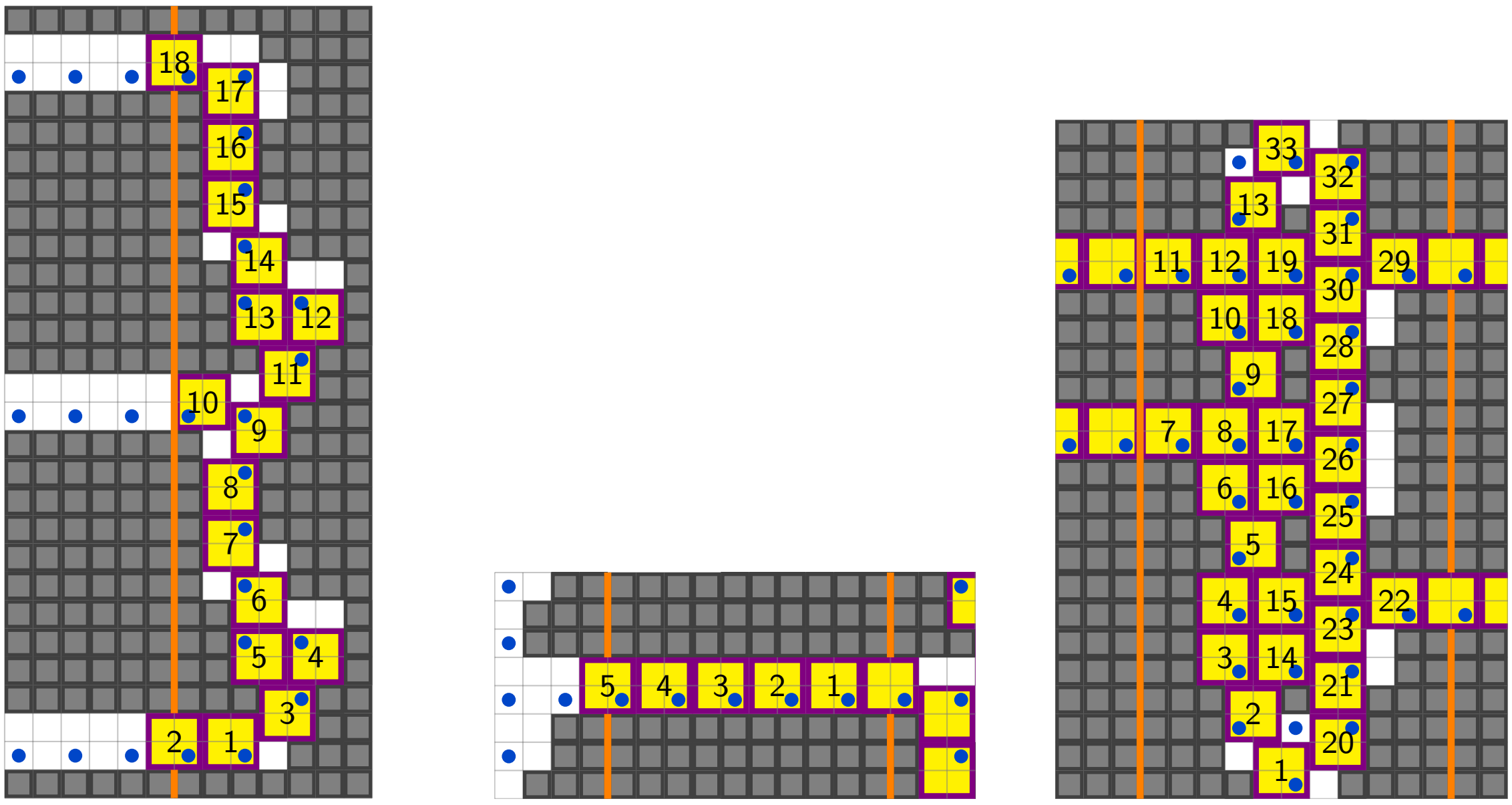

First we place the pieces covering the pins of the clause gadgets, each gadget from the bottom up (Figure~\ref{fig:order}\subref{fig:order-1}).
Meanwhile the variable pile and the wires are empty, so a piece can descend the pile to the height of any wire and follow the wire into its clause gadget.
The part of a module at or below its wire is reached through that wire, and the part above it through the wire of the module above, descending through the shaft between them; placing the pieces from the bottom up keeps this way open.
The top module of a gadget lies at or below its wire (Figure~\ref{fig:clause-parts}), so nothing is left out.
Next we place the pieces covering the pins of the wires, each wire from its clause end back towards the variable pile, which is still empty (Figure~\ref{fig:order}\subref{fig:order-2}).
Finally we place the pieces covering the pins of the variable pile, one gadget at a time from the bottom up, and within each gadget in the order of Figure~\ref{fig:order}\subref{fig:order-3}.
In every phase each piece comes to rest on a wall or on a piece already placed, and so it locks where it should.

This completes the proof of Theorem~\ref{thm:main}.

 \section{Open problems}
\label{sec:open}

The hardness proofs of~\cite{MBDHL26} for the other six piece types rely on the Super Rotation System, in particular on kicks that let a piece climb a staircase.
Our result holds under every conceivable rotation system:
the play we construct never rotates a piece, and the proof that a surviving play gives a satisfying assignment uses only packing arguments.
This brings into focus the question of which of the other six piece types remain hard when rotation is forbidden, or under a rotation system other than SRS.
Demaine et al.~\cite{DDE17} settle some cases for unrotatable $k$-ominoes, but not for each tetromino type on its own.

Two further questions lie outside the reach of our construction.
Clearing with only \ilo~pieces, the other open problem of~\cite{MBDHL26}, remains open, and so does survival with \ilo~pieces under other dropping models, such as hard drops only or $20$G~\cite{MDHL24}.
 \subsection*{Acknowledgement}
The authors used Anthropic Claude for assistance with English-language editing and figure drawing.
\bibliographystyle{plain}

\end{document}